\documentclass[vecphys]{svmult}

\usepackage{makeidx}         
\usepackage{graphicx}        
\usepackage{multicol}        
\usepackage[bottom]{footmisc}
\usepackage{amsmath}
\usepackage{exscale}
\usepackage{float}	
\usepackage{here}
\usepackage{textcomp}
\usepackage{epsfig}

\makeindex             

\begin{document}

\title*{Global Properties of Nucleus-Nucleus Collisions}
\author{Michael Kliemant\inst{1}\and
Raghunath Sahoo\inst{2}\and
Tim Schuster\inst{1}\and
Reinhard Stock\inst{1}}
\institute{Goethe-Universit\"{a}t Frankfurt, Germany
\texttt{stock@ikf.uni-frankfurt.de}
\and SUBATECH-Ecole des Mines de Nantes, France \texttt{raghu@subatech.in2p3.fr}}
%
%
\maketitle

\section{Introduction}
\label{chap:Introduction} 

QCD as a theory of extended, strongly interacting matter is familiar from big bang evolution which, within the time interval from electro-weak
decoupling ($10^{-12}$ s) to hadron formation ($5\cdot10^{-6}$ s), is dominated by the expansion of quark-gluon matter, a color conducting plasma that is deconfined. In the 1970's deconfinement was predicted
\cite{36,2,37} to arise from the newly discovered ``asymptotic freedom''
property of QCD; i.e.\ the plasma was expected to be a solution of perturbative QCD at asymptotically high square-momentum transfer $Q^2$, or temperature $T$. Thus the Quark-Gluon-Plasma (QGP) was seen as a dilute gas
of weakly coupled partons.
This picture may well hold true at temperatures in the GeV to TeV range.
However it was also known since R. Hagedorns work \cite{38} that hadronic matter features a phase boundary at a very much lower temperature, $T(H) = 170$ MeV. As it was tempting to identify this temperature with that of the cosmological hadronization transition, thus suggesting $T(H) = T(\mathrm{crit})$, the
QGP state must extend downward to such a low temperature, with $Q^2 << 1 \mathrm{GeV}^2$, and far into the non-perturbative sector of QCD, and very far from asymptotic freedom.
The fact that, therefore, the confinement-deconfinement physics of QCD,
occuring at the parton-hadron phase boundary, had to be explained in terms other than a dilute perturbative parton gas, was largely ignored until rather recently, when laboratory experiments concerning of the QGP had reached maturity.

In order to recreate matter at the corresponding high
energy density in the terrestial laboratory one collides heavy
nuclei (also called ``heavy ions") at ultrarelativistic energies. Quantum Chromodynamics
predicts \cite{2,3,4} a phase transformation to occur between
deconfined quarks and confined hadrons. At near-zero net baryon
density (corresponding to big bang conditions) non-perturbative
Lattice-QCD places this transition at an energy density of about $1
\: GeV/fm^3$, and at a critical temperature, $T_{crit}$ $\approx$ $170 \: MeV$
\cite{4,5,6,7,8}. The ultimate goal of the physics with
ultrarelativistic heavy ions is to locate this transition, elaborate
its properties, and gain insight into the detailed nature of the
deconfined QGP phase that should exist above. What is meant by the
term ''ultrarelativistic'' is defined by the requirement that the
reaction dynamics reaches or exceeds the critical density $\epsilon
\approx 1 \: GeV/fm^3$. Required beam energies turn out \cite{8} to
be $\sqrt{s} \ge 10\: GeV$, and various experimental programs have
been carried out or are being prepared at the CERN SPS (up to about
$20 \: GeV$), at the BNL RHIC collider (up to $200 \:GeV)$ and
finally reaching up to $5.5 \:TeV$ at the LHC of CERN.

QCD confinement-deconfinement is of course not limited to the domain
that is relevant to cosmological expansion dynamics, at very
small excess of baryon over anti-baryon number density and, thus,
near zero baryo-chemical potential $\mu_B$. In fact, modern QCD
suggests \cite{9,10,11} a detailed phase diagram of QCD matter and
its states, in the plane of $T$ and baryo-chemical potential $\mu_B$. 
For a map of the QCD matter phase diagram we are thus employing the terminology of the grand canonical Gibbs ensemble that describes an extended volume $V$ of partonic or hadronic matter at temperature $T$. 
In it, total particle number is not conserved at relativistic energy, due to particle production-annihilation processes occurring at the microscopic level. However, the probability distributions (partition functions)
describing the particle species abundances have to respect the presence of certain, to be conserved net quantum numbers ($i$), notably non-zero net baryon number and zero net strangeness and charm.
Their global conservation is achieved by a thermodynamic trick, adding to the system Lagrangian a so-called Lagrange multiplier term, for each of such quantum number conservation tasks. This procedure enters a "chemical potential" $\mu_i$ that modifies the partition function via an extra term $\exp{\left(-\mu_i/T\right)}$ occuring in the phase space integral (see section~\ref{chap:hadronization} for detail). It modifies the canonical "punishment factor" $\exp{\left(-E/T\right)}$, where
$E$ is the total particle energy in vacuum, to arrive at an analogous grand canonical factor for the extended medium,of $\exp{\left(-E/T - \mu_i/T\right)}$. 
This concept is of prime importance for a description of the state of matter created in heavy ion collisions, where net-baryon number (valence quarks) carrying objects are considered --- extended "fireballs" of QCD matter.
The same applies to the matter in the interior of neutron stars. The corresponding conservation of net baryon number is introduced into the grand canonical statistical model of QCD matter via the "baryo-chemical potential" $\mu_B$.

\begin{figure}[h]   
\begin{center}
\includegraphics[scale=0.5]{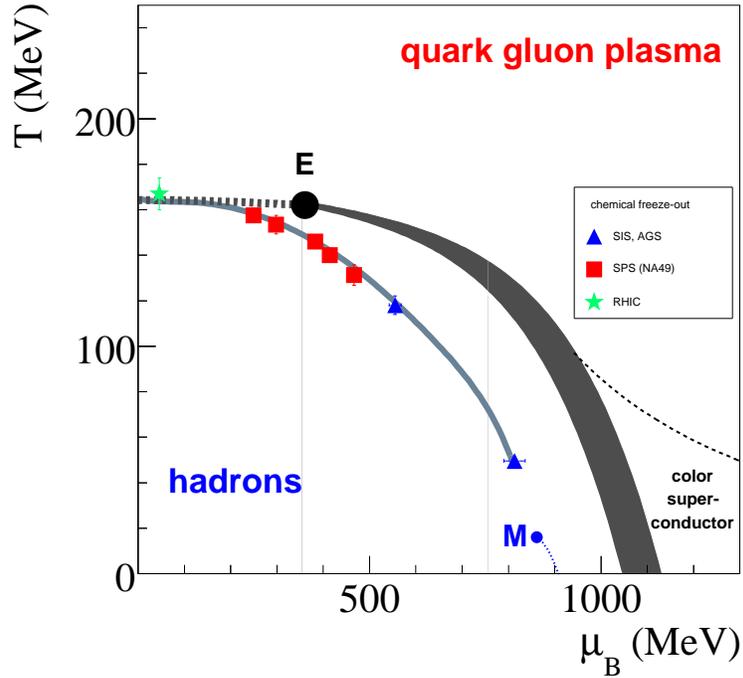}
\caption{Sketch of the QCD matter phase diagram in the plane of temperature
$T$ and baryo-chemical potential $\mu_B$. The parton-hadron phase
transition line from lattice QCD  \cite{8,9,10,11} ends in a
critical point $E$. A cross-over transition occurs at smaller $\mu_B$. Also shown are the points of hadro-chemical
freeze-out from the grand canonical statistical model.}
\label{fig:Figure1}
\end{center}
\end{figure}

We employ this terminology to draw a phase diagram of QCD matter in figure~\ref{fig:Figure1}, in the variables $T$ and $\mu_B$. Note that $\mu_B$ is high at low energies of collisions creating a matter fireball. In a head-on collision
of two mass 200 nuclei at $\sqrt{s}=15 GeV$ the fireball contains about
equal numbers of newly created quark-antiquark pairs (of zero net baryon number), and of initial valence quarks. The accomodation of the latter,
into created hadronic species, thus requires a formidable redistribution task of net baryon number, reflecting in a high value of $\mu_B$. Conversely, at LHC energy ($\sqrt{s}$=5.5TeV in Pb+Pb collisions), the initial valence quarks constitute a mere 5\% fraction of the total quark density, correspondingly requiring a small value of $\mu_B$. In the extreme, big bang matter evolves toward hadronization (at $T$=170 MeV) featuring a quark over antiquark density excess of $10^{-9}$ only, resulting in $\mu_B \approx 0$.

Note that the limits of
existence of the hadronic phase are not only reached by temperature
increase, to the so-called Hagedorn value $T_H$ (which coincides
with $T_{crit}$ at $\mu_B \rightarrow 0$), but also by density
increase to $\varrho > (5-10)\: \varrho_0$: ''cold compression''
beyond the nuclear matter ground state baryon density $\varrho_0$ of
about 0.16 $B/fm^3$. We are talking about the deep interior sections
of neutron stars or about neutron star mergers \cite{12,13,14}. A
sketch of the present view of the QCD phase diagram \cite{9,10,11}
is given in Fig.~\ref{fig:Figure1}. It is dominated by the parton-hadron phase
transition line that interpolates smoothly between the extremes of
predominant matter heating (high $T$, low $\mu_B$) and predominant
matter compression ($T \rightarrow 0, \: \mu_B > 1 \: GeV$). Onward
from the latter conditions, the transition is expected to be of
first order \cite{15} until the critical point of QCD matter is
reached at $200 \le \mu_B \: (E)\: \le 500 \: MeV$. The relatively
large position uncertainty reflects the preliminary character of
Lattice QCD calculations at finite $\mu_B$ \cite{9,10,11}. Onward from the critical point, E, the phase transformation at lower $\mu_B$ is a cross-over\cite{11}.\\
 \\
We note, however, that these estimates represent a major recent
advance of QCD lattice theory which was, for two decades, believed to be
restricted to the $\mu_B=0$ situation. Onward from the critical
point, toward lower $\mu_B$, the phase transformation should acquire
the properties of a rapid cross-over \cite{16}, thus also including
the case of primordial cosmological expansion. This would finally
rule out former ideas, based on the picture of a violent first order
''explosive'' cosmological hadronization phase transition, that
might have caused non-homogeneous conditions, prevailing during
early nucleo-synthesis \cite{17}, and fluctuations of global matter
distribution density that could have served as seedlings of galactic
cluster formation \cite{18}. However, it needs to be stressed that
the conjectured order of phase transformation, occuring along the
parton - hadron phase boundary line, has not been unambiguously
confirmed by experiment, as of now.

On the other hand, the {\it position} of the QCD phase boundary at
low $\mu_B$ has, in fact, been located by the hadronization points
in the $T, \: \mu_B$ plane that are also illustrated in Fig.~\ref{fig:Figure1}. They
are obtained from statistical model analysis \cite{19} of the
various hadron multiplicities created in nucleus-nucleus collisions, which results in a [$T, \: \mu_B$]
determination at each incident energy, which ranges from SIS
via AGS and SPS to RHIC energies, i.e. $3\le \sqrt{s} \le 200 \:
GeV$. Toward low $\mu_B$ these hadronic freeze-out points merge with
the lattice QCD parton-hadron coexistence line: hadron formation
coincides with hadronic species freeze-out. These points also
indicate the $\mu_B$ domain of the phase diagram which is accessible
to relativistic nuclear collisions. The domain at $\mu_B \ge 1.5 \:
GeV$ which is predicted to be in a further new phase of QCD
featuring color-flavor locking and color superconductivity \cite{20}
will probably be accessible only to astrophysical observation.

One may wonder how states and phases of matter in
thermodynamical equilibrium - as implied by a description in grand
canonical variables - can be sampled via the dynamical evolution of
relativistic nuclear collisions. Employing heavy nuclei, $A  \approx
200$, as projectiles/targets or in colliding beams (RHIC, LHC),
transverse dimensions of the primordial interaction volume do not
exceed about $8 \: fm$, and strong interaction ceases after about
$20 \: fm/c$. We note, for now, that the time and
dimension scale of primordial perturbative QCD interaction at the
microscopic partonic level amounts to subfractions of $1 \: fm/c$,
the latter scale, however, being representative of non perturbative
processes (confinement, ''string'' formation etc.). The A+A
 fireball size thus exceeds, by far, the elementary non perturbative scale. 
An equilibrium quark gluon plasma represents an extended non-perturbative
QCD object, and the question whether its relaxation time scale can be
provided by the expansion time scale of an A+A collision, needs
careful examination. Reassuringly, however, the hadrons that are
supposedly created from such a preceding non-perturbative QGP phase
at top SPS and RHIC energy, do in fact exhibit perfect hydrodynamic and hadrochemical
equilibrium, the derived [$T, \: \mu_B$] values \cite{19} thus
legitimately appearing in the phase diagram, Fig.~\ref{fig:Figure1}.

In the present book we will order the physics observables to be
treated, with regard to their origin from successive stages that
characterize the overall dynamical evolution of a relativistic
nucleus-nucleus collision. In rough outline this evolution can be
seen to proceed in three major steps. An initial period of matter
compression and heating occurs in the course of interpenetration of
the projectile and target baryon density distributions. Inelastic
processes occuring at the microscopic level convert initial beam
longitudinal energy to new internal and transverse degrees of
freedom, by breaking up the initial baryon structure functions.
Their partons thus acquire virtual mass, populating transverse phase
space in the course of inelastic perturbative QCD shower
multiplication. This stage should be far from thermal equilibrium,
initially. However, in step two, inelastic interaction between the
two arising parton fields (opposing each other in longitudinal phase
space) should lead to a pile-up of partonic energy density centered
at mid-rapidity (the longitudinal coordinate of the overall center
of mass). Due to this mutual stopping down of the initial target and
projectile parton fragmentation showers, and from the concurrent
decrease of parton virtuality (with decreasing average square
momentum transfer $Q^2$) there results a slowdown of the time scales
governing the dynamical evolution. Equilibrium could be approached
here, the system ''lands'' on the $T, \: \mu$ plane of Fig.~\ref{fig:Figure1}, at
temperatures of about 300 and $200 \: MeV$ at top RHIC and top SPS
energy, respectively. The third step, system expansion and decay,
thus occurs from well above the QCD parton-hadron boundary line.
Hadrons and hadronic resonances then form, which decouple swiftly
from further inelastic transmutation so that their yield ratios
become stationary (''frozen-out''). A final expansion period dilutes
the system to a degree such that strong interaction ceases all
together.

It is important to note that the above description, in terms of successive global stages of evolution, is only valid at very high energy, e.g. at and above top RHIC energy of $\sqrt{s}=200$GeV. At this energy the target-projectile interpenetration time $2R/\gamma=0.12\mathrm{fm}/c$, and thus the interpenetration phase is over when the supposed next phase (perturbative QCD shower formation at the partonic level by primordial,
"hard" parton scattering) settles, at about $0.25\mathrm{fm}/c$. "Hard" observables (heavy flavour production, jets, high $p_T$ hadrons) all originate from this primordial interaction phase. On the other hand it is important to realize that at top SPS energy, $\sqrt{s}=17.3$GeV, global interpenetration takes as long as 1.5fm/$c$, much longer than microscopic shower formation time. There is thus no global, distinguishable phase of hard QCD mechanisms: they are convoluted with the much longer interpenetration time. During that it is thus impossible to consider a global physics of the interaction volume, or any equilibrium. Thus we can think of the dynamical evolution in terms of global "states" of the system's dynamical evolution (such as local or global equilibrium) only after about 2-3fm/$c$, just before bulk
hadronization sets in. Whereas at RHIC, and even more ideally so at the LHC, the total interaction volume is "synchronized" at times below 0.5fm/$c$, such that a hydrodynamic description becomes possible: we can expect that "flow" of partons sets in at this time, characterized by extremely high parton density. The dynamics at such early time can thus be accessed in well defined variables (e.g. elliptic flow or jet quenching).

In order to verify in detail this qualitative overall model, and to
ascertain the existence (and to study the properties) of the
different states of QCD that are populated in sequence, one seeks
observable physics quantities that convey information imprinted
during distinct stages of the dynamical evolution, and
''freezing-out'' without significant obliteration by subsequent
stages. Ordered in sequence of their formation in the course of the
dynamics, the most relevant such observables are briefly
characterized below:
\begin{enumerate}
\item Suppression of $J/\Psi$ and $Y$ production by Debye-screening
in the QGP. These vector mesons result from primordial, pQCD
production of $c\overline{c}$ and $b\overline{b}$ pairs that would
hadronize unimpeded in elementary collisions but are broken up if
immersed into a npQCD deconfined QGP, at certain characteristic
temperature thresholds.
\item Suppression of dijets which arise from primordial
$q\overline{q}$ pair production fragmenting into partonic showers
(jets) in vacuum but being attenuated by QGP-medium induced gluonic
bremsstrahlung: Jet quenching in A+A collisions.
\begin{enumerate}
\item A variant of this: {\it any} primordial hard parton suffers a
high, specific loss of energy when traversing a deconfined medium:
High $p_T$ suppression in A+A collisions.
\end{enumerate}
\item Hydrodynamic collective motion develops with the onset of
(local) thermal equilibrium. It is created by partonic pressure
gradients that reflect the initial collisional impact geometry via
non-isotropies in particle emission called ''directed'' and
''elliptic'' flow. The latter reveals properties of the QGP, seen
here as an ideal partonic fluid.
\begin{enumerate}
\item Radial hydrodynamical expansion flow (''Hubble expansion'')
is a variant of the above that occurs in central, head on collisions
with cylinder symmetry, as a consequence of an isentropic expansion.
It should be sensitive to the mixed phase conditions characteristic
of a first order parton-hadron phase transition.
\end{enumerate}
\item Hadronic ''chemical'' freeze-out fixes the abundance ratios of
the hadronic species into an equilibrium distribution. Occuring very
close to, or at hadronization, it reveals the dynamical evolution
path in the [$T, \:\mu_B$] plane and determines the critical
temperature and density of QCD. The yield distributions in A+A
collisions show a dramatic strangeness enhancement effect,
characteristic of an extended QCD medium.
\item Fluctuations, from one collision event to another (and even
within a single given event) can be quantified in A+A collisions due
to the high charged hadron multiplicity density (of up to 600 per
rapidity unit at top RHIC energy). Such event-by-event (ebye)
fluctuations of pion rapidity density and mean transverse momentum
(event ''temperature''), as well as event-wise fluctuations of the
strange to non-strange hadron abundance ratio (may) reflect the
existence and position of the conjectured critical point of QCD
(Fig.~\ref{fig:Figure1}).
\item Two particle Bose-Einstein-Correlations are the analog of the
Hanbury-Brown, Twiss (HBT) effect of quantum optics. They result
from the last interaction experienced by mesons, i.e. from the
global decoupling stage. Owing to a near isentropic hadronic
expansion they reveal information on the overall
space-time-development  of the ''fireball'' evolution.
\end{enumerate}

In an overall view the first group of observables (1 to 2a) is
anchored in established pQCD physics that is well known from
theoretical and experimental analysis of elementary collisions
($e^+e^-$ annihilation, $pp$ and $p\overline{p}$ data). In fact, the
first generation of high $Q^2$ baryon collisions, occuring at the
microscopic level in A+A collisions, should closely resemble such
processes. However, their primary partonic products do not escape
into pQCD vacuum but get attenuated by interaction with the
concurrently developing extended high density medium , thus serving
as diagnostic tracer probes of that state. The remaining observables
capture snapshots of the bulk matter medium itself. After initial
equilibration we may confront elliptic flow data with QCD during the
corresponding partonic phase of the dynamical evolution employing 
thermodynamic \cite{21} and hydrodynamic \cite{22}
models of a high temperature parton plasma. The hydro-model stays
applicable well into the hadronic phase. Hadron formation
(confinement) occurs in between these phases (at about 5
microseconds time in the cosmological evolution). In fact
relativistic nuclear collision data may help to finally pin down the
mechanism(s) of this fascinating QCD process \cite{23,24,25} as we
can vary the conditions of its occurence, along the parton-hadron
phase separation line of Fig.~\ref{fig:Figure1}, by proper choice of collisional
energy $\sqrt{s}$, and system size A, while maintaining the overall
conditions of an extended imbedding medium of high energy density
within which various patterns \cite{9,10,11,15,16} of the
hadronization phase transition may establish. The remaining physics
observables (3a, 5 and 6 above) essentially provide for auxiliary
information about the bulk matter system as it traverses (and
emerges from) the hadronization stage, with special emphasis placed
on manifestations of the conjectured critical point.

The observables from 1 to 4 above will all be treated, in detail, in this book. We shall focus here on the bulk matter expansion processes of the primordially formed collisional volume, as reflected globally in the population patterns of transverse and longitudinal (rapidity) phase space
(Section 3), and on the transition from partons to hadrons and on hadronic hadro-chemical decoupling, resulting in the observed abundance systematics of the hadronic species (Section 4).These Sections will be preceded by a detailed recapitulation of relativistic kinematics, notably rapidity, to which we shall turn now.

\section{Relativistic Kinematics}
\label{sec:1}

\subsection{Description of Nucleus-Nucleus Collisions in terms of Light-Cone 
Variables}

In relativistic nucleus-nucleus collisions, it is convenient to use kinematic
variables which take simple forms under Lorentz transformations for the change
of frame of reference. A few of them are the light cone variables $x_+$ and 
$x_-$, the rapidity and pseudorapidity variables, $y$ and $\eta$. A particle 
is characterized by its 4-momentum, $p_{\mu}=(E,{\bf p})$. In fixed target 
and collider experiments where the beam(s) define reference frames, boosted 
along their direction, it is important to express the 4-momentum in terms of 
more practical kinematic variables.

\begin{figure}
\centering
\includegraphics[height=8cm]{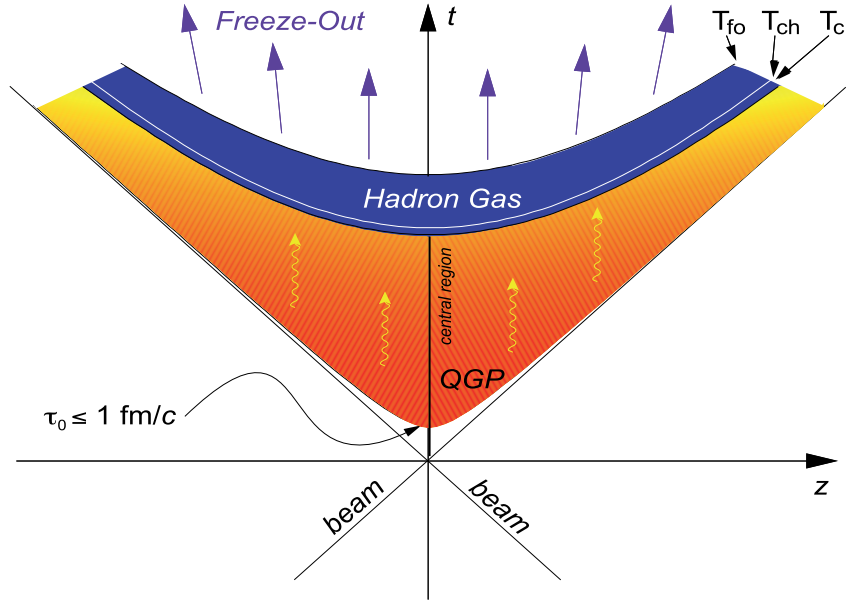}
%
\caption{Description of heavy-ion collisions in one space ($z$) and one time
($t$) dimension.}
\label{lightcone}
\end{figure}

Figure \ref{lightcone} shows the collision of two Lorentz contracted nuclei
approaching each other with velocities nearly equal to velocity of light.
The vertical axis represents the time direction with the lower half representing
time before the collision and the upper half, time after the collision. The
horizontal axis represents the spatial direction. Both the nuclei collide 
at $(t,z) = (0,0)$ and then the created fireball expands in time going through
various processes till the created particles freeze-out and reach the detectors.
The lines where $t^2 - z^2 ~=~ 0$ (note that $\sqrt{t^2 - z^2} ~\equiv~ \tau$, 
 $\tau$ being the proper time of the particle) along the path of the colliding 
nuclei define the light cone. The upper part of the light-cone, where 
$t^2 - z^2 ~ >~ 0$, is
the time-like region. In nucleus-nucleus collisions, particle production occurs
in the upper half of the $(t,z)$-plane within the light-cone. The region 
outside the light cone for which $t^2 - z^2 ~ <~ 0$ is called space-like region.
The space-time rapidity is defined as
\begin{equation}
\eta_s ~=~ \frac{1}{2}~ln\left(\frac{t+z}{t-z}\right)
\end{equation}
It could be seen that $\eta_s$ is not defined in the space-like region. It takes
the value of positive and negative infinity along the beam directions for which
$t = \pm z$ respectively. A particle is "light-like" along the beam direction. Inside the light-cone which
is time-like, $\eta_s$ is properly defined.

For a particle with 4-momentum $p~(p_0,{\bf p_T},p_z)$, the light-cone momenta
are defined by
\begin{eqnarray}
p_+ &=& p_0 + p_z\\
p_- &=& p_0 - p_z
\end{eqnarray}
$p_+$ is called ``{\it forward light-cone momentum}'' and $p_-$ is called
``{\it backward light-cone momentum}''.\\For a particle traveling along the
beam direction, has higher value of forward light-cone momentum and traveling
opposite to the beam direction has lower value of forward light-cone momentum.
The advantages of using light-cone variables to study particle production
are the following.\\
1. The forward light-cone momentum of any particle in one frame is related to
the forward light-cone momentum of the same particle in another boosted Lorentz
frame by a constant factor.\\
2. Hence, if a daughter particle $c$ is fragmenting from a parent particle
$b$, then the ratio of the forward light-cone momentum of $c$ relative to that
of $b$ is independent of the Lorentz frame.\\
Define
\begin{eqnarray}
x_+ &=& \frac{p_0^c + p_z^c}{p_0^b + p_z^b}\\
&=& \frac{c_+}{b_+} . \nonumber
\end{eqnarray}
The forward light-cone variable $x_+$ is always positive because $c_+$ can't be
greater than $b_+$. Hence the upper limit of $x_+$ is $1$. $x_+$ is Lorentz
invariant. \\
3. The Lorentz invariance of $x_+$ provides a tool to measure the momentum
of any particle in the scale of the momentum of any reference particle. 

\subsection{The Rapidity Variable}
The co-ordinates along the beam line (conventionally along the $z$-axis) is 
called {\it longitudinal} and perpendicular to it is called {\it transverse} 
({\it x-y}). The 3-momentum can be decomposed into the longitudinal ($p_z$) 
and the transverse (${\bf p}_T)$, ${\bf p}_T$ being a vector quantity which is 
invariant under a Lorentz boost along the longitudinal direction. The 
variable rapidity ``$y$'' is defined by
\begin{equation}
y = \frac{1}{2}ln\left(\frac{E+p_z}{E-p_z}\right).
\end{equation}
It is a dimensionless quantity related to the ratio of forward light-cone to 
backward light-cone momentum. The rapidity changes by an additive constant 
under longitudinal Lorentz boosts.
 
For a free particle which is on the mass shell (for which $E^2=p^2+m^2$), 
the 4-momentum has only three degrees of freedom and can be represented by 
$(y,{\bf p}_T)$. $(E,{\bf p}_T)$ could be expressed in terms of 
$(y,{\bf p}_T)$ as
\begin{eqnarray}
E &=& m_T ~ cosh ~y\\
p_z &=& m_T ~ sinh ~y
\end{eqnarray}
$m_T$ being the transverse mass which is defined as
\begin{equation}
m_T^2 = m^2+{\bf p}_T^2.
\end{equation}
The advantage of rapidity variable is that the shape of the rapidity 
distribution remains unchanged under a longitudinal Lorentz boost. When we go 
from CMS to LS, the rapidity distribution is the same, with the $y$-scale 
shifted by an amount equal to $y_{cm}$. This is shown below.

\subsubsection{Rapidity of Center of Mass in the Laboratory System}
The total energy in the CMS system is $E_{cm}=\sqrt{s}$. The energy and 
momentum of the CMS in the LS are $\gamma_{cm}\sqrt{s}$ and $\beta_{cm}
\gamma_{cm}\sqrt{s}$ respectively. The rapidity of the CMS in the LS is 
\begin{eqnarray}
y_{cm} &=& \frac{1}{2} ~ ln \left[\frac{\gamma_{cm}\sqrt{s} + \beta_{cm}
\gamma_{cm}\sqrt{s}}{\gamma_{cm}\sqrt{s} - \beta_{cm}\gamma_{cm}\sqrt{s}} 
\right]  \nonumber\\
&=& \frac{1}{2} ~ ln \left[\frac{1+\beta_{cm}}{1-\beta_{cm}}\right]
\end{eqnarray}
It is a constant for a particular Lorentz transformation.

\subsubsection{Relationship between Rapidity of a particle in LS and 
rapidity in CMS}
The rapidities of a particle in the LS and CMS of the collision are 
respectively,
$y = \frac{1}{2}~ln\left(\frac{E+p_z}{E-p_z}\right)$ 
and $y^* = \frac{1}{2}~ln\left(\frac{E^*+p_z^*}{E^*-p_z^*}\right)$. 
Inverse Lorentz transformations on $E$ and $p_z$ give
\begin{eqnarray}
y &=& \frac{1}{2} ~ ln \left[\frac{\gamma(E^*+\beta p_z^*) + 
\gamma(\beta E^*+p_z^*)}
{\gamma(E^*+\beta p_z^*) - \gamma(\beta E^*+p_z^*)}\right]  \nonumber\\
&=& \frac{1}{2} ~ ln \left[\frac{E^*+p_z^*}{E^*-p_z^*}\right] 
+ \frac{1}{2} ~ ln \left[\frac{1+\beta}{1-\beta}\right]\\
\Rightarrow y &=& y^* + y_{cm}.
\end{eqnarray}
Hence the rapidity of a particle in the laboratory system is equal to the sum 
of the rapidity of the particle in the center of mass system and the rapidity 
of the center of mass in the laboratory system. It can also be state that the 
rapidity of a particle in a moving (boosted) frame is equal to the rapidity 
in its own rest frame minus the rapidity of the moving frame. In the 
non-relativistic limit, this is like the subtraction of velocity of the
moving frame. However, this is not surprising because, non-relativistically, 
the rapidity $y$ is equal to longitudinal velocity $\beta$. Rapidity is a 
relativistic measure of the velocity. This simple property of 
the rapidity variable under Lorentz transformation makes it a suitable choice to 
describe the dynamics of relativistic particles.

\subsubsection{Relationship between Rapidity and Velocity}
Consider a particle traveling in $z$-direction with a longitudinal velocity 
$\beta$. The energy $E$ and the longitudinal momentum $p_z$ of the particle are 
\begin{eqnarray}
E &=& \gamma m\\
p_z &=& \gamma \beta m
\end{eqnarray}
where $m$ is the rest mass of the particle. Hence the rapidity of the particle 
traveling in $z$-direction with velocity $\beta$ is
\begin{eqnarray}
y_{\beta} &=& \frac{1}{2} ~ ln \left[\frac{E+p_z}{E-p_z}\right] 
= \frac{1}{2} ~ ln \left[\frac{\gamma m + \gamma \beta m}{\gamma m - \gamma 
\beta m}\right] \nonumber\\
&=& \frac{1}{2} ~ ln \left[\frac{1+\beta}{1-\beta}\right] 
\end{eqnarray} 
Note here that $y_{\beta}$ is independent of particle mass. In the non-relativistic 
limit when $\beta$ is small, expanding $y_{\beta}$ in terms of $\beta$ leads to
\begin{equation}
y_{\beta} = \beta + {\it O}(\beta^3)
\end{equation}
Thus the rapidity of the particle is the relativistic realization of its 
velocity.

\subsubsection{Beam Rapidity}
We know, \\
$E ~=~ m_T ~ cosh ~y$, $p_z ~=~m_T ~ sinh ~y$ and $m_T^2 = m^2+{\bf p}_T^2 $.\\
For the beam particles, $p_T = 0$. \\
Hence, $E ~=~ m_b ~ cosh ~y_b$ and $p_z ~=~ m_b ~ sinh ~y_b$,\\
where $m_b$ and $y_b$ are the rest mass and rapidity of the beam particles.
\begin{eqnarray}
y_b &=& cosh^{-1} ~(E/m_b) \nonumber \\
&=& cosh^{-1} ~\left[\frac{\sqrt{s_{NN}}}{2~ m_n}\right]
\end{eqnarray}
and
\begin{equation}
 y_b = sinh^{-1} ~(p_z/m_b)
\end{equation}
Here $m_n$ is the mass of the nucleon. Note that the beam energy 
$E =\sqrt{s_{NN}}/2$.\\

\begin{example}
For the nucleon-nucleon center of mass energy $\sqrt{s_{NN}} = 9.1$ GeV, 
the beam rapidity
$y_b ~=~ cosh^{-1} \left(\frac{9.1}{2 \times 0.938}\right)~=~2.26$\\
For p+p collisions with lab momentum 100 GeV/c,\\ 
$y_b ~=~ sinh^{-1} \left(\frac{p_z}{m_b}\right)
~=~sinh^{-1} \left(\frac{100}{0.938}\right)~=~5.36$\\
and for Pb+Pb collisions at SPS with lab energy 158 AGeV, $y_b ~=~ 2.92$.
\end{example}

\subsubsection{Rapidity of the CMS in terms of Projectile and Target Rapidities}
Let us consider the beam particle ``$b$'' and the target particle 
``$a$''.\\ $b_z ~=~ m_T~ sinh~y_b ~=~ m_b~sinh~y_b$. This is because $p_T$ 
of beam particles is zero. Hence
\begin{equation}
y_{b} ~=~ sinh^{-1}~ (b_z/m_b).
\end{equation}
The energy of the beam particle in the laboratory frame is \\
$b_0 ~=~m_T~cosh~y_b ~=~ m_b~cosh~y_b$.\\ 
Assuming target particle $a$ has longitudinal momentum $a_z$, its rapidity in the
laboratory frame is given by
\begin{equation}
y_{a} ~=~ sinh^{-1}~ (a_z/m_a)
\end{equation}
and its energy
\begin{equation}
a_{0} ~=~ m_a~cosh~y_a .
\end{equation}
The CMS is obtained by boosting the LS by  a velocity of the center-of-mass 
frame $\beta_{cm}$ such that the longitudinal momentum of the beam particle 
$b_z^*$ and of the target particle $a_z^*$ are equal and opposite. Hence 
$\beta_{cm}$ satisfies the condition,\\
$a_z^* ~=~ \gamma_{cm}(a_z-\beta_{cm}a_0) ~=~ -b_z^* ~=~ 
-\gamma_{cm}(b_z-\beta_{cm}b_0)$,
where $\gamma_{cm} ~=~ \frac{1}{\sqrt{1-\beta_{cm}^2}}$.
Hence,
\begin{equation}
\beta_{cm} ~=~ \frac{a_z+b_z}{a_0+b_0} .
\label{bcm}
\end{equation}
We know the rapidity of the center of mass is 
\begin{equation}
y_{cm} ~=~ \frac{1}{2}~ln \left[\frac{1+\beta_{cm}}{1-\beta_{cm}}\right]
\label{ycm}
\end{equation}
Using equations \ref{bcm} and \ref{ycm}, we get
\begin{equation}
y_{cm} ~=~ \frac{1}{2}~ln \left[\frac{a_0+a_z+b_0+b_z}{a_0-a_z+b_0-b_z}\right].
\end{equation}
Writing energies and momenta in terms of rapidity variables in the LS,
\begin{eqnarray}
y_{cm} &=& \frac{1}{2}~ln \left[\frac{m_a~cosh~y_a + m_a~sinh~y_a + 
    m_b~cosh~y_b + m_b~sinh~y_b}
  {m_a~cosh~y_a - m_a~sinh~y_a + m_b~cosh~y_b - m_b~sinh~y_b}\right] 
\nonumber \\
&=& \frac{1}{2} (y_a + y_b) ~+~ \frac{1}{2}~ln \left[ \frac{m_a~e^{y_a} + 
    m_b~e^{y_b}}
  {m_a~e^{y_b} + m_b~e^{y_a}}\right]
\end{eqnarray}
For a symmetric collision (for $m_a~=~m_b$),
\begin{equation}
  y_{cm} ~=~ \frac{1}{2}(y_a+y_b)
\end{equation}
Rapidities of $a$ and $b$ in the CMS are
\begin{equation}
y_a^* ~=~ y_a - y_{cm} ~=~ -\frac{1}{2} (y_b-y_a)
\end{equation}
\begin{equation}
y_b^* ~=~ y_b - y_{cm} ~=~ \frac{1}{2} (y_b-y_a).
\end{equation}
Given the incident energy, the rapidity of projectile particles and the 
rapidity of the target particles can thus be determined. The greater is the 
incident energy, the greater is the separation between the projectile and 
target rapidity.
\subparagraph{Central Rapidity} 
The region of rapidity mid-way between the projectile and target 
rapidities is called central rapidity.\\
\begin{example}
In p+p collisions at a laboratory momentum of 100 $GeV/c$, beam rapidity 
$y_b=5.36$, target rapidity $y_a=0$ and the central rapidity $\approx 2.7$.
\end{example}

\subsubsection{Mid-rapidity in Fixed target and Collider Experiments}
In fixed-target experiments (LS), $y_{target} ~=~ 0$.\\
$y_{lab} ~=~ y_{target} ~+ ~y_{projectile} ~=~y_{beam}$
Hence mid-rapidity in fixed-target experiment is given by,
\begin{equation}
y_{mid}^{LS} ~=~ y_{beam}/2 .
\end{equation}
In collider experiments (center of mass system),\\
$y_{projectile} ~=~ -y_{target} ~= ~y_{CMS} ~=~y_{beam}/2$.\\
Hence, mid-rapidity in CMS system is given by
\begin{equation}
y_{mid}^{CMS} ~=~ (y_{projectile}+y_{target})/2 ~=~ 0 .
\end{equation}
This is valid for a symmetric energy collider.
The rapidity difference is given by $y_{projectile}-y_{target}=2y_{CMS}$ and 
this increases with energy for a collider as $y$ increases with energy.

\subsubsection{Light-cone variables and Rapidity}
Consider a particle having rapidity $y$ and the beam rapidity is $y_b$. The
particle has forward light-cone variable $x_+$ with respect to the beam 
particle
\begin{eqnarray}
x_+ &=& \frac{p_0^c + p_z^c}{p_0^b + p_z^b}\nonumber \\
&=& \frac{m_T^c}{m^b}e^{y-y_b} 
\end{eqnarray}
where $m_T^c$ is the transverse mass of $c$. Note that the transverse momentum
of the beam particle is zero. Hence,
\begin{eqnarray}
y &=& y_b + ln~x_+ + ln\left(\frac{m_b}{m_T^c}\right)
\end{eqnarray}
Similarly, relative to the target particle $a$ with a target rapidity $y_a$, the
backward light-cone variable of the detected particle $c$ is $x_-$. $x_-$ is
related to $y$ by
\begin{equation}
x_- = \frac{m_T^c}{m^b}e^{y_a-y} 
\end{equation}
and conversely,
\begin{equation}
y = y_a - ln~x_- - ln\left(\frac{m_a}{m_T^c}\right).
\end{equation}
In general, the rapidity of a particle is related to its light-cone momenta by
\begin{equation}
y = \frac{1}{2}~ln\left(\frac{p_+}{p_-}\right)
\end{equation}
Note that in situations where there is a frequent need to work with boosts
along z-direction, it's better to use $(y,{\bf p_T})$ for a particle rather
than using it's 3-momentum, because of the simple transformation rules for
$y$ and ${\bf p_T}$ under Lorentz boosts.

\subsection{The Pseudorapidity Variable}
Let us assume that a particle is emitted at an angle $\theta$ relative to the 
beam axis. Then its rapidity can be written as\\
$y ~=~ \frac{1}{2}~ln \left(\frac{E+P_L}{E-P_L}\right) 
~=~ \frac{1}{2}~ln \left[\frac{\sqrt{m^2+p^2} + p~cos~\theta}{\sqrt{m^2+p^2} 
- p~cos~\theta}\right]$. 
At very high energy, $p \gg m$ and hence
\begin{eqnarray}
y &=& \frac{1}{2}~ln \left[\frac{p+p~cos~\theta}{p-p~cos~\theta}\right] 
\nonumber \\
  &=& -ln ~ tan~\theta/2 \equiv \eta
\end{eqnarray}
$\eta$ is called the pseudorapidity. 
Hence at very high energy,
\begin{equation}
y ~\approx ~ \eta ~=~ -ln ~ tan~\theta/2.
\end{equation}
In terms of the momentum, $\eta$ can be re-written as 
\begin{equation}
\eta ~=~ \frac{1}{2}~ln \left[\frac{|{\bf p}| + p_z}{|{\bf p}| - p_z}\right].
\label{eta}
\end{equation}
$\theta$ is the only quantity to be measured for the determination of 
pseudorapidity, independent of any particle identification mechanism. 
Pseudorapidity is defined for any value of mass, momentum and energy of the 
collision. This also could be measured with or without momentum information 
which needs a magnetic field.

\subsubsection{Change of variables from $(y,{\bf p}_T)$ to $(\eta,{\bf p}_T)$}

\begin{figure}
\centering
\includegraphics[height=6cm]{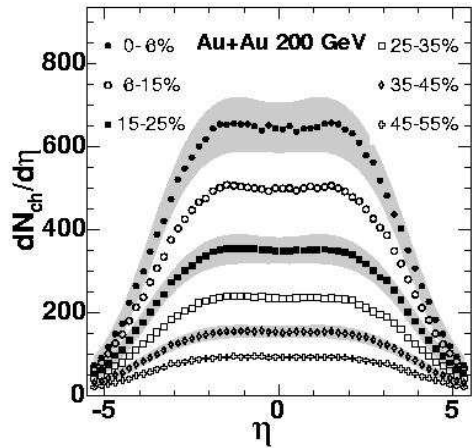}
\caption{The mid-rapidity $dN_{ch}/d\eta$ for Au+Au collisions at $\sqrt{s_{NN}}=$
200 GeV \cite{phobos}. }
\label{dndeta}
\end{figure}

By equation \ref{eta},
\begin{eqnarray}
e^{\eta} &=& \sqrt{\frac{|{\bf p}| + p_z}{|{\bf p}| - p_z}}\\
e^{-\eta} &=& \sqrt{\frac{|{\bf p}| - p_z}{|{\bf p}| + p_z}}
\end{eqnarray}
Adding both of the equations, we get
\begin{equation}
|{\bf p}| ~=~p_T ~ cosh~\eta
\end{equation}
${\bf p}_T ~=~ \sqrt{|{\bf p}|^2-p_z^2}$. 
By subtracting the above equations, we get
\begin{equation}
p_z ~=~ p_T ~ sinh~\eta
\end{equation}
Using these equations in the definition of rapidity, we get
\begin{equation}
y ~=~ \frac{1}{2}~ln \left[\frac{\sqrt{p_T^2~cosh^2~\eta + m^2} + p_T~sinh~\eta}
{\sqrt{p_T^2~cosh^2~\eta + m^2} - p_T~sinh~\eta}\right]
\end{equation}
Similarly $\eta$ could be expressed in terms of $y$ as,
\begin{equation}
\eta ~=~ \frac{1}{2}~ln \left[\frac{\sqrt{m_T^2~cosh^2~y - m^2} + m_T~sinh~y}
{\sqrt{m_T^2~cosh^2~y - m^2} - m_T~sinh~y}\right]
\end{equation}
The distribution of particles as a function of rapidity is related to the 
distribution as a function of pseudorapidity by the formula
\begin{equation}
\frac{dN}{d\eta d{\bf p}_T} ~=~ \sqrt{1-\frac{m^2}{m_T^2~cosh^2~y}}~
\frac{dN}{dyd{\bf p}_T} .
\end{equation}
In the region $y \gg 0$, the pseudorapidity distribution ($dN/d\eta$) and the 
rapidity distribution ($dN/dy$) which are essentially the ${\bf p}_T$-integrated
 values of $\frac{dN}{d\eta d{\bf p}_T}$ and $\frac{dN}{dy d{\bf p}_T}$ 
respectively, are approximately the same. 
In the  region $y \approx 0$, there is a small ``{\it depression}'' in 
$dN/d\eta$ distribution compared to $dN/dy$ distribution due to the above 
transformation. At very high energies where $dN/dy$ has a mid-rapidity plateau, 
this transformation gives a small dip in $dN/d\eta$ around $\eta \approx 0$
(see Figure \ref{dndeta}). However, for a massless particle like photon, the 
dip in $dN/d\eta$ is not expected (which is clear from the above equation). 
Independent of the frame of reference where $\eta$ is measured, the difference 
in the maximum magnitude of $dN/d\eta$ appears due to the above transformation. 
In the CMS, the maximum of the distribution is located at
$y \approx \eta \approx 0$ and the $\eta$-distribution is suppressed by a factor
$\sqrt{1-m^2/<m_T^2>}$ with reference to the rapidity distribution. In the 
laboratory frame, however the maximum is located around half of the beam 
rapidity $\eta \approx y_b/2$ and the suppression factor is 
$\sqrt{1-m^2/<m_T^2>~cosh^2~(y_b/2)}$, which is about
unity. Given the fact that the shape of the rapidity distribution is 
independent of frame of reference, the peak value of the pseudorapidity 
distribution in the CMS frame is lower than its value in LS. This suppression 
factor at SPS energies is $\sim 0.8 ~ - ~0.9$.

\subsection{The Invariant Yield}
First we show $\frac{d^3p}{E}$ is Lorentz invariant. The differential 
of Lorentz boost in longitudinal direction is given by
\begin{equation}
dp_z^* ~=~\gamma(dp_z-\beta dE).
\label{dp}
\end{equation}
Taking the derivative of the equation $E^2 ~=~ p^2 + m^2$, we get
\begin{equation}
EdE ~=~p_zdp_z.
\label{de}
\end{equation}
Using equations \ref{dp} and \ref{de} we get
\begin{eqnarray}
dp_z^* &=& \gamma(dp_z -\beta \frac{p_zdp_z}{E}) \nonumber \\
&=& \frac{dp_z}{E}~E^* .
\end{eqnarray}
As ${\bf p}_T$ is Lorentz invariant, multiplying ${\bf p}_T$ on both the sides 
and re-arranging gives
\begin{equation}
\frac{d^3p^*}{E^*} ~=~\frac{d^2 {\bf p}_T~dp_z}{E} ~=~ \frac{d^3p}{E}.
\end{equation}
In terms of experimentally measurable quantities, $\frac{d^3p}{E}$ could be 
expressed as
\begin{eqnarray}
\frac{d^3p}{E} &=& d{\bf p}_T~dy \nonumber \\
&=& p_Tdp_Td\phi dy \\
&=& m_Tdm_Td\phi dy .
\end{eqnarray}
The Lorentz invariant differential cross-section
$\frac{Ed^3\sigma}{dp^3} ~=~ \frac{Ed^3N}{dp^3}$ is the invariant yield. In 
terms of experimentally measurable quantities this could be expressed as
\begin{eqnarray}
\frac{Ed^3\sigma}{dp^3} &=& \frac{1}{m_T} ~\frac{d^3N}{dm_Td\phi dy} 
\nonumber \\
&=&  \frac{1}{2\pi~m_T} ~\frac{d^2N}{dm_T dy} \nonumber \\
&=&  \frac{1}{2\pi~p_T} ~\frac{d^2N}{dp_T dy}  .
\label{inv}
\end{eqnarray}
To measure the invariant yields of identified particles equation \ref{inv} is 
used experimentally.

\subsection{Inclusive Production of Particles and the Feynman Scaling 
variable $x_F$}
In a reaction of type\\
$beam ~+~ target ~ \longrightarrow ~ A ~+~ anything$\\
where $A$ is known is called an ``{\it inclusive reaction}''. The cross-section 
for particle production could be written separately as functions of ${\bf p}_T$ 
and $p_L$:\\
\begin{equation}
\sigma ~=~f({\bf p}_T) g(p_L).
\end{equation}
This factorization is empirical and convenient because each of these factors 
has simple parameterizations which fit well to experimental data.\\

Similarly the differential cross-section could be expressed by
\begin{equation}
\frac{d^3\sigma}{dp^3} ~=~ \frac{d^2\sigma}{{\bf p}_T^2}~ \frac{d\sigma}{dp_L}
\end{equation}
Define the variable 
\begin{eqnarray}
x_F &=& \frac{p_L^*}{p_L^*(max)}  \\
&=& \frac{2p_L^*}{\sqrt{s}}  
\end{eqnarray}
$x_F$ is called the {\it Feynman scaling variable}: longitudinal component of 
the cross-section when measured in CMS of the collision, would scale {\it i.e.} 
would not depend on the
energy $\sqrt{s}$. Instead of $\frac{d\sigma}{dp_L^*}$, $\frac{d\sigma}{dx_F}$ 
is measured which wouldn't depend on energy of the reaction, $\sqrt{s}$. This 
Feynman's assumption is valid approximately.\\

  The differential cross-section for the inclusive production of a particle is 
then written as 
\begin{equation}
\frac{d^3\sigma}{dx_Fd^2{\bf p}_T} ~=~ F(s,~x_F,~{\bf p}_T)
\end{equation}
Feynman's assumption that at high energies the function $F(s,~x_F,~{\bf p}_T)$ 
becomes asymptotically independent of the energy means:\\
$lim_{s \rightarrow \infty} F(s,~x_F,~{\bf p}_T)~=~ F(x_F, ~{\bf p}_T)~=~ 
f({\bf p}_T)~g(x_F)$

\subsection{The ${\bf p}_T$-Distribution}
The distribution of particles as a function of ${\bf p}_T$ is called 
${\bf p}_T$-distribution.
Mathematically, 
\begin{equation}
\frac{dN}{d{\bf p}_T} ~=~ \frac{dN}{2\pi~|{\bf p}_T|d|{\bf p}_T|}
\end{equation}
where $dN$ is the number of particles in a particular ${\bf p}_T$-bin. People 
usually plot $\frac{dN}{p_Tdp_T}$ as a function of $p_T$ taking out the factor 
$1/2\pi$ which is a constant. Here $p_T$ is a scalar quantity. The low-$p_T$ 
part of the $p_T$-spectrum is well described by an exponential function having 
thermal origin. However, to describe the whole range of the $p_T$, one uses 
the Levy function which has an exponential part to describe low-$p_T$ and a 
power-law function to describe the hight-$p_T$ part which is dominated by hard 
scatterings (high momentum transfer at early times of the collision). The 
inverse slope parameter of $p_T$-spectra is called the effective temperature 
($T_{eff}$), which has a thermal contribution because of the random kinetic 
motion of the produced particles and a contribution from the collective motion 
of the particles. This will be described in details in the section of 
freeze-out properties and how to determine the chemical and kinetic freeze-out 
temperatures experimentally.\\

The most important parameter is then the mean $p_T$ which carries the 
information of the effective temperature of the system. Experimentally, 
$\left<p_T\right>$ is studied as a function of $\frac{dN_{ch}}{d\eta}$ which 
is the measure of the entropy density of the system. This is like studying the 
temperature as a function of entropy to see the signal of phase transition. 
The phase transition is of 1st order if a plateau is observed in the spectrum 
signaling the existence of latent heat of the system. This was first proposed
by L. Van Hove \cite{vanHove}.\\

The average of any quantity $A$ following a particular probability distribution 
$f(A)$ can be written as
\begin{equation}
\left<A\right> ~=~ \frac{\int A~f(A)~dA}{\int f(A)~dA}.
\end{equation}
Similarly,
\begin{eqnarray}
\left<p_T\right> &=& \frac{\int_0^{\infty} p_T~(\frac{dN}{dp_T})~dp_T}
{\int_0^{\infty} (\frac{dN}{dp_T})~dp_T} \nonumber \\ 
&=&  \frac{\int_0^{\infty} p_T~dp_T~~p_T(\frac{dN}{p_Tdp_T})}
{\int_0^{\infty}p_T~dp_T (\frac{dN}{p_Tdp_T})} \nonumber \\ 
&=&  \frac{\int_0^{\infty} p_T~dp_T~~p_T~f(p_T)}{\int_0^{\infty}p_T~dp_T f(p_T)}
\end{eqnarray}
where $2\pi~p_T~dp_T$ is the phase space factor and the $p_T$-distribution 
function is given by 
\begin{equation}
f(p_T) ~=~ \frac{dN}{d{\bf p}_T} ~=~ \frac{dN}{p_Tdp_T} .
\end{equation}

\begin{example}
Experimental data on $p_T$-spectra are sometimes fitted to the exponential 
Boltzmann type function given by
\begin{equation}
f(p_T) ~=~ \frac{1}{p_T}\frac{dN}{dp_T} ~\simeq~ C~ e^{-m_T/T_{eff}}.
\end{equation}
\end{example}
The $\left<m_T\right>$ could be obtained by
\begin{eqnarray}
\left<m_T\right> &=& \frac{\int_0^{\infty}p_T ~dp_T~m_T~exp.(-m_T/T_{eff})}
{\int_0^{\infty}p_T ~dp_T~exp.(-m_T/T_{eff})} \nonumber \\
&=& \frac{2T_{eff}^2+2m_0T_{eff}+m_0^2}{m_0+T_{eff}}
\end{eqnarray} 
where $m_0$ is the rest mass of the particle. It can be seen from the above 
expression that for a massless particle
 \begin{equation}
\left<m_T\right> ~=~ \left<p_T\right>  ~=~ 2T_{eff} .
\end{equation}
This also satisfies the principle of equipartition of energy which is expected 
for a massless Boltzmann gas in equilibrium. \\

However, in experiments the higher limit of $p_T$ is a finite quantity. In 
that case the integration will involve an incomplete gamma function.

\subsection{Energy in CMS and LS}
\subsubsection{For Symmetric Collisions ($A+A$)}
Consider the collision of two particles. In LS, the projectile with momentum 
${\bf p}_1$, energy $E_1$ and mass $m_1$ collides with a particle of mass 
$m_2$ at rest. The 4-momenta of the particles are\\
$p_1 ~=~ (E_1,{\bf p}_1),~~~~~~~~~~$ ~$p_2 ~=~ (m_2, {\bf 0})$ \\
In CMS, the momenta of both the particles are equal and opposite, the 4-momenta 
are\\
$p_1^* ~=~ (E_1^*,{\bf p}_1^*),~~~~~~~~~~$ ~$p_2^* ~=~ (E_2^*, -{\bf p}_1^*)$ \\
The total 4-momentum of the system is a conserved quantity in the collision.\\
In CMS,\\
$(p_1+p_2)^2 ~=~ (E_1+E_2)^2 - ({\bf p}_1+{\bf p}_2)^2 $\\ 
$~=~ (E_1+E_2)^2 ~=~ E_{cm}^2 ~\equiv ~s$ .\\
$\sqrt{s}$ is the total energy in the CMS which is the invariant mass of the 
CMS.\\
In LS, \\
$(p_1+p_2)^2 ~=~ m_1^2+m_2^2 + 2E_1m_2$ .\\
Hence 
\begin{equation}
E_{cm} ~=~ \sqrt{s} ~=~ \sqrt{m_1^2+m_2^2 + 2E_{proj}m_2}
\end{equation}
where $E_1 = E_{proj}$, the projectile energy in LS. Hence it is evident here 
that the CM frame with an invariant mass $\sqrt{s}$ moves in the laboratory in 
the direction of ${\bf p}_1$ with a velocity corresponding to:\\
Lorentz factor, 
\begin{eqnarray}
\gamma_{cm} &=& \frac{E_1+m_2}{\sqrt{s}}  \\
\Rightarrow \sqrt{s} &=& \frac{E_{lab}}{\gamma_{cm}}, 
\end{eqnarray}
this is because $E~=~ \gamma m$
and 
\begin{equation}
y_{cm} ~=~ cosh^{-1} ~\gamma_{cm}.
\end{equation}

\begin{note}
We know
\begin{equation}
s~=~E_{cm}^2 ~=~ m_1^2 + m_2^2 + 2(E_1+E_2+{\bf p}_1.{\bf p}_2) 
\end{equation}
For a head-on collision with $m_1, ~m_2 ~\ll~ E_1,E_2$
\begin{equation}
E_{cm}^2 ~\simeq ~ 4E_1E_2
\end{equation}
For two beams crossing at an angle $\theta$, 
\begin{equation}
E_{cm}^2 ~= ~ 2E_1E_2 (1+cos~\theta)
\end{equation}
The CM energy available in a collider with equal energies ($E$) for new particle
production rises linearly with $E$ {\it i.e.} 
\begin{equation}
E_{cm} ~\simeq ~ 2E
\end{equation}
For a fixed-target experiment the CM energy rises as the square root of the 
incident energy:
\begin{equation}
E_{cm} ~\simeq ~ \sqrt{2m_2E_1}
\end{equation}
Hence the highest energy available for new particle production is 
achieved at collider experiments. For example, at SPS fixed-target experiment 
to achieve a CM energy of 17.3 AGeV the required incident beam energy is 
158 AGeV.  
\end{note}
\begin{note}
Most of the times the energy of the collision is expressed in terms of 
nucleon-nucleon center of mass energy. In the nucleon-nucleon CM frame, two 
nuclei approach each other with the same boost factor $\gamma$. The 
nucleon-nucleon CM is denoted by $\sqrt{s_{NN}}$ and is related to the total 
CM energy by
\begin{equation}
\sqrt{s} ~= ~ A~\sqrt{s_{NN}}
\end{equation}
This is for a symmetric collision with number of nucleons in each nuclei as $A$.
The colliding nucleons approach each other with energy $\sqrt{s_{NN}}/2$ and 
with equal and opposite momenta. The rapidity of the nucleon-nucleon center 
of mass is\\
 $y_{NN} = 0$
and taking $m_1=m_2=m_N $, the projectile and target nucleons are at equal 
and opposite rapidities.
\begin{equation}
y_{proj}~=~ -y_{target} ~=~ cosh^{-1}~\frac{\sqrt{s_{NN}}}{2m_N} ~=~ y_{beam}.
\end{equation}
\end{note}

\begin{note}:
Lorentz Factor\\
\begin{eqnarray}
\gamma &=& \frac{E}{M} \nonumber
=\frac{\sqrt{s}}{2A~m_N} \nonumber \\
&=&\frac{A~\sqrt{s_{NN}}}{2A~m_N} \nonumber
=\frac{\sqrt{s_{NN}}}{2~m_N} \\
&=&\frac{E_{beam}^{CMS}}{m_N} 
\end{eqnarray}
where E and M are Energy and Mass in CMS respectively. Assuming mass of the nucleon
$m_N \sim 1$ GeV, the Lorentz factor is of the order of beam energy in CMS for a
symmetric collision.

\end{note}

\subsubsection{For Asymmetric Collisions ($A+B$)}

During the early phase of relativistic nuclear collision research, the projectile mass was limited by accelerator-technical conditions ($^{38}$Ar at the Bevalac, $^{28}$Si at the AGS, $^{32}$S at the SPS). Nevertheless, collisions with mass $\approx$ 200 nuclear targets were investigated. Analysis of such collisions is faced with the problem of determining an "effective" center of mass frame, to be evaluated from the numbers of projectile and target participant nucleons, respectively. Their ratio - an thus the effective CM rapidity - depends on impact parameter. Moreover, this effective CM frame refers to soft hadron production only, whereas hard processes are still referred to the frame of nucleon-nucleon collisions. The light projectile on heavy target kinematics are described in \cite{asymmetric}.

\subsection{Luminosity}
The luminosity is an important parameter in collision experiments. The reaction 
rate in a collider is given by
\begin{equation}
R ~=~ \sigma L 
\end{equation}
where, \\
$\sigma \equiv$ interaction cross-section\\
$L \equiv$ luminosity (in $cm^{-2} s^{-1}$)
\begin{equation}
L ~=~ fn ~\frac{N_1N_2}{A}
\end{equation}
where,\\
$f \equiv$ revolution frequency\\
$N_1, ~N_2 \equiv$ number of particles in each bunch\\
$n \equiv$ number of bunches in one beam in the storage ring\\
$A \equiv$ cross-sectional area of the beams\\
$L$ is larger if the beams have small cross-sectional area.


\subsection{Collision Centrality}
\begin{figure}
\centering
\includegraphics[height=8cm,width=8cm]{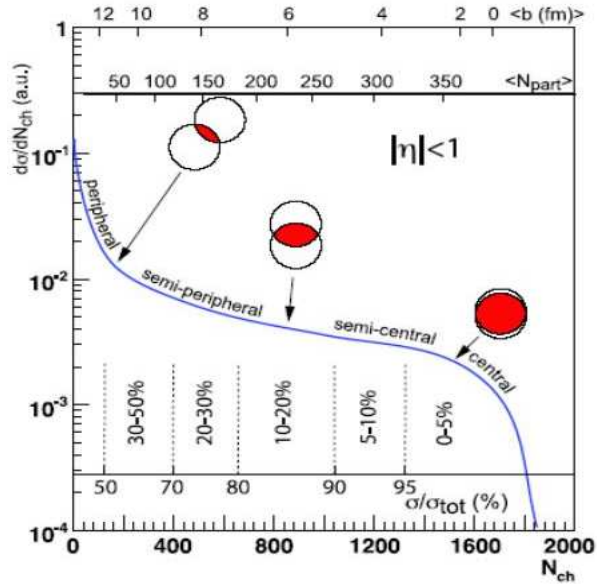}
\caption{A cartoon showing the centrality definition from the final state particle
multiplicity and its correlation with the impact parameter ($b$) and the number 
of participating nucleons ($N_{part}$) in the collisions.}
\label{centrality}
\end{figure}

In a collision of two nuclei, the impact parameter ($b$) can carry values from 
$0$ to $R_1+R_2$, where $R_1$ and $R_2$ are the diameters of the two nuclei. 
When $b=0$, it is called {\it head-on collision}. When collisions with 
$0 \leq b \leq (R_1+R_2)$ are allowed, it is called {\it minimum-bias 
collision}.  In heavy-ion collisions, initial geometric quantities such as 
impact parameter and the collision geometry can not be directly measured 
experimentally. Contrary, it is however possible to relate the particle 
multiplicity, transverse energy and the number of spectactor nucleons 
(measured by a ``zero-degree calorimeter'' ZDC) to the centrality of the collisions.\\
It is straight forward to assume that on the average,\\
1) the energy released in a collision is proportional to the number of nucleons
participating in the collisions.\\
2) the particle multiplicity is proportional to the participating nucleon number.\\
Hence the particle multiplicity is proportional to the energy released in the 
collision. One can measure the particle multiplicity distribution or the transverse 
energy ($E_T$) distribution for minimum-bias collisions. Here the high values of 
particle multiplicity or $E_T$ correspond to central collisions and lower 
values correspond to more peripheral collisions. Hence the minimum-bias $E_T$ or 
multiplicity distribution could be used for centrality determination in an 
collision experiment. Figure \ref{centrality} shows the minimum-bias multiplicity
($N_{ch}$) distribution used for the selection of collision centrality.
The minimum-bias yield has been cut into successive intervals starting from
the maximum value of $N_{ch}$. The first $5\%$ of the high $N_{ch}$ events 
correspond to top $5\%$ central collisions. The correlation of centrality 
and the impact parameter with the number of participating nucleons has also been
elaborated, in detail, by Glauber-type Monte Carlo calculations employing Woods-Saxon nuclear density distributions.
 
\subsection{Number of Participants and Number of Binary Collisions}
Experimentally there is no direct way to estimate the number of participating 
nucleons ($N_{part}$) and the number of binary collisions ($N_{bin}$) in any
event, for an given impact parameter. The Glauber model calculation is performed
to estimate the above two quantities as a function of the impact parameter. 
The Glauber model treats a nucleus-nucleus collision as a superposition of 
many independent nucleon-nucleon ($N-N$) collisions. This model depends on 
the nuclear density profile (Woods-Saxon) and the non-diffractive inelastic 
$N+N$ cross-sections. The Woods-Saxon distribution is given by
\begin{equation}
\rho(r) ~=~ \frac{\rho_0}{1+exp(\frac{r-r_0}{c})}
\end{equation}
where, $r$ is the radial distance from the center of the nucleus, $r_0$ is the mean
radius of the nucleus, $c$ is the skin depth of the nucleus and $\rho_0$ is the
nuclear density constant. The parameters $r_0$ and $c$ are measured in 
electron-nucleus scattering experiments. $\rho_0$ is determined from the overall
normalization condition
\begin{equation}
\int \rho(r)~d^3r ~=~ A
\end{equation}
where $A$ is the mass number of the nucleus.

There are two separate implementations of Glauber 
approach: Optical and Monte Carlo (MC). In the Optical Glauber approach, $N_{part}$ 
and $N_{bin}$ are estimated by an analytic integration of overlapping 
Woods-Saxon distributions.

The MC Glauber calculation proceeds in two steps. First the nucleon position
in each nucleus is determined stochastically. Then the two nuclei are ``collided'',
assuming the nucleons travel in a straight line along the beam axis (this is 
called eikonal approximation). The position of each nucleon in the nucleus is
determined according to a probability density function which is typically taken
to be uniform in azimuth and polar angles. The radial probability function
is modeled from the nuclear charge densities extracted from electron scattering
experiments. A minimum inter-nucleon separation is assumed between the positions
of nucleons in a nucleus, which is the characteristic length of the repulsive
nucleon-nucleon force. Two colliding nuclei are simulated by distributing
 $A$ nucleons of nucleus $A$ and $B$ nucleons of nucleus $B$ in 3-dimensional
co-ordinate system according to their nuclear density distribution. A random
impact parameter $b$ is chosen from the distribution $d\sigma/db ~=~ 2\pi b$.
A nucleus-nucleus collision is treated as a sequence of independent nucleon-nucleon
collisions with a collision taking place if their distance $D$ in the transverse
plane satisfies
\begin{equation}
D < \sqrt{\sigma_{inel}^{NN}/\pi}
\end{equation}
where $\sigma_{inel}^{NN}$ is the total inelastic nucleon-nucleon cross-section.
An arbitrary number of such nucleus-nucleus collisions are performed by the
monte carlo and the resulting distributions of $d\sigma/N_{part}$ and 
$d\sigma/N_{bin}$, $d\sigma/db$ are determined. Here $N_{part}$ is defined as 
the total number of nucleons that underwent at least one interaction and 
$N_{bin}$ is the total number of interactions in an event. These histograms 
are binned according to fractions of the total cross-sections. This determines 
the mean values of $N_{part}$ and $N_{bin}$ for each centrality class. The 
systematic uncertainties in these values are estimated by varying the 
Wood-Saxon parameters, by varying the value of $\sigma_{inel}^{NN}$ and from the
uncertainty in the determination of total nucleus-nucleus cross-section. These
sources of uncertainties are treated as fully correlated in the final systematic
uncertainty in the above measured variables.

When certain cross-sections scale with number of participants, those are said 
to be associated with {\it ``soft''} processes: small momentum transfer processes.
The low-$p_T$ hadron production which accounts for almost $95\%$ of the bulk
hadron multiplicity comes in the ``soft processes''. These soft processes are
described by phenomenological non-perturbative models. Whereas, in {\it ``hard''}
QCD processes like jets, charmonia, other heavy flavor and processes associated
with high-$p_T$ phenomena, the cross-section scales with the number of primordial
target/projectile parton collisions. This is estimated in the above Glauber 
formalism as the total number of inelastic participant-participant collisions.
For the hard processes the interaction is at partonic level with large momentum
transfer and is governed by pQCD. $N_{coll}$ is always higher than $N_{part}$: 
when $N_{part}$ grows like $A$, $N_{coll}$ grows like $A^{4/3}$.

 Sometimes, to study the contribution of soft and hard processes to any
cross-section one takes a two-component model like:\\

\begin{equation}
cross-section ~=~ (1-f)~ N_{part} ~+~f~ N_{coll}
\end{equation}
where $f$ is the fractional contribution from hard processes.




\section {Bulk Hadron Production in A+A Collisions}
\label{sec:Bulk_Hadron_Prod}
We will now take an overall look at bulk hadron production in
nucleus-nucleus collisions. In view of the high total c.m. energies
involved at e.g. top SPS $(E ^{tot}_{cm} \approx  3.3 \:TeV$) and
top RHIC (38 $TeV$) energies, in central Pb+Pb (SPS) and Au+Au
(RHIC) collisions, one can expect an extraordinarily high spatial
density of produced particles. The average number of produced particles at SPS energies is $\approx$ 1600, while at RHIC multiplicities of $\approx$ 4000 are reached. Thus, as an overall idea of analysis,
one will try to relate the observed flow of energy into transverse
and longitudinal phase space and particle species to the high energy
density contained in the primordial interaction volume, thus to
infer about its contained matter.

Most of the particles under investigation correspond to ''thermal'' pions ($p_T$ up to $2 \:GeV$) and, in
general, such thermal hadrons make up for about 95\% of the observed
multiplicity: the bulk of hadron production. Their distributions in
phase space will be illustrated in the subsections below. This will
lead to a first insight into the overall reaction dynamics, and also
set the stage for consideration of the rare signals, imbedded in
this thermal bulk production: direct photons, jets, heavy flavors, which are the subject of later chapters in this volume.

\subsection {Particle Multiplicity and Transverse Energy Density}
\label{subsec:Particle_Multiplicity}
\begin{figure}
\begin{center}
\includegraphics[scale=1.1]{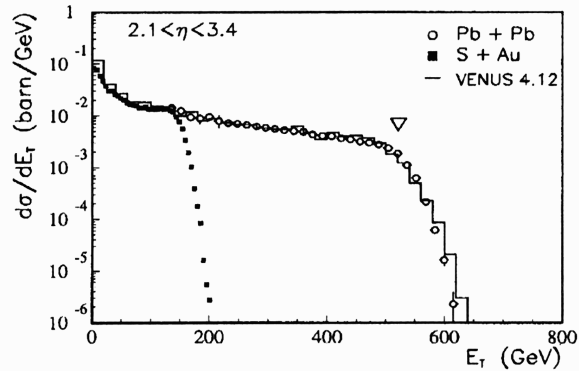}
\caption{Minimum bias distribution of total transverse energy in Pb+Pb
collisions at $\sqrt{s}=17.3 \: GeV$, and S+Au collisions at
$\sqrt{s}=20 \: GeV$, in the rapidity interval $2.1 < y < 3.4$, from
\cite{43}.}
\label{fig:Figure3}
\end{center}
\end{figure}

Particle production can be assessed globally by the total created
transverse energy, the overall result of the collisional creation of
{\it transverse} momentum $p_T$ or transverse mass ($m_T=\sqrt{
p^2_T + m^2_0}$), at the microscopic level. Fig.~\ref{fig:Figure3} shows the
distribution of total transverse energy $E_T=\sum\limits_i \:
E(\theta_i) \cdot sin \theta$ resulting from a calorimetric
measurement of energy flow into calorimeter cells centered at angle
$\theta_i$ relative to the beam \cite{43}, for $^{32}S + ^{197}$Au
collisions at $\sqrt{s} = 20 \: GeV$, and for $^{208}Pb+^{208}$Pb
collisions at $\sqrt{s}=17.3\: GeV$. \\
 \\
The shape is characteristic of the impact parameter probability distribution (for equal size
spheres in the Pb+Pb case). The turnoff at $E_T=520 \: GeV$
indicates the point where geometry runs out of steam, i.e. where $b
\rightarrow 0$, a configuration generally referred to as a ''central
collision''. The adjacent shoulder results from genuine event by
event fluctuations of the actual number of participant nucleons from
target and projectile (recall the diffuse Woods-Saxon nuclear
density profiles), and from experimental factors like calorimeter
resolution and limited acceptance. The latter covers 1.3 units of
pseudo-rapidity and contains mid-rapidity $\eta_{mid}=2.9$.
Re-normalizing \cite{43} to $\Delta \: \eta=1$ leads to $dE_T/d\eta
(mid)= 400 \:GeV$, in agreement with the corresponding WA80 result
\cite{44}. Also, the total transverse energy of central Pb+Pb
collisions at $\sqrt{s}=17.3 \: GeV$ turns out to be about $1.2
\:TeV$. As the definition of a central collision, indicated in
Fig.~\ref{fig:Figure3}, can be shown \cite{42} to correspond to an average nucleon
participant number of $N_{part}=370$ one finds an average total
transverse energy per nucleon pair, of $E_T/\left<0.5 \: N_{part}\right>=6.5 \:
GeV$. After proper consideration of the baryon pair rest mass (not
contained in the calorimetric $E_T$ response but in the
corresponding $\sqrt{s}$) one concludes \cite {43} that the observed
total $E_T$ corresponds to about 0.6 $E_T^{max}$, the maximal $E_T$
derived from a situation of ''complete stopping'' in which the
incident $\sqrt{s}$ gets fully transformed into internal excitation
of a single, ideal isotropic fireball located at mid-rapidity. The
remaining fraction of $E_T^{max}$ thus stays in longitudinal motion,
reflecting the onset, at SPS energy, of a transition from a central
fireball to a longitudinally extended ''fire-tube'', i.e. a
cylindrical volume of high primordial energy density. In the limit
of much higher $\sqrt{s}$ one may extrapolate to the idealization of
a boost invariant primordial interaction volume, introduced by
Bjorken \cite{45}.

\begin{figure}   
\begin{center}
\includegraphics[scale=0.5]{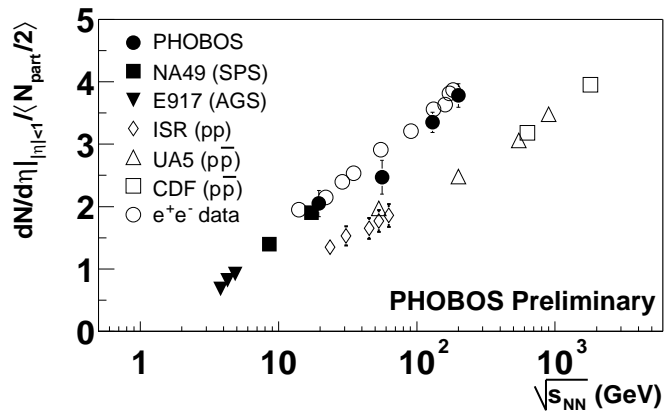}
\caption{Charged hadron rapidity density at mid-rapidity vs. $\sqrt{s}$,
compiled from $e^+e^-, \: pp, \: p \overline{p}$ and A+A collisions
\cite{53}.}
\label{fig:Figure4}
\end{center}
\end{figure}

\begin{figure}   
\begin{center}
\includegraphics[scale=0.4]{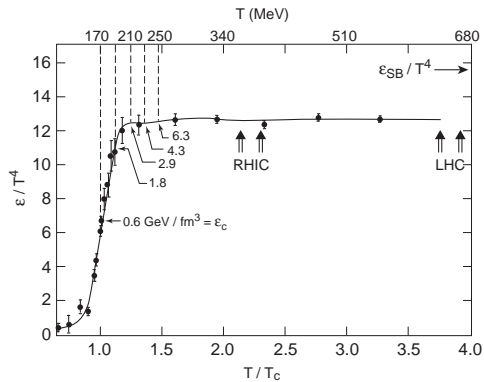}
\caption{Lattice QCD results at zero baryon potential for energy density
$\epsilon/T^4$ versus $T/T_c$ with three light quark flavors,
compared to the Stefan-Boltzmann-limit $\epsilon_{SB}$ of an ideal
quark-gluon gas \cite{48}.}
\label{fig:Figure5}
\end{center}
\end{figure}

We shall show below (section~\ref{subsec:Rap_Distribution}) that the charged particle rapidity
distributions, from top SPS to top RHIC energies, do in fact
substantiate a development toward a boost-invariant situation. One
may thus employ the Bjorken model for an estimate of the primordial
spatial energy density $\epsilon$, related to the energy density in
rapidity space via the relation \cite{45}
\begin{equation}
\epsilon(\tau_0) = \frac{1}{\pi R^2} \: \frac{1}{\tau_0} \:
\frac{dE_T}{dy}
\label{eq:equation1}
\end{equation}
where the initially produced collision volume is considered as a
cylinder of length $dz=\tau_0 dy$ and transverse radius $R \propto
A^{1/3}$. Inserting for $\pi R^2$ the longitudinally projected
overlap area of Pb nuclei colliding near head-on (''centrally''),
and assuming that the evolution of primordial pQCD shower
multiplication (i.e. the energy transformation into internal degrees
of freedom) proceeds at a time scale $\tau_0 \le 1 fm/c$, the above
average transverse energy density, of $dE_T/dy=400 \: GeV$ at top
SPS energy \cite{43,44} leads to the estimate
\begin{equation}
\epsilon (\tau_0=1fm)=3.0 \pm 0.6 \:GeV/fm^3,
\label{eq:equation2}
\end{equation}
thus exceeding, by far, the estimate of the critical energy density
$\epsilon_0$ obtained from lattice  QCD (see below), of about 1.0
$GeV/fm^3$. Increasing the collision energy to $\sqrt{s}=200 \: GeV$
for Au+Au at RHIC, and keeping the same formation time, $\tau_0=1 \:
fm/c$ (a conservative estimate as we shall show in section~\ref{subsec:Gluon_Satu_in_AA_Coll}), the
Bjorken estimate grows to $\epsilon \approx 6.0 \pm 1 \: GeV/fm^3$.
This statement is based on the increase of charged particle
multiplicity density at mid-rapidity with $\sqrt{s}$, as illustrated
in Fig.~\ref{fig:Figure4}. 
From top SPS to top RHIC energy \cite{46} the density per
participant nucleon pair almost doubles. However, at $\sqrt{s}=200
\: GeV$ the formation or thermalization time $\tau_0$, employed in
the Bjorken model \cite{45}, was argued \cite{47} to be shorter by a
factor of about 4. We will return to such estimates of $\tau_0$ in
section~\ref{subsec:Transvers_phase_space} but note, for now, that the above choice of $\tau_0=1 \:
fm/c$ represents a conservative upper limit at RHIC energy.\\
 \\
These Bjorken-estimates of spatial transverse energy density are
confronted in Fig.~\ref{fig:Figure5} with lattice QCD results obtained for three
dynamical light quark flavors \cite{48}, and for zero baryo-chemical
potential (as is realistic for RHIC energy and beyond but still
remains a fair approximation at top SPS energy where $\mu_B \approx
250 \: MeV)$. The energy density of an ideal, relativistic parton
gas scales with the fourth power of the temperature,
\begin{equation}
\epsilon = gT^4
\label{eq:equation3}
\end{equation}
where $g$ is related to the number of degrees of freedom. For an
ideal gluon gas, $g=16 \: \pi^2/30$; in an interacting system the
effective $g$ is smaller. The results of Fig.~\ref{fig:Figure5} show, in fact, that
the Stefan-Boltzmann limit $\epsilon_{SB}$ is not reached, due to
non perturbative effects, even at four times the critical
temperature $T_c=170 \: MeV$. The density $\epsilon/T^4=g$ is seen
to ascend steeply, within the interval $T_c \pm 25 \: MeV$. At $T_c$
the critical QCD energy density $\epsilon=0.6-1.0 \: GeV/fm^3$.
Relating the thermal energy density with the Bjorken estimates
discussed above, one arrives at an estimate of the initial
temperatures reached in nucleus-nucleus collisions, thus implying
thermal partonic equilibrium to be accomplished at time scale
$\tau_0$ (see section~\ref{subsec:Transvers_phase_space}). For the SPS, RHIC and LHC energy domains
this gives an initial temperature in the range 190 $\le T^{SPS} \le
220 \: MeV, \: 220 \le T^{RHIC} \le 400 \: MeV$ (assuming \cite{47}
that $\tau_0$ decreases to about 0.3 $fm/c$ here) and $T^{LHC} \ge
600 \: MeV$, respectively. From such estimates one tends to conclude
that the immediate vicinity of the phase transformation is sampled
at SPS energy, whereas the dynamical evolution at RHIC and LHC
energies dives deeply into the ''quark-gluon-plasma'' domain of QCD.
We shall return to a more critical discussion of such ascertations
in section~\ref{subsec:Transvers_phase_space}. \\ \\     
One further aspect of the mid-rapidity charged particle densities
per participant pair requires attention: the comparison with data
from elementary collisions. Fig.~\ref{fig:Figure4} shows a compilation of $pp, \:
p\overline{p}$ and $e^+e^-$ data covering the range from ISR to LEP and 
Tevatron energies. \\
\begin{figure}
\begin{center}
\includegraphics[scale=0.32]{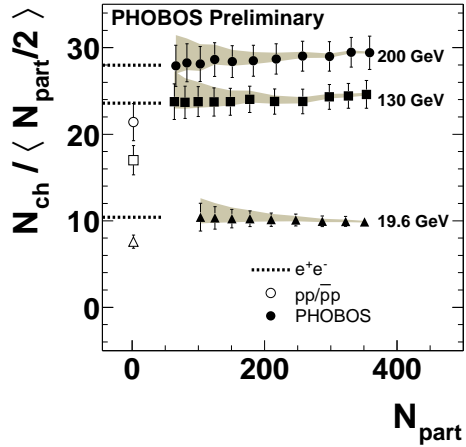}
\caption{The total number of charged hadrons per participant pair shown as a
function of $N_{part}$ in Au+Au collisions at three RHIC energies
\cite{53}.}
\label{fig:Figure6}
\end{center}
\end{figure}\\
The data from $e^+e^-$ represent $dN_{ch}/dy$,
the rapidity density along the event thrust axis, calculated
assuming the pion mass \cite{49} (the difference between $dN/dy$ and
$dN/d\eta$ can be ignored here). Remarkably, they superimpose with the central A+A collision data, 
whereas $pp$ and $p\overline{p}$
show similar slope but amount to only about 60\% of the AA and
$e^+e^-$ values. This difference between $e^+e^-$ annihilation to
hadrons, and $pp$ or $p\overline{p}$ hadro-production has been
ascribed \cite{50} to the characteristic leading particle effect of
minimum bias hadron-hadron collisions which is absent in $e^+e^-$.
It thus appears to be reduced in AA collisions due to subsequent
interaction of the leading parton with the oncoming thickness of the
remaining target/projectile density distribution. This naturally
leads to the scaling of total particle production with $N_{part}$
that is illustrated in Fig.~\ref{fig:Figure6}, for three RHIC energies and minimum
bias Au+Au collisions; the close agreement with $e^+e^-$
annihilation data is obvious again. One might conclude that,
analogously, the participating nucleons get ''annihilated'' at high
$\sqrt{s}$, their net quantum number content being spread out over
phase space (as we shall show in the next section).
\begin{figure}
\begin{center}
\includegraphics[scale=0.61]{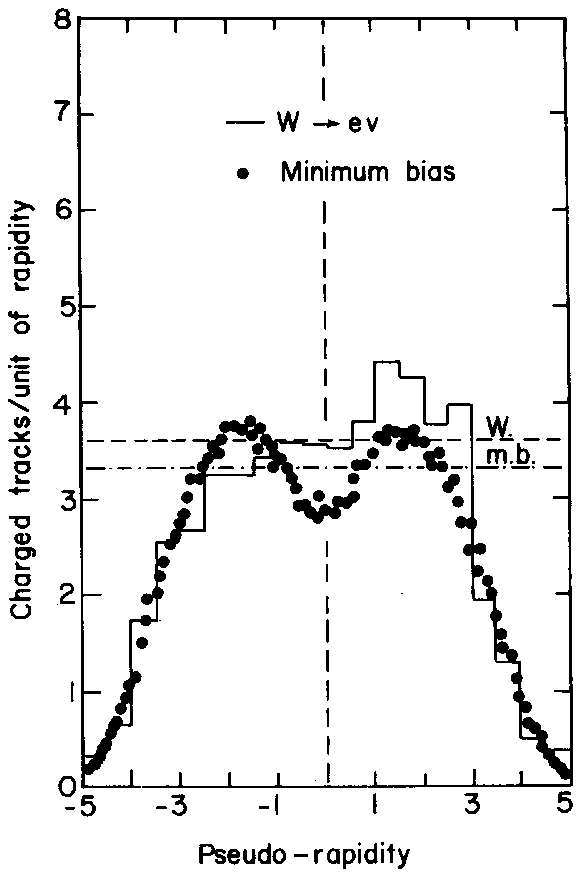}
\includegraphics[scale=0.3]{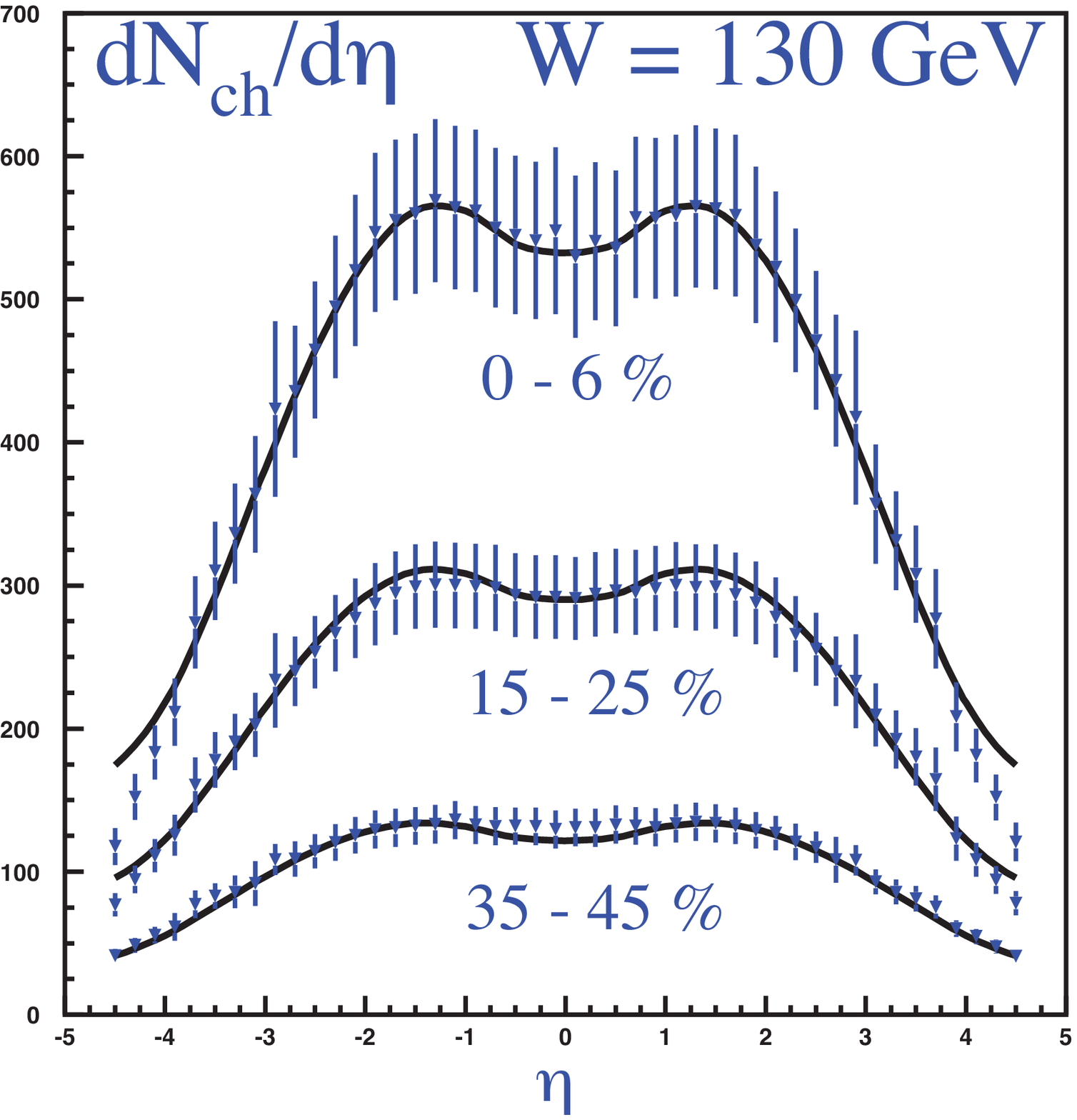}
\caption{Left panel: charged particle pseudo-rapidity distribution in $p
\overline{p}$ collisions at $\sqrt{s}=540 \: GeV$ \cite{51}. Right
panel: same in RHIC Au+Au collisions at $\sqrt{s}=130 \: GeV$ at
different centralities \cite{52}. Closed lines represent fits with
the color glass condensate model \cite{64}.}
\label{fig:Figure7}
\end{center}
\end{figure}

\subsection {Rapidity Distributions}
\label{subsec:Rap_Distribution}
Particle production number in A+A collisions depends globally on
$\sqrt{s}$ and collision centrality, and differentially on $p_T$ and
rapidity $y$, for each particle species $i$. Integrating over $p_T$
results in the rapidity distribution $dN_i/dy$. Particle rapidity
$y=sin h^{-1} \: p_L/M_T$ (where $M_T=\sqrt{m^2 + p_T^2}$), requires mass identification. If that is
unknown one employs pseudo-rapidity ($\eta = - \mbox{ln} \:
[\mbox{tan}(\theta/2)]$) instead. This is also chosen if the joint
rapidity distribution of several unresolved particle species is considered:
notably the charged hadron distribution. We show two 
examples in Fig.~\ref{fig:Figure7}. The left panel illustrates charged particle
production in $p\overline{p}$ collisions studied by UA1 at
$\sqrt{s}=540 \:GeV$ \cite{51}. Whereas the minimum bias
distribution (dots) exhibits the required symmetry about the center
of mass coordinate, $\eta=0$, the rapidity distribution
corresponding to events in which a $W$ boson was produced
(histogram) features, both, a higher average charged particle yield,
and an asymmetric shape. The former effect can be seen to reflect
the expectation that the $W$ production rate increases with the
''centrality'' of $p\overline{p}$ collisions, involving more
primordial partons as the collisional overlap of the partonic
density profiles gets larger, thus also increasing the overall,
softer hadro-production rate. The asymmetry should result from a
detector bias favoring $W$ identification at negative rapidity: the
transverse $W$ energy, of about $100 \: GeV$ would {\it locally}
deplete the energy store available for associated soft production.
If correct, this interpretation suggests that the wide rapidity gap
between target and projectile, arising at such high $\sqrt{s}$, of
width $\Delta y \approx 2 \: ln \:(2 \gamma_{CM})$, makes it
possible to define local sub-intervals of rapidity within which the
species composition of produced particles varies.

\begin{figure}   
\begin{center}
\includegraphics[scale=0.3]{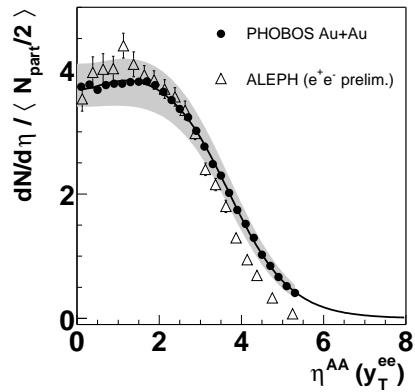}
\caption{Pseudo-rapidity distribution of charged hadrons produced in central
Au+Au collisions at $\sqrt{s}=200 \: GeV$ compared with $e^+e^-$
data at similar energy. The former data normalized by $N_{part}/2$.
From ref. \cite{53}.}
\label{fig:Figure8}
\end{center}
\end{figure}

\begin{figure}   
\begin{center}
\includegraphics[scale=0.27]{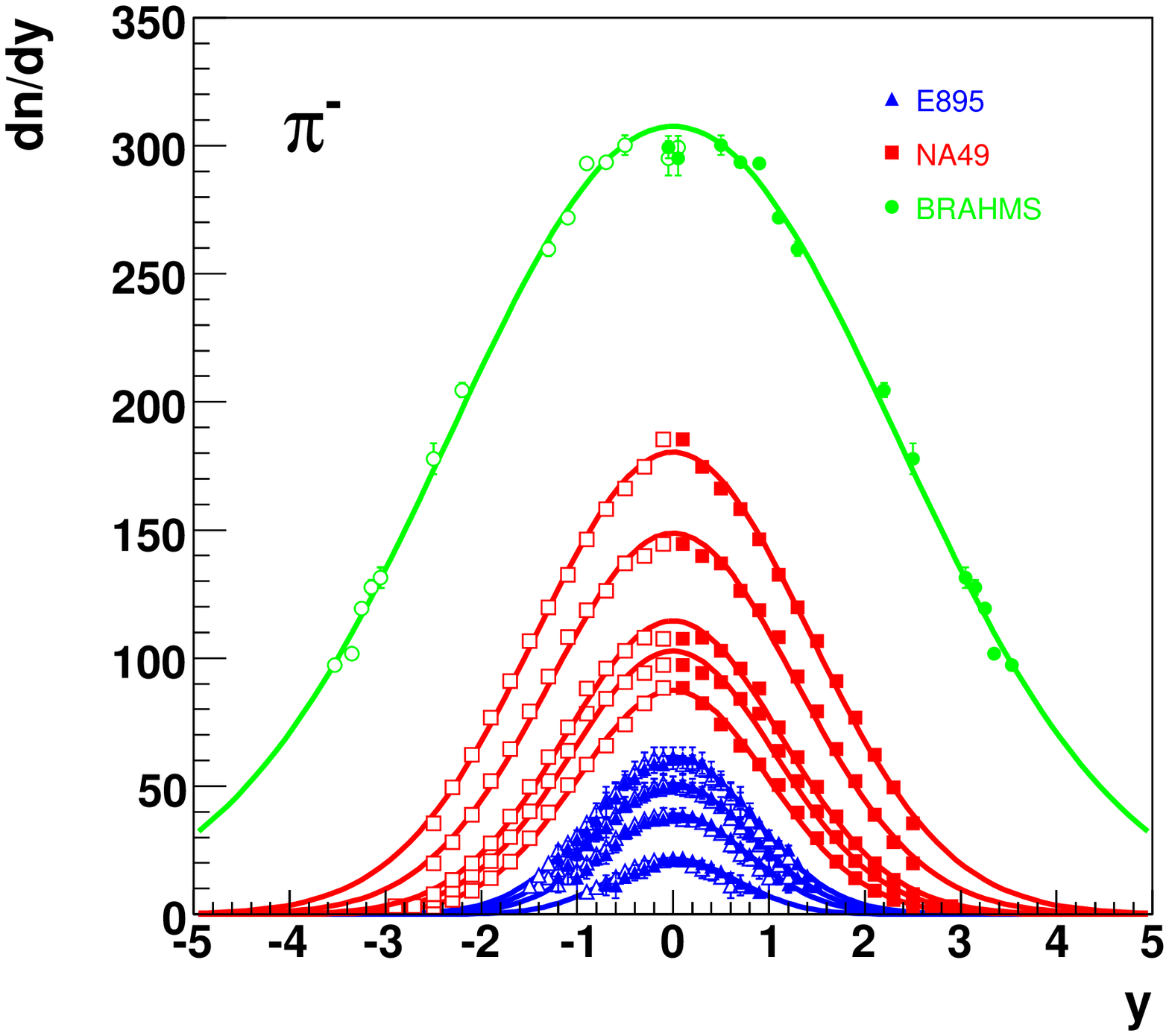}
\includegraphics[scale=0.27]{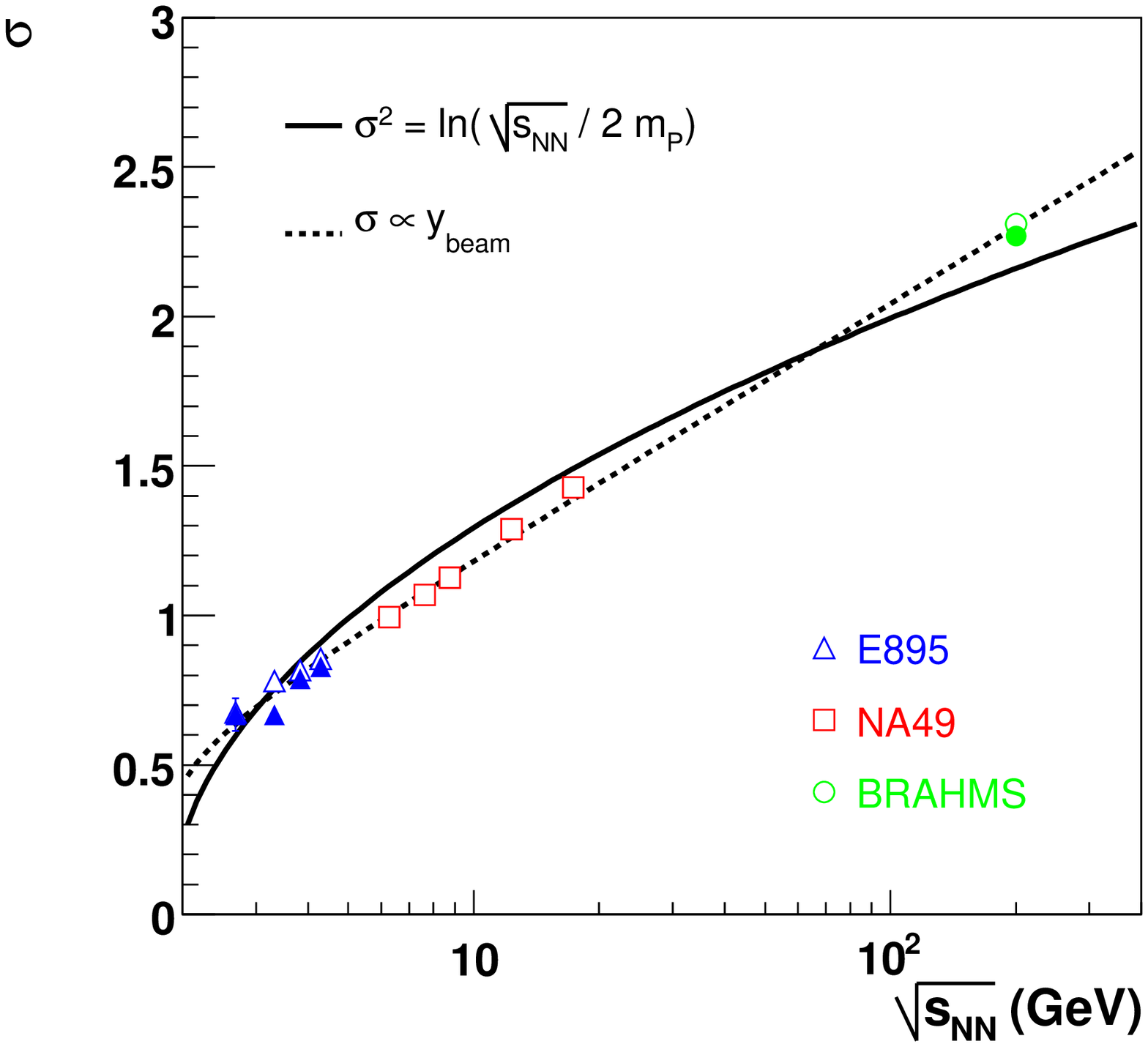}
\caption{Left panel: negative pion rapidity distributions in central Au+Au
and Pb+Pb collisions from AGS via SPS to RHIC energies \cite{54}.
Right panel: the Gaussian rapidity width of pions versus $\sqrt{s}$,
confronted by Landau model predictions (solid line) \cite{54}.}
\label{fig:Figure9}
\end{center}
\end{figure}

The right panel of Fig.~\ref{fig:Figure7} shows charged particle pseudo-rapidity
density distributions for Au+Au collisions at $\sqrt{s}=130 \: GeV$
measured by RHIC experiment PHOBOS \cite{52} at three different
collision centralities, from ''central'' (the 6\% highest charged
particle multiplicity events) to semi-peripheral (the corresponding
35-45\% cut). We will turn to centrality selection in more detail
below. Let us first remark that the slight dip at mid-rapidity and,
moreover, the distribution shape in general, are common to
$p\overline{p}$ and Au+Au. This is also the case for $e^+e^-$ annihilation as is shown in Fig.~\ref{fig:Figure8} which compares the ALEPH rapidity distribution along the mean $p_T$ (``thrust") axis
of jet production in $e^+e^-$ at $\sqrt{s}=200 \: GeV$ \cite{49}
with the scaled PHOBOS-RHIC distribution of central Au+Au at the
same $\sqrt{s}$ \cite{53}. Note that the mid-rapidity values contained in Figs.~\ref{fig:Figure7} and~\ref{fig:Figure8} have been employed already in Fig.~\ref{fig:Figure4}, which
showed the overall $\sqrt{s}$ dependence of mid-rapidity charged
particle production. What we concluded there was a perfect scaling
of A+A with $e^+e^-$ data at $\sqrt{s} \ge 20 \: GeV$ and a 40\%
suppression of the corresponding $pp, \: p\overline{p}$ yields. We
see here that this observation holds, semi-quantitatively, for the
entire rapidity distributions. These are not ideally boost invariant
at the energies considered here but one sees in $dN_{ch}/dy$ a
relatively smooth ''plateau'' region extending over $\mid y \mid \le
1.5 -2.5$. \\ \\
The production spectrum of charged hadrons is, by far, dominated by
soft pions ($p_T \le 1 \: GeV/c)$ which contribute about 85\% of the
total yield, both in elementary and nuclear collisions. The
evolution of the $\pi^-$ rapidity distribution with $\sqrt{s}$ is
illustrated in Fig.~\ref{fig:Figure9} for central Au+Au and Pb+Pb collisions from AGS
via SPS to RHIC energy, $2.7 \le \sqrt{s} \le 200 \: GeV$ \cite{54}. \\
\\
At lower $\sqrt{s}$ the distributions are well described by single
Gaussian fits \cite{54} with $\sigma (y)$ nearly linearly
proportional to the total rapidity gap $\Delta y \propto \: ln
\sqrt{s}$ as shown in the right hand panel of Fig.~\ref{fig:Figure9}. Also
illustrated is the prediction of the schematic hydrodynamical model
proposed by Landau \cite{55},
\begin{equation}
\sigma^2 \propto ln \: (\frac{\sqrt{s}}{2m_p})
\label{eq:equation4}
\end{equation}
which pictures hadron production in high $\sqrt{s}$ $pp$ collisions
to proceed via a dynamics of initial complete ''stopping down'' of
the reactants matter/energy content in a mid-rapidity fireball that
would then expand via 1-dimensional ideal hydrodynamics. Remarkably,
this model that has always been considered a wildly extremal
proposal falls rather close to the lower $\sqrt{s}$ data for central
A+A collisions but, as longitudinal phase space widens approaching
boost invariance we expect that the (non-Gaussian) width of the
rapidity distribution grows linearly with the rapidity gap $\Delta
y$. LHC data will finally confirm this expectation, but Figs.~\ref{fig:Figure7} to \ref{fig:Figure9}
clearly show the advent of boost invariance, already at $\sqrt{s} =
200 \: GeV$.

\begin{figure}
\begin{center}
\includegraphics[scale=0.3]{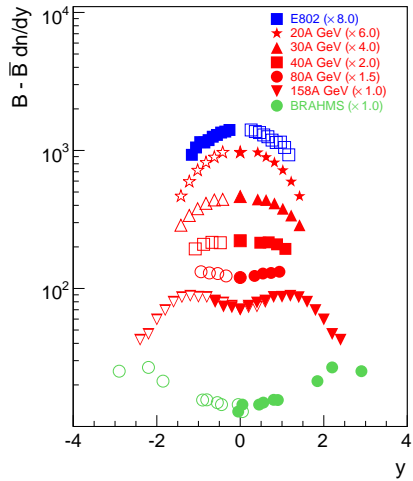}
\caption{Net proton rapidity distributions in central Au+Au/Pb+Pb collisions
at AGS, SPS and RHIC energies \cite{56, 57}.}
\label{fig:Figure10}
\end{center}
\end{figure}

A short didactic aside: At low $\sqrt{s}$ the total rapidity gap
$\Delta y = 2-3$ does closely resemble the total rapidity width
obtained for a thermal pion velocity distribution at temperature
$T=120-150 \: MeV$, of a single mid-rapidity fireball, the
y-distribution of which represents the longitudinal component according to
the relation \cite{19}
\begin{equation}
\frac{dN}{dy} \propto (m^2T \:+ \frac{2mT^2}{coshy} \: +
\frac{2T^2}{cosh^2y}) \: exp \: [-m \cdot coshy/T]
\label{eq:equation5}
\end{equation}
where $m$ is the pion mass. {\it Any} model of preferentially
longitudinal expansion of the pion emitting source, away from a
trivial single central ''completely stopped'' fireball, can be
significantly tested only once $\Delta y > 3$ which occurs upward from SPS
energy. The agreement of the Landau model prediction with the data
in Fig.~\ref{fig:Figure9} is thus fortuitous, below $\sqrt{s} \approx 10 \: GeV$, as \emph{any} created fireball occupies the entire rapidity gap with pions.\\
\\
The Landau model offers an extreme view of the mechanism of
''stopping'', by which the initial longitudinal energy of the
projectile partons or nucleons is inelastically transferred to
produced particles and redistributed in transverse and longitudinal
phase space, of which we saw the total transverse fraction in Fig.~\ref{fig:Figure3}.
Obviously $e^+e^-$ annihilation to hadrons represents the extreme
stopping situation. Hadronic and nuclear collisions offer the
possibility to analyze the final distribution in phase space of
their non-zero net quantum numbers, notably net baryon number. Figure~\ref{fig:Figure10} shows
the net-proton rapidity distribution (i.e. the proton rapidity
distribution subtracted by the antiproton distribution) for central
Pb+Pb/Au+Au collisions at AGS ($\sqrt{s}=5.5 \: GeV$), SPS
($\sqrt{s} \le 17.3 \: GeV$) and RHIC ($ \sqrt{s}=200 \: GeV$)
\cite{56}. With increasing energy we see a central (but
non-Gaussian) peak developing into a double-hump structure that
widens toward RHIC leaving a plateau about mid-rapidity. The RHIC-BRAHMS experiment
acceptance for $p, \: \overline{p}$ identification does
unfortunately not reach up to the beam fragmentation domain at
$y_p=5.4$ (nor does any other RHIC experiment) but only to $y
\approx 3.2$, with the consequence that the major fraction of
$p^{net}$ is not accounted for. However the mid-rapidity region is
by no means net baryon free. At SPS energy the NA49 acceptance
covers the major part of the total rapidity gap, and we observe in
detail a net $p$ distribution shifted down from $y_p=2.9$ by an
average rapidity shift \cite{56} of $\left<\delta y\right> = 1.7$. From Fig.~\ref{fig:Figure10}
we infer that $\left<\delta y\right>$ can not scale linearly with $y_p \approx
ln (2\gamma_{CM}) \approx ln \sqrt{s}$ for ever  - as it does up to
top SPS energy where $\left<\delta y\right>=0.58 \: y_p$ \cite{56}. Because
extrapolating this relation to $\sqrt{s}=200 \: GeV$ would result in
$\left<\delta y\right>=3.1$, and with $y_p \approx 5.4$ at this energy we would
expect to observe a major fraction of net proton yield in the
vicinity of $y=2.3$ which is not the case. A saturation must thus
occur in the $\left<\delta y\right>$ vs. $\sqrt{s}$ dependence.\\
\\
The re-distribution of net baryon density over longitudinal phase
space is, of course, only partially captured by the net proton yield
but a recent study \cite{57} has shown that proper inclusion of
neutron\footnote{Neutrons are not directly measured in the SPS and
RHIC experiments but their production rate, relative to protons,
reflects in the ratio of tritium to $^3He$ production measured by
NA49 \cite{57}, applying the isospin mirror symmetry of the
corresponding nuclear wave functions.} and hyperon production data
at SPS and RHIC energy scales up, of course, the $dN/dy$
distributions of Fig.~\ref{fig:Figure10} but leaves the peculiarities of their shapes
essentially unchanged. As the net baryon rapidity density
distribution should resemble the final valence quark distribution
the Landau model is ruled out as the valence quarks are seen to be
streaming from their initial position at beam rapidity toward
mid-rapidity (not vice versa). It is remarkable, however, to see
that some fraction gets transported very far, during the primordial
partonic non-equilibrium phase. We shall turn to its theoretical
description in section~\ref{subsec:Gluon_Satu_in_AA_Coll} but note, for now, that $pp$ collisions
studied at the CERN ISR \cite{58} lead to a qualitatively similar
net baryon rapidity distribution, albeit characterized by a smaller
$\left<\delta y\right>$.

The data described above suggest that the stopping mechanism universally resides in the primordial, first generation of collisions at the microscopic level. 
The rapidity distributions of charged particle
multiplicity, transverse energy and valence quarks exhibit
qualitatively similar shapes (which also evolve similarly with
$\sqrt{s}$) in $pp, \: p\overline{p}, \: e^+e^-$ reactions, on the
one hand, and in central or semi-peripheral collisions of $A \approx
200$ nuclei, on the other. Comparing in detail we formulate a
nuclear modification factor for the bulk hadron rapidity distributions,
\begin{equation}
R^{AA}_y \:  \equiv \: \frac{dN^{ch}/dy \: (y) \:\:\:\: \mathrm{in} \:\: A+A}{0.5
\:  N_{part} \:\: dN^{ch}/dy \:\:\:\: \mathrm{in} \:\:  pp}
\label{eq:equation6}
\end{equation}
where $N_{part} < 2A$ is the mean number of ''participating
nucleons'' (which undergo at least one inelastic collision with another nucleon) which
increases with collision centrality. For identical nuclei colliding
$\left<N^{proj}_{part}\right> \simeq \left<N^{targ}_{part}\right>$ and thus $0.5 \: N_{part}$
gives the number of opposing nucleon pairs. $R^{AA}=1$ if each such
''opposing'' pair contributes the same fraction to the total A+A
yield as is produced in minimum bias $pp$ at similar $\sqrt{s}$.
From Figs.~\ref{fig:Figure4} and~\ref{fig:Figure6} we infer that for $\mid \eta \mid < 1, \: \: 
R^{AA}=1.5$ at top RHIC energy, and for the pseudo-rapidity
integrated total $N^{ch}$ we find $R^{AA} = 1.36$, in central Au+Au
collisions. AA collisions thus provide for a higher stopping power than $pp$
(which is also reflected in the higher rapidity shift $\left<\delta y\right>$
of Fig.~\ref{fig:Figure10}). The observation that their stopping  power resembles the
$e^+e^-$ inelasticity suggests a substantially reduced leading
particle effect in central collisions of heavy nuclei. This might
not be surprising. In a Glauber-view of
successive minimum bias nucleon collisions occuring during
interpenetration, each participating nucleon is struck $\nu > 3$
times on average, which might saturate the possible inelasticity,
removing the leading fragment.

This view naturally leads to the scaling of the total particle
production in nuclear collisions with $N_{part}$, as seen clearly in
Fig.~\ref{fig:Figure6}, reminiscent of the ''wounded nucleon model'' \cite{59} but
with the scaling factor determined by $e^+e^-$ rather than $pp$
\cite{60}. Overall we conclude from the still rather close
similarity between nuclear and elementary collisions that the
mechanisms of longitudinal phase space population occur
primordially, during interpenetration which is over after $0.15 \:
fm/c$ at RHIC, and after $1.5 \: fm/c$ at SPS energy. I.e. it is the
primordial non-equilibrium pQCD shower evolution that accounts for
stopping, and its time extent should be a lower limit to the
formation time $\tau_0$ employed in the Bjorken model \cite{45},
equation~\ref{eq:equation1}. Equilibration at the partonic level might begin at
$t>\tau_0$ only (the development toward a quark-gluon-plasma phase),
but the primordial parton redistribution processes set the stage for
this phase, and control the relaxation time scales involved in
equilibration \cite{61}. More about this in section~\ref{subsec:Transvers_phase_space}. We infer
the existence of a saturation scale \cite{62}
controlling the total inelasticity: with ever higher reactant
thickness, proportional to $A^{1/3}$, one does not get a total
rapidity or energy density proportional to $A^{4/3}$ (the number of
''successive binary collisions'') but to $A^{1.08}$ only \cite{63}.
Note that the lines shown in Fig.~\ref{fig:Figure7} (right panel) refer to such a
saturation theory: the color glass condensate (CGC) model \cite{64}
developed by McLerran and Venugopulan. The success of these models
demonstrates that ''successive binary baryon scattering'' is not an
appropriate picture at high $\sqrt{s}$. One can free the partons
from the nucleonic parton density distributions only {\it once},
and their corresponding transverse areal density sets the
stage for the ensuing QCD parton shower evolution \cite{62}.
Moreover, an additional saturation effect appears to modify this evolution at high transverse areal parton density (see section~\ref{subsec:Gluon_Satu_in_AA_Coll}).

\subsection {Dependence on system size}
\label{subsec:Dep_on_sys_size}
We have discussed above a first attempt toward a variable
($N_{part}$) that scales the system size dependence in A+A
collisions. Note that one can vary the size either by centrally
colliding a sequence of nuclei, $A_1 + A_1, \: A_2 + A_2$ etc., or
by selecting different windows in $N_{part}$ out of minimum bias
collision ensembles obtained for heavy nuclei for which BNL employs
$^{197}Au$ and CERN $^{208}Pb$. The third alternative, scattering a
relatively light projectile, such as $^{32}S$, from increasing $A$
nuclear targets, has been employed initially both at the AGS and SPS
but got disfavored in view of numerous disadvantages, of both
experimental (the need to measure the entire rapidity distribution,
i.e. lab momenta from about 0.3-100 $GeV$/c, with uniform efficiency)
and theoretical nature (different density distributions of projectile and target; 
occurence of an''effectiv'' center of mass, different for hard and soft collisions, 
and depending on impact parameter).

The determination of $N_{part}$ is of central interest, and thus we
need to look at technicalities, briefly. 
The approximate linear scaling with $N_{part}$ that we observed
in the total transverse energy and the total charged particle number
(Figs.~\ref{fig:Figure3},~\ref{fig:Figure6}) is a reflection of the primordial redistribution of
partons and energy. Whereas all observable properties that refer to
the system evolution at later times, which are of interest as
potential signals from the equilibrium, QCD plasma ''matter'' phase,
have different specific dependences on $N_{part}$, be it
suppressions (high $p_T$ signals, jets, quarkonia production) or
enhancements (collective hydrodynamic flow, strangeness production).
$N_{part}$ thus emerges as a suitable common reference scale.

$N_{part}$ captures the number of potentially directly
hit nucleons. It is estimated from an eikonal straight trajectory Glauber
model as applied to the overlap region arising, in dependence of
impact parameter $b$, from the superposition along beam direction of
the two initial Woods-Saxon density distributions of the interacting
nuclei. To account for the dilute surfaces of these distributions
(within which the intersecting nucleons might not find an
interaction partner) each incident nucleon trajectory gets equipped
with a transverse radius  that represents the total inelastic NN
cross section at the corresponding $\sqrt{s}$. The formalism is
imbedded into a Monte Carlo simulation (for detail see \cite{66})
starting from random microscopic nucleon positions within the
transversely projected initial Woods-Saxon density profiles. 
Overlapping cross sectional tubes of target and projectile nucleons 
are counted as a participant nucleon pair. Owing to the statistics of nucleon initial position
sampling each considered impact parameter geometry thus results in a
probability distribution of derived $N_{part}$. Its width $\sigma$
defines the resolution $\Delta (b)$ of impact parameter $b$
determination within this scheme via the relation
\begin{equation}
\frac{1}{\Delta (b)} \: \sigma (b) \approx \frac{d\left<N_{part}
(b)\right>}{db}
\label{eq:equation7}
\end{equation}
which, at A=200, leads to the expectation to determine $b$ with
about 1.5 $fm$ resolution \cite{66}, by measuring $N_{part}$.

\begin{figure}   
\begin{center}
\includegraphics[scale=0.85]{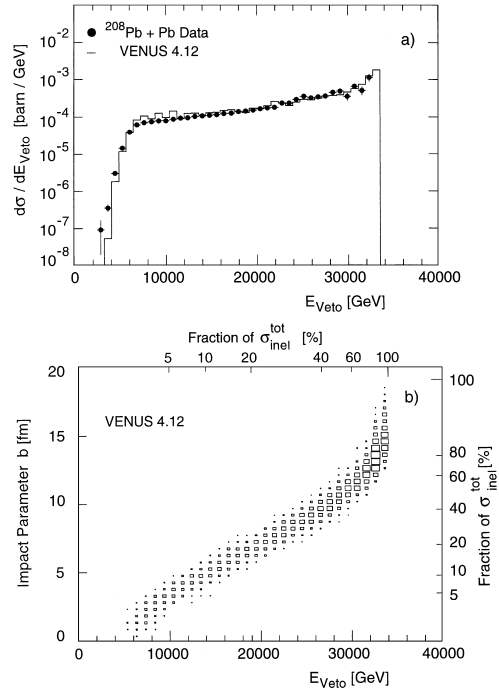}
\caption{(a) Energy spectrum of the forward calorimeter in Pb+Pb collisions
at 158$A$ GeV; (b) impact parameter and fraction of
total inelastic cross section related to forward energy from the
VENUS model \cite{67,68}.}
\label{fig:Figure11}
\end{center}
\end{figure}

How to measure $N_{part}$? In fixed target experiments one can
calorimetrically count all particles with beam momentum per nucleon
and superimposed Fermi momentum distributions of nucleons, i.e. one
looks for particles in the beam fragmentation domain $y_{beam} \pm
0.5, \:\: p_T\le 0.25 \: GeV/c$. These are identified as spectator
nucleons, and $N^{proj}_{part}=A-N^{proj}_{spec}$. For identical
nuclear collision systems $\left<N_{part}^{proj}\right> = \left<N_{part}^{targ}\right>$,
and thus $N_{part}$ gets approximated by 2 $N_{part}^{proj}$. This
scheme was employed in the CERN experiments NA49 and WA80, and
generalized \cite{67} in a way that is illustrated in Fig.~\ref{fig:Figure11}. \\
\\
The top panel shows the minimum bias distribution of total energy
registered in a forward calorimeter that covers the beam fragment domain in 
Pb+Pb collisions at lab. energy of 158 $GeV$ per projectile nucleon, $\sqrt{s}=17.3 \:
GeV$. The energy spectrum extends from about $3 \: TeV$ which
corresponds to about 20 projectile spectators (indicating a
''central'' collision), to about 32 $TeV$ which is close to the
total beam energy and thus corresponds to extremely peripheral
collisions. Note that the shape of this forward energy spectrum is
the mirror image of the minimum bias transverse energy distribution
of Fig.~\ref{fig:Figure3}, both recorded by NA49. From both figures we see that the
{\it ideal} head-on, $b \rightarrow 0$ collision can not be selected
from these (or any other) data, owing to the facts that $b=0$
carries zero geometrical weight, and that the diffuse Woods-Saxon
nuclear density profiles lead to a fluctuation of participant
nucleon number at given finite $b$. Thus the $N_{part}$ fluctuation
at finite weight impact parameters overshadows the genuinely small
contribution of near zero impact parameters. Selecting ''central''
collisions, either by an on-line trigger cut on minimal forward
energy or maximal total transverse energy or charged particle
rapidity density, or by corresponding off-line selection, one thus faces a compromise between event
statistics and selectivity for impact parameters near zero. In the
example of Fig.~\ref{fig:Figure11} these considerations suggest a cut at about $8 \:
TeV$ which selects the 5\% most inelastic events, from among the
overall minimum bias distribution, then to be labeled as ''central''
collisions. This selection corresponds to a soft cutoff at $b \le 3
\: fm$.

\begin{figure}
   \begin{minipage}[t]{55mm}
      \includegraphics*[width=5.34cm, height=4.71cm]{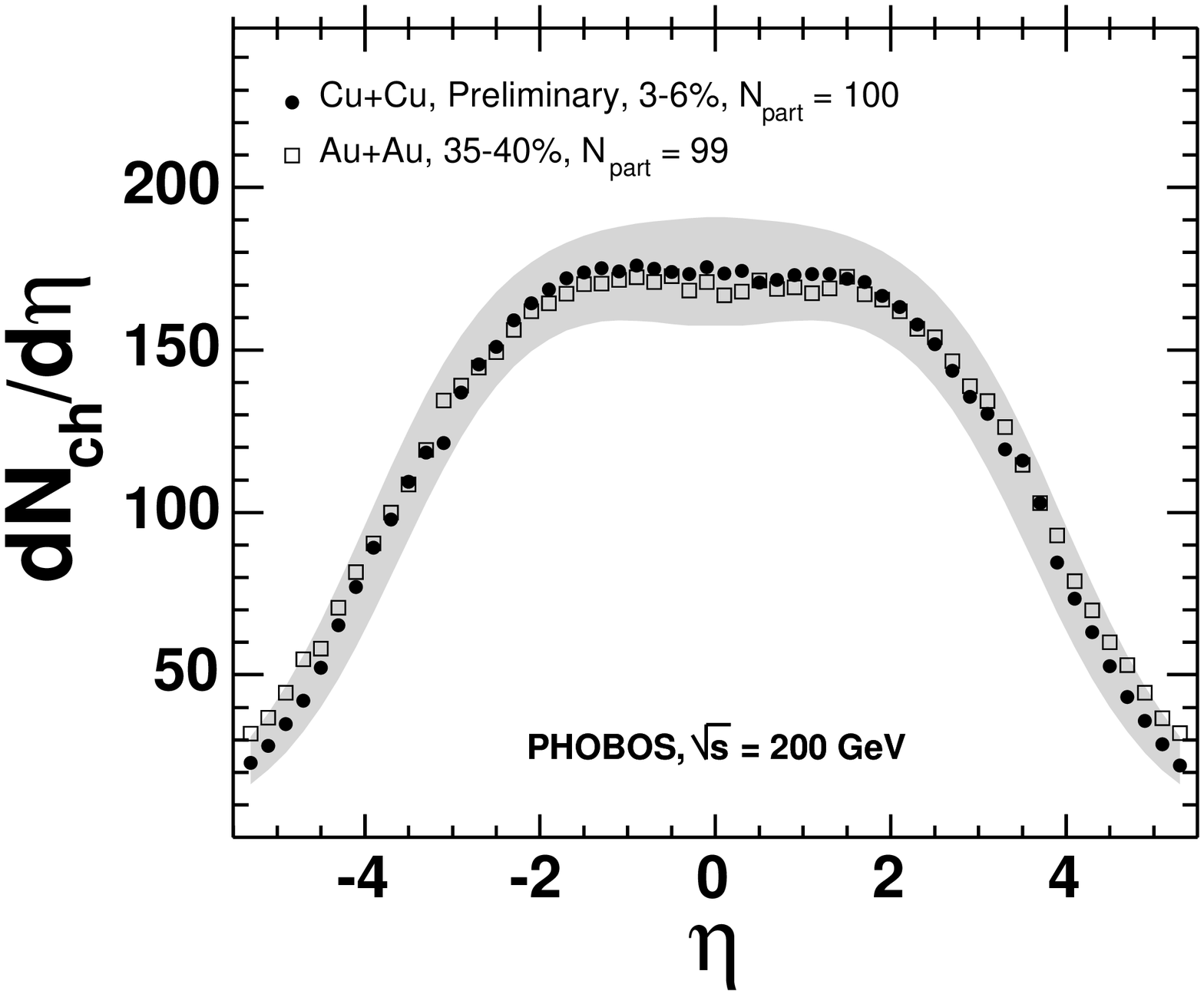}    
      \caption{Charged hadron pseudo-rapidity distributions in Cu+Cu and Au+Au
collisions at $\sqrt{s}=200 \: GeV$, with similar $N_{part} \approx
100$ \cite{52}.}
      \label{fig:Figure12}
   \end{minipage}
      \hspace{\fill}
  \begin{minipage}[t]{55mm}
      \includegraphics[width=5.19cm]{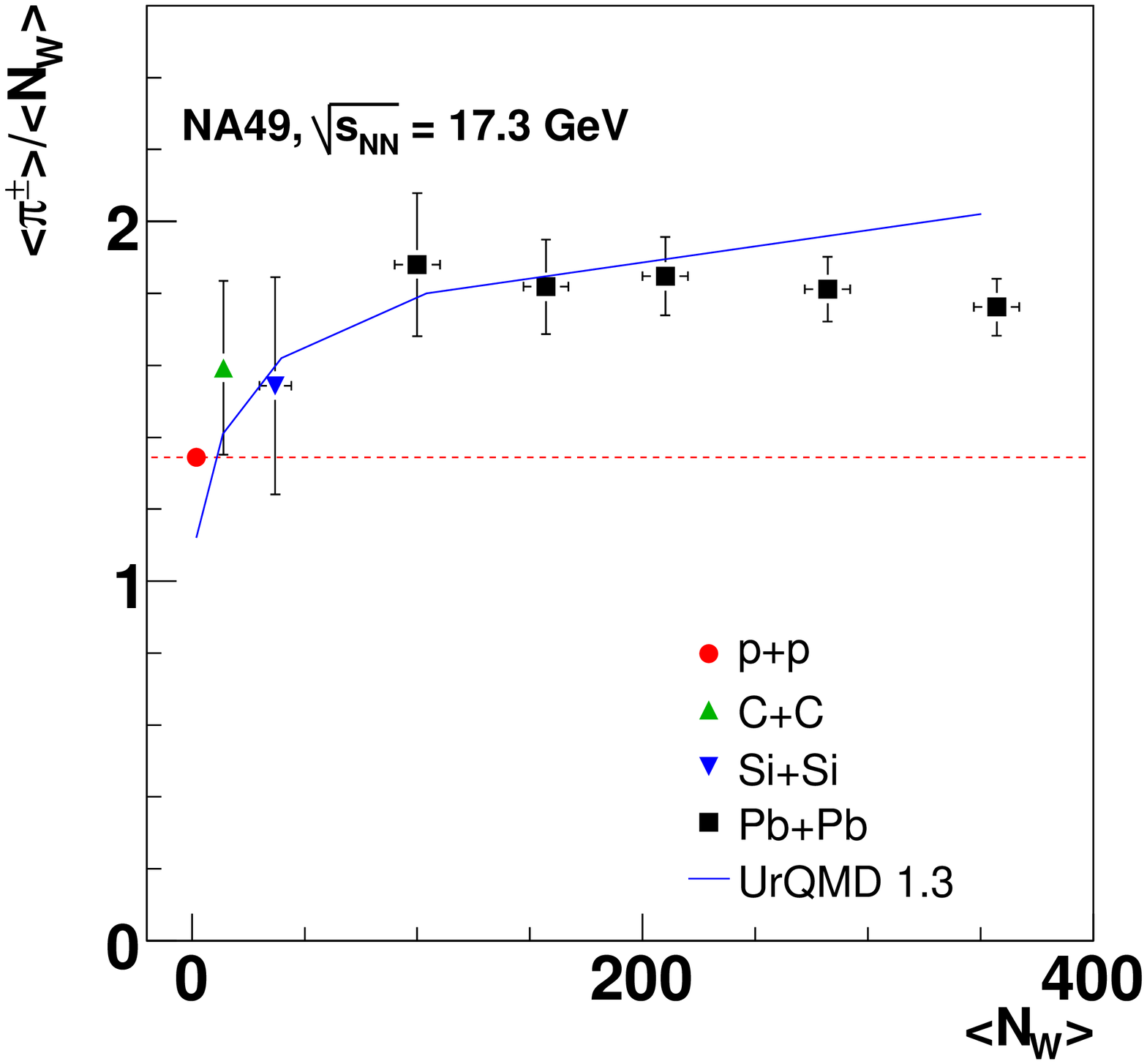}
      \caption{Charged pion multiplicity normalized by $N_W$ vs. centrality in p+p,
C+C, Si+Si and Pb+Pb collisions at $\sqrt{s}=17.3 \: GeV$
\cite{67,69}.}
      \label{fig:Figure13}
   \end{minipage}
\end{figure}

The selectivity of this, or of other less stringent cuts on
collision centrality is then established by comparison to a Glauber
or cascade model. The bottom panel of Fig.~\ref{fig:Figure11} employs the VENUS
hadron/string cascade model \cite{68} which starts from a Monte
Carlo position sampling of the nucleons imbedded in Woods-Saxon
nuclear density profiles but (unlike in a Glauber scheme with
straight trajectory overlap projection) following the cascade of
inelastic hadron/string multiplication, again by Monte Carlo
sampling. It reproduces the forward energy data reasonably well and
one can thus read off the average impact parameter and participant
nucleon number corresponding to any desired cut on the percent
fraction of the total minimum bias cross section. Moreover, it is
clear that this procedure can also be based on the total minimum
bias transverse energy distribution, Fig.~\ref{fig:Figure3}, which is the mirror image
of the forward energy distribution in Fig.~\ref{fig:Figure11}, or on the total, and
even the mid-rapidity charged particle density (Fig.~\ref{fig:Figure6}). The latter
method is employed by the RHIC experiments STAR and PHENIX. \\ \\
How well this machinery works is illustrated in Fig.~\ref{fig:Figure12} by RHIC-PHOBOS
results at $\sqrt{s}=200 \: GeV$ \cite{52}. The charged particle
pseudo-rapidity density distributions are shown for central (3-6\%
highest $N_{ch}$ cut) Cu+Cu collisions, with $\left<N_{part}\right>=100$, and
semi-peripheral Au+Au collisions selecting the cut window (35-40\%)
such that the same $\left<N_{part}\right>$ emerges. The distributions are
nearly identical. In extrapolation to $N_{part}=2$ one would expect
to find agreement between min. bias $p+p$, and ''super-peripheral''
A+A collisions, at least at high energy where the nuclear Fermi
momentum plays no large role. Fig.~\ref{fig:Figure13} shows that this expectation is
correct \cite{69}. As it is technically difficult to select
$N_{part}=2$ from A=200 nuclei colliding, NA49 fragmented the
incident SPS Pb beam to study $^{12}C+^{12}C$ and $^{28}Si+^{28}Si$
collisions \cite{67}. These systems are isospin symmetric, and
Fig.~\ref{fig:Figure13} thus plots $0.5(\left<\pi^+\right>+\left<\pi^-\right>)/\left<N_W\right>$ including
$p+p$ where $N_W=2$ by definition. We see that the pion multiplicity of A+A collisions
interpolates to the p+p data point. \\
\\
Note that NA49 employs the term ''wounded nucleon'' number ($N_{W}$) to count the 
nucleons that underwent at least one inelastic {\it nucleon-nucleon} collision. This is what
the RHIC experiments (that follow a Glauber model) call $N_{part}$
whereas NA49 reserves this term for nucleons that underwent {\it
any} inelastic collision.  Thus $N_W$ in
Fig.~\ref{fig:Figure13} has the same definition as $N_{part}$ in Figs.~\ref{fig:Figure4},~\ref{fig:Figure6},~\ref{fig:Figure8},~\ref{fig:Figure12}. We
see that a smooth increase joins the $p+p$ data, via the light A+A
central collisions, to a saturation setting in with semi-peripheral
Pb+Pb collisions, the overall, relative increase amounting to about
40\% (as we saw in Fig.~\ref{fig:Figure4}).

There is nothing like an $N_{part}^{1/3}$ increase (the
thickness of the reactants) observed here, pointing to the saturation
mechanism(s) mentioned in the previous section, which are seen from
Fig.~\ref{fig:Figure13} to dampen the initial, fast increase once the primordial
interaction volume contains about 80 nucleons. In the Glauber model
view of successive collisions (to which we attach only symbolical
significance at high $\sqrt{s}$) this volume corresponds to $\left<\nu\right>
\approx 3$, and within the terminology of such models we might thus
argue, intuitively, that the initial {\it geometrical} cross
section, attached to the nucleon structure function as a whole, has
disappeared at $\left<\nu\right> \approx 3$, all constituent partons being
freed.

\subsection {Gluon Saturation in A+A Collisions}
\label{subsec:Gluon_Satu_in_AA_Coll}
We will now take a closer look at the saturation phenomena of high
energy QCD scattering, and apply results obtained for deep inelastic
electron-proton reactions to nuclear collisions, a procedure that
relies on a universality of high energy scattering. This
arises at high $\sqrt{s}$, and at relatively low momentum
transfer squared $Q^2$ (the condition governing bulk charged particle
production near mid-rapidity at RHIC, where Feynman $x \approx 0.01$
and $Q^2 \le 5 \: GeV^2$). Universality comes
about as the transverse resolution becomes higher and higher, with
$Q^2$, so that within the small area tested by the collision there
is no difference whether the partons sampled there belong to the
transverse gluon and quark density projection of any hadron species,
or even of a nucleus. And saturation arises once the areal
transverse parton density exceeds the resolution, leading to
interfering QCD sub-amplitudes that do not reflect in the total
cross section in a manner similar to the mere summation of separate\footnote{Note that QCD considers interactions only of single charges or charge-anticharge pairs.},
resolved color charges~\cite{61,62,63,64,65,70,71}.

\begin{figure}   
\begin{center}
\includegraphics[scale=0.26]{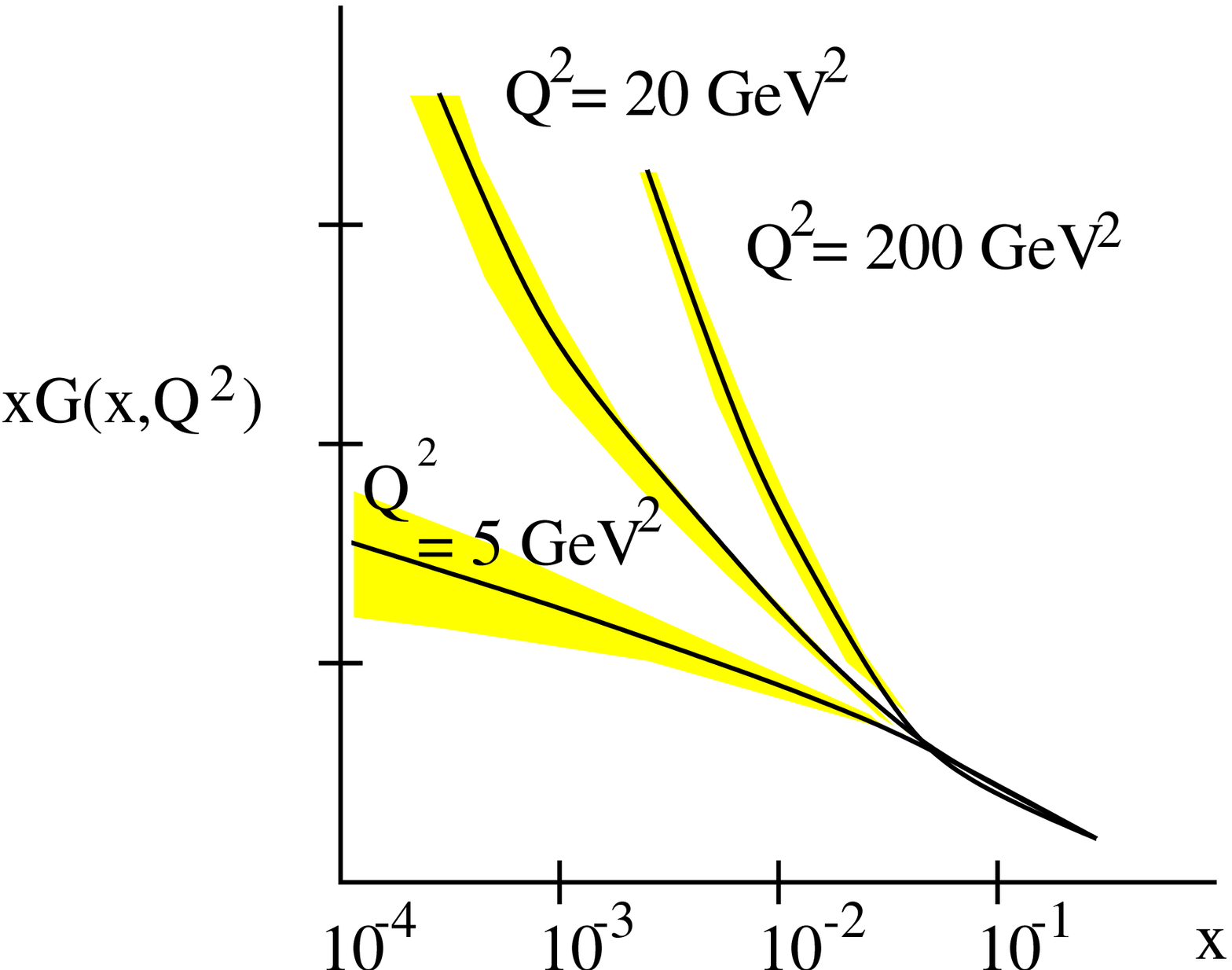}
\includegraphics[scale=0.2]{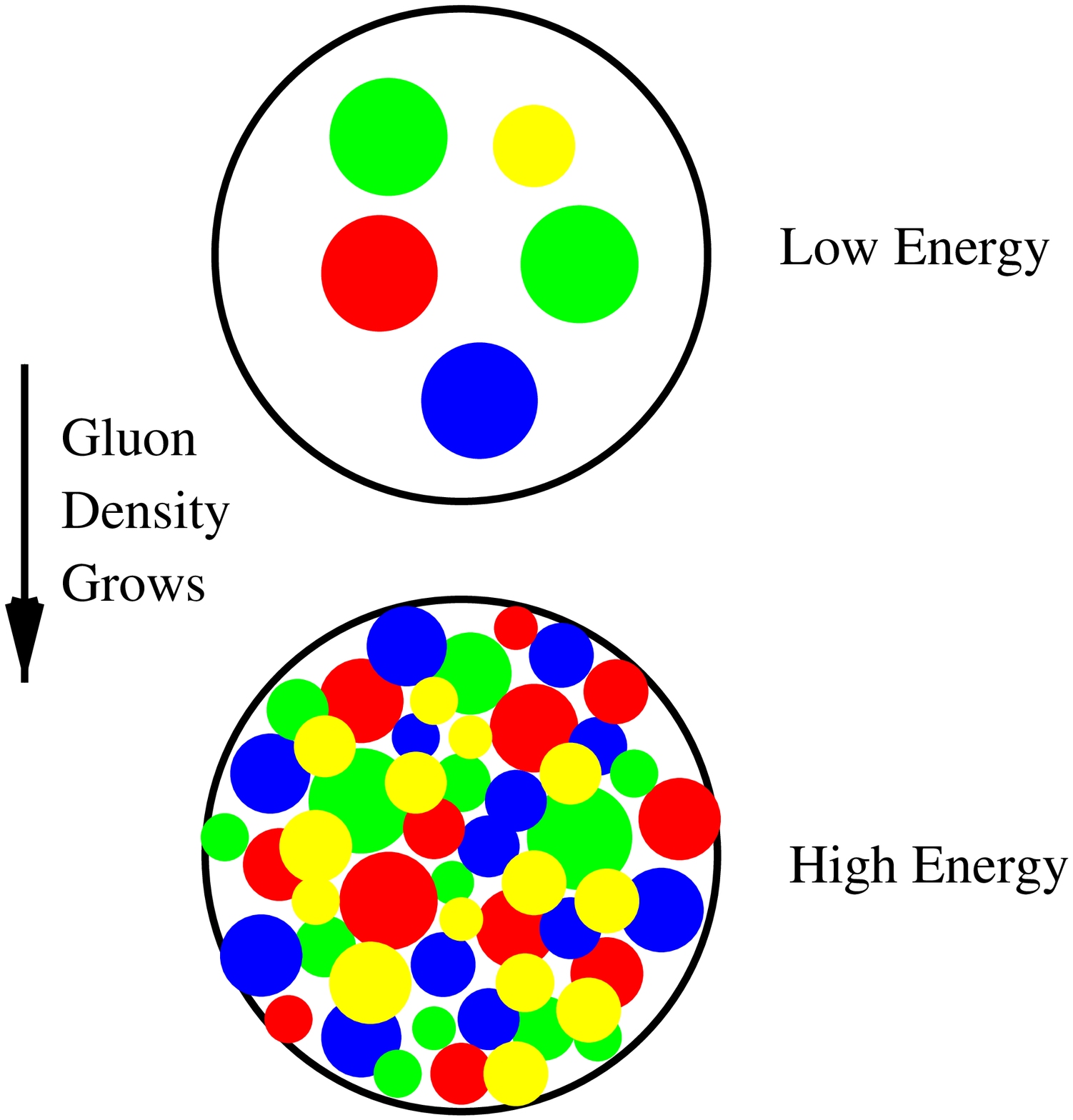}
\caption{(left) The HERA data for the gluon distribution function as a function of
fractional momentum $x$ and square momentum transfer $Q^2$ \cite{72}. (right)
Saturation of gluons in a hadron; a head on view as $x$ decreases
\cite{75}.}
\label{fig:Figure14}
\end{center}
\end{figure}

The ideas of saturation and universality are motivated by HERA deep
inelastic scattering (DIS) data \cite{72} on the gluon distribution
function shown in Fig.~\ref{fig:Figure14} (left side). The gluon rapidity density,
$xG(x,Q^2)=\frac{dN^{gluon}}{dy}$ rises rapidly as a function of
decreasing fractional momentum, $x$, or increasing resolution,
$Q^2$. The origin of this rise in the gluon density is, ultimately,
the non-abelian nature of QCD. Due to the intrinsic non-linearity of
QCD \cite{70,71}, gluon showers generate more gluon showers,
producing an avalanche toward small $x$. As a consequence of this
exponential growth the spatial density of gluons (per unit
transverse area per unit rapidity) of any hadron or nucleus must
increase as $x$ decreases \cite{65}. This follows because
the transverse size, as seen via the total cross section, rises more
slowly toward higher energy than the number of gluons. This is
illustrated in Fig.~\ref{fig:Figure14} (right side). In a head-on view of a hadronic
projectile more and more partons (mostly gluons) appear as $x$
decreases. This picture reflects a representation of the hadron in
the ''infinite momentum frame'' where it has a large light-cone
longitudinal momentum $P^+ \gg M$. In this frame one can describe
the hadron wave function as a collection of constituents carrying a
fraction $p^+=xP^+, \: 0 \le x<1$, of the total longitudinal
momentum \cite{73} (''light cone quantization'' method~\cite{74}). In 
DIS at large $\sqrt{s}$ and $Q^2$ one measures the quark distributions $dN_q/dx$ at small
$x$, deriving from this the gluon distributions $xG(x, Q^2)$ of
Fig.~\ref{fig:Figure14}.

It is useful \cite{75} to consider the rapidity distribution implied
by the parton distributions, in this picture. Defining $y=y_{hadron}
- ln(1/x)$ as the rapidity of the potentially struck parton, the
invariant rapidity distribution results as
\begin{equation}
dN/dy = x \:dN/dx = xG(x,Q^2).
\label{eq:equation8}
\end{equation}
At high $Q^2$ the measured quark and gluon structure functions are
thus simply related to the number of partons per unit rapidity,
resolved in the hadronic wave function.\\
\\
The above textbook level \cite{74,75} recapitulation leads, however,
to an important application: the $dN/dy$ distribution of constituent
partons of a hadron (or nucleus), determined by the DIS experiments,
is similar to the rapidity distribution of produced particles in
hadron-hadron or A+A collisions as we expect the initial gluon
rapidity density to be represented in the finally observed, produced
hadrons, at high $\sqrt{s}$. Due to the longitudinal boost
invariance of the rapidity distribution, we can apply the above
conclusions to hadron-hadron or A+A collisions at high $\sqrt{s}$,
by replacing the infinite momentum frame hadron rapidity by the center of mass
frame projectile rapidity, $y_{proj}$, while retaining the result
that the rapidity density of potentially interacting partons grows
with increasing distance from $y_{proj}$ like
\begin{equation}
\Delta y \equiv y_{proj} - y = ln(1/x).
\label{eq:equation9}
\end{equation}
At RHIC energy, $\sqrt{s}= 200 \: GeV$, $\Delta y$ at mid-rapidity
thus corresponds to $x<10^{-2}$ (well into the domain of growing
structure function gluon density, Fig.~\ref{fig:Figure14}), and the two intersecting
partonic transverse density distributions thus attempt to
resolve each other given the densely packed situation that is
depicted in the lower circle of Fig.~\ref{fig:Figure14} (right panel). At given $Q^2$
(which is modest, $Q^2 \le 5 \: GeV^2$, for bulk hadron production
at mid-rapidity) the packing density at mid-rapidity will increase
toward higher $\sqrt{s}$ as
\begin{equation}
\Delta y^{midrap} \approx ln(\sqrt{s}/M), \: i.e. \: 1/x \approx
\sqrt{s}/M
\label{eq:equation10}
\end{equation}
thus sampling smaller $x$ domains in Fig.~\ref{fig:Figure14} according to equation~\ref{eq:equation9}. It
will further increase in proceeding from hadronic to nuclear
reaction partners A+A. Will it be in proportion to $A^{4/3}$?
We know from the previous sections (\ref{subsec:Rap_Distribution} and \ref{subsec:Dep_on_sys_size}) that this is not the case, the data indicating an increase with $A^{1.08}$. This observation is, in fact caused by the parton saturation effect, to which we turn now.

For given transverse resolution $Q^2$ and increasing $1/x$ the
parton density of Fig.~\ref{fig:Figure14} becomes so large that one can not neglect
their mutual interactions any longer. One expects such interactions
to produce ''{\it shadowing}'', a decrease of the scattering cross
section relative to incoherent independent scattering \cite{70,71}.
As an effect of such shadowed interactions there occurs \cite{75} a
{\it saturation} \cite{61,62,63,64,65,70,71,75} of the cross section
at each given $Q^2$, slowing the increase with $1/x$ to become
logarithmic once $1/x$ exceeds a certain critical value $x_s(Q^2)$.
Conversely, for fixed $x$, saturation occurs for transverse momenta
below some critical $Q^2(x)$,
\begin{equation}
Q_s^2(x) = \alpha_s N_c \: \frac{1}{\pi R^2} \: \frac{dN}{dy}
\label{eq:equation11}
\end{equation}
where $dN/dy$ is the $x$-dependent gluon density (at $y=y_{proj} -
ln(1/x))$. $Q^2_s$ is called the {\it saturation scale}. In equation~\ref{eq:equation11}
$\pi R^2$ is the hadron area (in transverse projection), and $\alpha
_s N_c$ is the color charge squared of a single gluon. More
intuitively, $Q^2_s (x)$ defines an inversely proportional
resolution area $F_s(x)$ and at each $x$ we have to choose $F_s(x)$
such that the ratio of total area $\pi R^2$ to $F_s(x)$ (the number
of resolved areal pixels) equals the number of single gluon charge
sources featured by the total hadron area. As a consequence the
saturation scale $Q^2_s(x)$ defines a critical areal resolution,
with two different types of QCD scattering theory defined, at each
$x$, for $Q^2 > Q^2_s$ and $Q^2 < Q^2_s$, respectively
\cite{62,65,75}.

As one expects a soft transition between such theories, to occur
along the transition line implied by $Q^2_s(x)$, the two types of
QCD scattering are best studied with processes featuring typical
$Q^2$ well above, or below $Q^2_s(x)$. Jet production at $\sqrt{s}
\ge 200 \: GeV$ in $p\overline{p}$ or AA collisions with typical
$Q^2$ above about $10^3 \: GeV^2$, clearly falls into the former
class, to be described e.g. by perturbative QCD DGLAP evolution of partonic
showers \cite{76}. The accronym DGLAP refers to the inventors of the perturbative QCD evolution
of parton scattering with the ''runing'' strong coupling constant $\alpha_{s}(Q^{2})$, Dokshitzer,
Gribov, Levine, Altarelli and Parisi. On the other hand, mid-rapidity bulk hadron
production at the upcoming CERN LHC facility ($\sqrt{s}=14 \: TeV$
for $pp$, and $5.5 \: TeV$ for A+A), with typical $Q^2 \le 5 \:
GeV^2$ at $x \le 10^{-3}$, will present a clear case for QCD
saturation physics, as formulated e.g. in the ''Color Glass
Condensate (CGC)'' formalism developed by McLerran, Venugopalan and
collaborators \cite{64,65,75,77}. This model develops a classical
gluon field theory for the limiting case of a high areal occupation
number density, i.e for the conceivable limit of the situation
depicted in Fig.~\ref{fig:Figure14} (right hand panel) where the amalgamating small
$x$ gluons would overlap completely, within any finite resolution
area at modest $Q^2$. Classical field theory captures, by
construction, the effects of color charge coherence, absent in DGLAP
parton cascade evolution theories \cite{75}. This model appears to
work well already at $\sqrt{s}$ as ''low'' as at RHIC, as far as
small $Q^2$ bulk charged particle production is concerned. We have
illustrated this by the CGC model fits \cite{64} to the PHOBOS
charged particle rapidity distributions, shown in Fig.~\ref{fig:Figure7}.

Conversely, QCD processes falling in the transition region between
such limiting conditions, such that typical $Q^2 \approx Q^2_s(x)$, should present observables that are functions of the ratio between
the transferred momentum $Q^2$ and the appropriate saturation scale,
expressed by $Q^2_s(x)$. As $Q^2$ defines the effective transverse
sampling area, and $Q^2_s(x)$ the characteristic areal size at which
saturation is expected to set in, a characteristic behaviour of
cross sections, namely that they are universal functions of
$Q^2/Q^2_s$, is called ''{\it geometric scaling}''. The HERA ep
scattering data obey this scaling law closely \cite{78}, and the
idea arises to apply the universality principle that we mentioned
above: at small enough $x$, all hadrons or nuclei are similar, their
specific properties only coming in via the appropriate saturation
scales $Q^2_s(x,h)$ or $Q^2_s(x,A)$. Knowing the latter for RHIC
conditions we will understand the systematics of charged particle
production illustrated in the previous section, and thus also be
able to extrapolate toward LHC conditions in $pp$ and AA collisions.\\
\begin{figure}
\begin{center}
\includegraphics[scale=0.45]{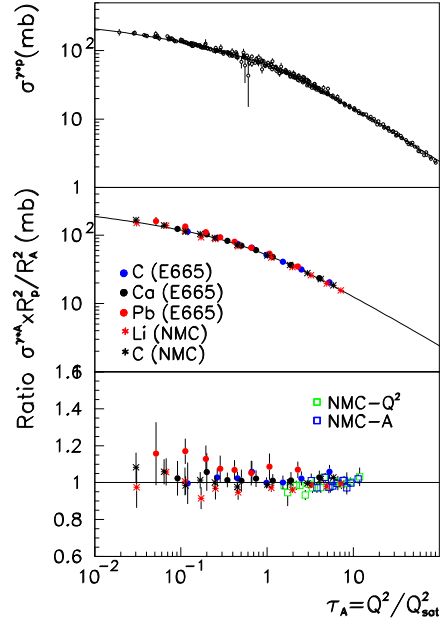}
\caption{(top) Geometric scaling of the virtual photo-absorption cross section
$\sigma ^{\gamma p}$ on protons; (middle) cross sections for nuclei
normalized according to equation~\ref{eq:equation13}; (bottom) the ratio of
$\sigma^{\gamma A}$ to a fit of $\sigma^{\gamma p}$ (see \cite{63}
for data reference).}
\label{fig:Figure15}
\end{center}
\end{figure}\\
All data for the virtual photo-absorption cross section $\sigma^{\gamma p}
(x,Q^2)$ in deep inelastic ep scattering with $x \le 0.01$
(which is also the RHIC mid-rapidity $x$-domain) have been found
\cite{78} to lie on a single curve when plotted against $Q^2/Q^2_s$,
with
\begin{equation}
Q_s^2(x) \sim  (\frac{x_0}{x})^\lambda \: 1 GeV^2
\label{eq:equation12}
\end{equation}
with $\lambda \simeq 0.3$ and $x_0 \simeq 10^{-4}$. This scaling
\cite{79} with $\tau=Q^2/Q^2_s$ is shown in Fig.~\ref{fig:Figure15} (top panel) to
interpolate all data. A chain of arguments, proposed by Armesto,
Salgado and Wiedemann \cite{63} connects a fit to these data with
photo-absorption data for (virtual) photon-A interactions \cite{80}
via the geometrical scaling ansatz
\begin{equation}
\frac{\sigma^{\gamma A} (\tau_A)}{\pi R^2_A} = \frac{\sigma^{\gamma
p} (\tau_p=\tau_A)}{\pi R^2_p}
\label{eq:equation13}
\end{equation}
assuming that the scale in the nucleus grows with the ratio of the
transverse parton densities, raised to the power $1/\delta$ (a free
parameter),
\begin{equation}
Q^2_{s,A}=Q^2_{s,p}\left(\frac{A \pi R^2_p}{\pi
R^2_A}\right)^{1/\delta}, \: \: \tau_A = \tau_h \left(\frac{\pi
R^2_A}{A \pi R^2_h}\right)^{1/\delta}.
\label{eq:equation14}
\end{equation}
Fig.~\ref{fig:Figure15} (middle and bottom panels) shows their fit to the nuclear
photo-absorbtion data which fixes $\delta=0.79$ and $\pi R^2_p=1.57
\: fm^2$ (see ref. \cite{63} for detail). The essential step in
transforming these findings to the case of A+A collisions is then
taken by the empirical ansatz
\begin{equation}
\frac{dN^{AA}}{dy} \: (at \: y \simeq 0) \: \propto \: Q^2_{s,A}(x)
\label{eq:equation15}
\pi R^2_A
\end{equation}
by which the mid-rapidity parton (gluon) density $dN/dy$ in equation~\ref{eq:equation11}
gets related to the charged particle mid-rapidity density at $y
\approx 0$ \cite{70,81}, measured in nucleus-nucleus collisions.
Replacing, further, the total nucleon number 2A in a collision of
identical nuclei of mass A by the number $N_{part}$ of participating
nucleons, the final result is \cite{63}
\begin{equation}
\frac{1}{N_{part}} \: \:  \frac{dN^{AA}}{dy} \: (at \: y \approx 0)
= N_0 (\sqrt{s})^{\lambda} \: N^{\alpha}_{part}
\label{eq:equation16}
\end{equation}
where the exponent $\alpha \equiv (1 - \delta)/3 \delta=0.089$, and
$N_0=0.47$. The exponent $\alpha$ is {\it far} smaller than 1/3, a
value that represents the thickness of the reactants, and would be
our naive guess in a picture of ''successive'' independent nucleon
participant collisions, whose average number $\left<\nu \right> \: \propto \:
(N_{part}/2)^{1/3}$. The observational fact (see Fig.~\ref{fig:Figure13}) that $\alpha < 1/3$ for
mid-rapidity low $Q^2$ bulk hadron production in A+A collisions
illustrates the importance of the QCD saturation effect. This is
shown \cite{63} in Fig.~\ref{fig:Figure16} where equation~\ref{eq:equation16} is applied to the RHIC
PHOBOS data for mid-rapidity charged particle rapidity density per
participant pair, in Au+Au collisions at $\sqrt{s}=19.6$, 130 and
200 $GeV$ \cite{82}, also including a {\it prediction} for LHC
energy. Note that the {\it factorization of energy and centrality
dependence}, implied by the RHIC data \cite{52}, is well captured by equation~\ref{eq:equation11}
 and the resulting fits in Fig.~\ref{fig:Figure16}. Furthermore, the steeper
slope, predicted for $N_{part} \le 60$ (not covered by the employed
data set), interpolates to the corresponding $pp$ and
$p\overline{p}$ data, at $N_{part}=2$. It resembles the pattern
observed in the NA49 data (Fig.~\ref{fig:Figure13}) for small $N_{part}$ collisions
of light A+A systems, at $\sqrt{s}=17-20 \: GeV$, and may be seen,
to reflect the onset of QCD saturation. Finally we note that the
conclusions of the above, partially heuristic approach~\cite{63},
represented by equations~\ref{eq:equation13} to~\ref{eq:equation16}, have been backed up by the CGC
theory of McLerran and Venugopulan \cite{64,65,75}, predictions of
which we have illustrated in Fig.~\ref{fig:Figure7}.\\
\begin{figure}
\begin{center}
\includegraphics[scale=0.45]{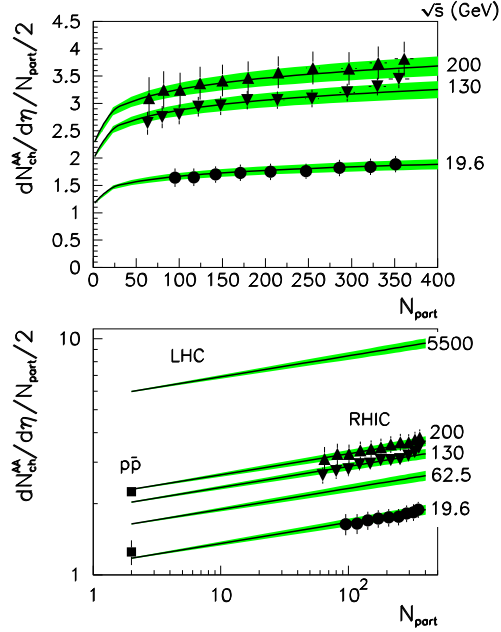}
\caption{Saturation model fit \cite{63} applied to RHIC charged hadron
multiplicity data at mid-rapidity normalized by number of
participant pairs, at various energies \cite{82}. Also shown is an
extrapolation to $p \overline{p}$ data and a prediction for minimum
bias Pb+Pb collisions at LHC energy, $\sqrt{s}=5500 \: GeV$.}
\label{fig:Figure16}
\end{center}
\end{figure}\\
Bulk hadron production in AA
collisions at high $\sqrt{s}$ can be related, via the assumption of
universality of high energy QCD scattering, to the phenomenon of
geometric scaling first observed in HERA deep inelastic ep cross
sections. The underlying feature is a QCD saturation effect arising
from the diverging areal parton density, as confronted with the
limited areal resolution $Q^2$, inherent in the considered
scattering process. The ''saturation scale'' $Q^2_s(x,A)$ captures
the condition that a single partonic charge source within the
transverse partonic density profile can just be resolved by a
sufficiently high $Q^2$. Bulk hadron production in A+A collisions falls below this scale.

\subsection {Transverse phase space: equilibrium and the QGP state}
\label{subsec:Transvers_phase_space}
At RHIC energy, $\sqrt{s}=200 \: GeV$, the Au+Au collision reactants
are longitudinally contracted discs. At a nuclear radius $R \approx
A^{1/3} \: fm$ and Lorentz $\gamma \approx 100$ their primordial
interpenetration phase ends at time $\tau_0 \le 0.15 \: fm/c$.
This time scale is absent in $e^+e^-$ annihilation at similar
$\sqrt{s}$ where $\tau_0 \approx 0.1 \: fm/c$ marks the end of the
primordial pQCD partonic shower evolution \cite{83} during which the
initially created $q \overline{q}$ pair, of ''virtually''
$Q=\sqrt{s}/2$ each, multiplies in the course of the QCD DGLAP
evolution in perturbative vacuum, giving rise to daughter partons of
far lower virtuality, of a few $GeV$. In A+A collisions this shower era
should last longer, due to the interpenetrational spread of primordial collision time.
It should be over by about 0.25 $fm/c$. The shower partons in $e^{+}e^{-}$ annihilation 
stay localized within back to back cone geometry reflecting the directions of the
primordial quark pair. The eventually observed ''jet'' signal,
created by an initial $Q^2$ of order $10^4 \: GeV^{2}$, is established by
then. Upon a slow-down  of the dynamical evolution time scale to
$\tau \approx 1 \: fm/c$ the shower partons fragment further,
acquiring transverse momentum and yet lower virtuality, then to
enter a non perturbative QCD phase of color neutralization during
which hadron-like singlet parton clusters are formed. Their net
initial pQCD virtuality, in pQCD vacuum, is recast in terms of
non-perturbative vacuum hadron mass. The evolution ends with
on-shell, observed jet-hadrons after about $3 \:fm/c$ of overall
reaction time.

\begin{figure}
\begin{center}
\includegraphics[scale=0.45]{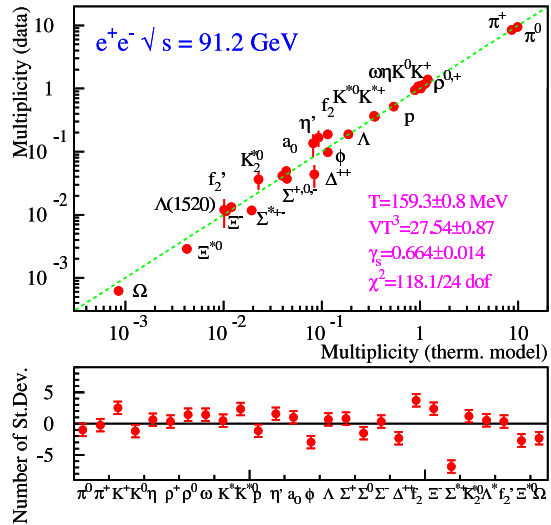}
\caption{Hadron multiplicities in LEP $e^+e^-$ annihilation at $\sqrt{s}=91.2
\: GeV$ confronted with the predictions of the canonical statistical
hadronization model \cite{84}.}
\label{fig:Figure17}
\end{center}
\end{figure}

Remarkably, even in this, somehow most elementary process of QCD
evolution, an aspect of equilibrium formation is observed, not in
the narrowly focussed final dijet momentum topology but in the
relative production rates of the various created hadronic species.
This so-called ''hadrochemical'' equilibrium among the hadronic
species is documented in Fig.~\ref{fig:Figure17}. The hadron multiplicities per
$e^+e^-$ annihilation event at $\sqrt{s}=91.2 \: GeV$  \cite{38} are
confronted with a Hagedorn \cite{38} canonical statistical Gibbs
ensemble prediction \cite {84} which reveals that the apparent
species equilibrium was fixed at a temperature of $T=165 \: MeV$,
which turns out to be the universal hadronization temperature of all
elementary and nuclear collisions at high $\sqrt{s}$ (Hagedorns
limiting temperature of the hadronic phase of matter). We shall
return to this topic in section~\ref{chap:hadronization} but note, for now, that reactions
with as few as 20 charged particles exhibit such statistical
equilibrium properties.\\
\\
What happens with parton (and hadron) dynamics in A+A collisions after
$\tau_0$? There will not be a QCD evolution in vacuum as the transverse radius of the interacting system is large. 
It may grow to about twice the nuclear radius, i.e. to about $15 \:
fm$ before interactions cease; i.e. the system needs about $15 \:
fm/c$ to decouple. This simple fact is the key to our expectation
that the expansive evolution of the initial high energy density
deposited in a cylinder of considerable diameter (about $10 \: fm$),
may create certain equilibrium properties that allow us to treat the
contained particles and energy in terms of thermodynamic phases of
matter, such as a partonic QGP liquid,  or a hadronic liquid or gas,
etc.. Such that the expansion dynamics makes contact to the phase
diagram illustrated in Fig.~\ref{fig:Figure1}. This expectation turns out to be
justified as we shall describe in section~\ref{chap:hadronization}. 
What results for the evolution after $\tau_0$ in a central A+A collision is sketched in
Fig.~\ref{lightcone} by means of a schematic 2-dimensional light cone diagram,
which is entered by the two reactant nuclei along $\pm z=t$ trajectories
where $z$ is the beam direction and Lorentz contraction has been
taken to an extreme, such that there occurs an idealized $t=z=0$
interaction ''point''. Toward positive $t$ the light cone proper
time profiles of progressing parton-hadron matter evolution are
illustrated. The first profile illustrated here corresponds to the
end of shower formation time $\tau_0$. From our above discussion of the
$e^+e^-$ annihilation process one obtains a first estimate, $\tau_0
\ge 0.25 \: fm/c$ (including interpenetration time of 0.15 $fm/c$ at RHIC)
which refers to processes of very high $Q^2 \ge 10^3 \: GeV^{2}$, 
far above the saturation scale $Q^2_s$ discussed in the previous section. 
The latter scale has to be taken into account for low $p_{T}$ hadron production.

It is the specific resolution scale $Q^2$ of a QCD
sub-process, as enveloped in the overall collision dynamics of two
slabs of given transverse partonic structure function density, that
determines which fraction of the constituent partons enters
interaction. In the simple case of extremely high $Q^2$ processes
the answer is that all constituents are resolved. However, at modest
$Q^2$ (dominating bulk hadron production) the characteristic QCD
saturation scale $Q^2_s (x)$ gains prominence, defined such that
processes with $Q^2 < Q^2_s$ do not exploit the initial transverse
parton densities at the level of independent single constituent
color field sources (see equation~\ref{eq:equation11}). For such processes the proper
formation time scale, $\tau_0$, is of order of the inverse
saturation momentum \cite{61}, $1/Q_s \sim 0.2 \: fm/c$ at
$\sqrt{s}=200 \: GeV$. The first profile of the time evolution, sketched in
Fig.~\ref{lightcone}, should correspond to proper time $t=\tau_0=0.25 \: fm/c$ at
RHIC energy. At top SPS energy, $\sqrt{s}=17.3 \: GeV$, we can not
refer to such detailed QCD considerations. A pragmatic approach
suggests to take the interpenetration time, at $\gamma \approx 8.5$,
for guidance concerning the formation time, which thus results as
$\tau_0 \approx 1.5 \: fm/c$.

In summary of the above considerations we assume that the initial
partonic color sources, as contained in the structure functions
(Fig.~\ref{fig:Figure14}), are spread out in longitudinal phase space after light
cone proper time $t=\tau_0 \approx 0.25 \: fm/c$, at top RHIC energy,
and after $\tau_0 \approx 1.4 \: fm/c$ at top SPS energy. No
significant transverse expansion has occured at this early stage, in
a central collision of $A \approx 200$ nuclei with transverse
diameter of about 12 $fm$. The Bjorken estimate \cite{45} of initial
energy density $\epsilon$ (equation~\ref{eq:equation1}) refers to exactly this
condition, after formation time $\tau_0$. In order to account for
the finite longitudinal source size and interpenetration time, at RHIC, we finally 
put the average $\tau_0\approx 0.3 \: fm$, at $\sqrt{s}=200 \: GeV$, indicating the
''initialization time'' after which all partons that have been
resolved from the structure functions are engaged in shower
multiplication. As is apparent from Fig.~\ref{lightcone}, this time scale is
Lorentz dilated for partons with a large longitudinal momentum, or
rapidity. This means that the slow particles are produced first
toward the center of the collision region, and the fast (large
rapidity) particles are produced later, away from the
collision region. This Bjorken ''inside-out'' correlation \cite{45}
between coordinate- and momentum-space is similar to the Hubble
expansion pattern in cosmology: more distant galaxies have higher
outward velocities. Analogously, the matter created in A+A
collisions at high $\sqrt{s}$ is born expanding, however with
the difference that the Hubble flow is initially one dimensional
along the collision axis. This pattern will continue, at
$\sqrt{s}=200 \: GeV$, until the system begins to feel the effects
of finite size in the transverse direction which will occur at some
time $t_0$ in the vicinity of $1 \: fm/c$. However, the tight
correlation between position and momentum initially imprinted on the
system will survive all further expansive evolution of the initial
''firetube'', and is well recovered in the expansion pattern of the
finally released hadrons of modest $p_T$ as we shall show when
discussing radial flow (see section~\ref{subsec:Bulk_hadron_transverse_spectra}).

In order to proceed to a more quantitative description of the
primordial dynamics (that occurs onward from $\tau_0$ for as long the time period
of predominantly longitudinal expansion might extend) we return to
the Bjorken estimate of energy density, corresponding to this
picture~\cite{45}, as implied by equation~\ref{eq:equation1}, which we now recast as
\begin{equation}
\epsilon = \left(\frac{dN_h}{dy} \right) \left<E^T_h\right> (\pi \: R^2_A \:
t_0)^{-1}
\label{eq:equation18}
\end{equation}
where the first term is the (average) total hadron multiplicity per
unit rapidity which , multiplied with the average hadron transverse
energy, equals the total transverse energy recorded in the
calorimetric study shown in Fig.~\ref{fig:Figure3}, as employed in~equation~\ref{eq:equation1}. 
The quantity $R_A$ is, strictly speaking, {\it not} the radius parameter of the
spherical Woods-Saxon nuclear density profile but the $rms$ of the
reactant overlap profiles as projected onto the transverse plane
(and thus slightly smaller than $R_A \approx A^{1/3} \:fm$).
Employing $A^{1/3}$ here (as is done throughout) leads to a
conservative estimate of $\epsilon$, a minor concern. However, the
basic assumption in~equation~\ref{eq:equation18} is to identify the primordial transverse
energy ''radiation'', of an interactional cylindric source of radius
$R_A$ and length $t_0$ (where $\tau_0 \le
t_0 \le 1 \: fm/c$, not Lorentz dilated at midrapidity), with the finally emerging bulk hadronic
transverse energy. We justify this assumption by the two
observations, made above, that
\begin{enumerate}
\item the bulk hadron multiplicity density per unit rapidity
$\frac{dN_h}{dy}$ resembles the parton density, primordially
released at saturation scale $\tau_0$ (Figs.~\ref{fig:Figure7},~\ref{fig:Figure16}) at $\sqrt{s}=200
\: GeV$, and that
\item the global emission pattern of bulk hadrons (in rapidity and
$p_T$) closely reflects the initial correlation between coordinate
and momentum space, characteristic of a primordial period of a
predominantly longitudinal expansion, as implied in the Bjorken
model.
\end{enumerate}
Both these observations are surprising, at first sight. The Bjorken
model was conceived for elementary hadron collisions where the
expansion proceeds into vacuum, i.e. directly toward observation.
Fig.~\ref{lightcone} proposes that, to the contrary, primordially produced partons
have to transform through further, successive stages of partonic and
hadronic matter, at decreasing but still substantial energy density,
in central A+A collisions. The very fact of high energy density, with implied short mean free path
of the constituent particles, invites a hydrodynamic description
of the expansive evolution. With initial conditions fixed between
$\tau_0$ and $t_0$, an ensuing 3-dimensional hydrodynamic
expansion would preserve the primordial Bjorken-type correlation
between position and momentum space, up to lower density conditions
and, thus, close to emission of the eventually observed hadrons. We
thus feel justified to employ~equation~\ref{eq:equation1} or~\ref{eq:equation18} for the initial
conditions at RHIC, obtaining \cite{61,84}
\begin{equation}
6 \: GeV/fm^3 \le  \epsilon \le 20 \: GeV/fm^3
\label{eq:equation19}
\end{equation}
for the interval $0.3 \:fm/c \le t_0 \le 1 \: fm/c$, in central
Au+Au collisions at $y \approx 0$ and $\sqrt{s}=200 \:GeV$. The
energy density at top SPS energy, $\sqrt{s}=17.3 \: GeV$, can
similarly be estimated \cite{43,44} to amount to about $3 \:
GeV/fm^3$ at a $t_0$ of 1 $fm/c$ but we can not identify conditions
at $\tau_0<t_0$ in this case as the mere interpenetration of two Pb
nuclei takes $1.4 \: fm/c$. Thus the commonly accepted $t_0=1 \:
fm/c$ may lead to a high estimate. An application of the
parton-hadron transport model of Ellis and Geiger \cite{85,86} to
this collision finds $\epsilon=3.3 \: GeV/fm^3$ at $t=1 \:fm/c$. A
primordial energy density of about $3 \: GeV/fm^3$ is twenty times
$\rho_0 \approx 0.15 \: GeV/fm^3$, the average energy density of
ground state nuclear matter, and it also exceeds, by far, the
critical QCD energy density, of $0.6 \le \epsilon_c \le 1 \:
GeV/fm^3$ according to lattice QCD \cite{48}. The initial dynamics
thus clearly proceeds in a deconfined  QCD system also at top SPS energy,
and similarly so with strikingly higher energy density, at RHIC,
where time scales below $1 \: fm/c$ can be resolved.

However, in order now to clarify the key question as to whether, and
when conditions of partonic dynamical equilibrium may arise under
such initial conditions, we need estimates both of the proper
relaxation time scale (which will, obviously, depend on energy
density and related collision frequency), and of the expansion time
scale as governed by the overall evolution of the collision volume.
Only if $\tau (relax.) < \tau (expans.)$ one may conclude that the
''deconfined partonic system'' can be identified with a ''deconfined
QGP {\it state} of QCD matter'' as described e.g. by lattice QCD,
and implied in the phase diagram of QCD matter suggested in Fig.~\ref{fig:Figure1}.

For guidance concerning the overall time-order of the system
evolution we consider information \cite{87} obtained from
Bose-Einstein correlation analysis of pion pair emission in momentum
space. Note that pions should be emitted at
{\it any} stage of the evolution, after formation time, from the
surface regions of the evolving ''fire-tube''. Bulk emission of
pions occurs, of course, after hadronization (the latest stages
illustrated in the evolution sketch given in Fig.~\ref{lightcone}). The dynamical
pion source expansion models by Heinz \cite{88} and Sinyukov
\cite{89} elaborate a Gaussian emission time profile, with mean
$\tau_f$ (the decoupling time) and width $\Delta \tau$ (the duration
of emission). \\
\begin{figure}
\begin{center}
\includegraphics[scale=0.6]{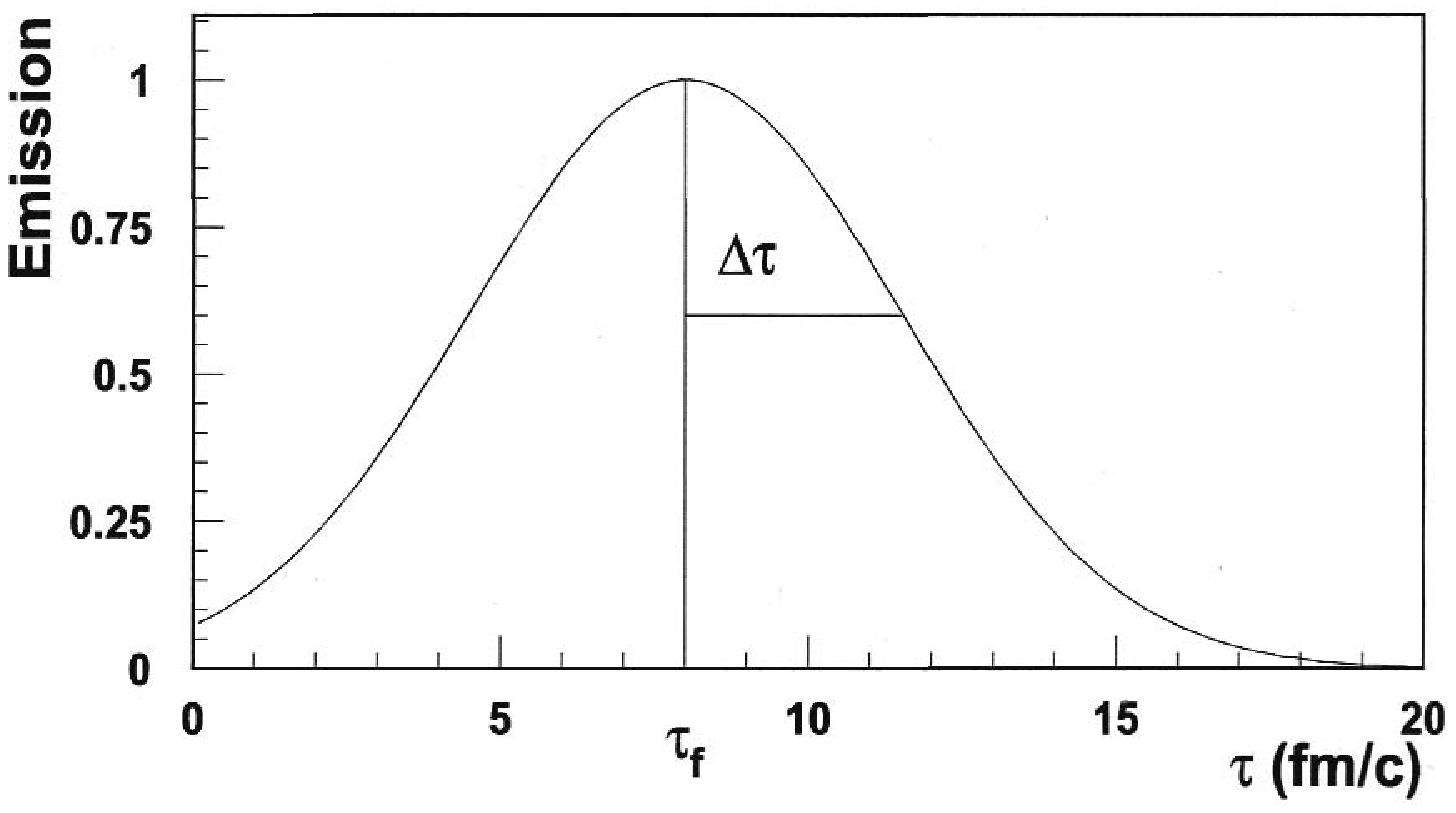}
\caption{Time profile of pion decoupling rate from the fireball in a central
Pb+Pb collision, with $\tau=0$ the end of the formation phase.
Bose-Einstein correlation of $\pi^- \pi^-$pairs yields an average
Gaussian decoupling profile with $\tau_f=8 \: fm/c$ and duration of
emission parameter $\Delta \tau=4 \: fm/c$ \cite{87,88}.}
\label{fig:Figure19}
\end{center}
\end{figure}\\
Fig.~\ref{fig:Figure19} shows an application of this analysis to
central Pb+Pb collision negative pion pair correlation data obtained
by NA49 at top SPS energy, $\sqrt{s}=17.3 \: GeV$ \cite{90}, where
$\tau_f \approx 8 \: fm/c$ and $\Delta \tau \approx 4\: fm/c$ (note that $\tau=0$
in Fig.~\ref{fig:Figure19} corresponds, not to interaction time $t=0$ but to $t
\approx 1.4 \: fm/c$, the end of the interpenetration phase). We
see, first of all, that the overall dynamical evolution of a central
Pb+Pb collision at $\sqrt{s}=17.3 \: GeV$ is ending at about $15
\:fm/c$; the proper time defines the position of the last,
decoupling profile illustrated in Fig.~\ref{lightcone}, for the SPS collisions
considered here. While the details of Fig.~\ref{fig:Figure19} will turn out to be
relevant to our later discussion of hadronization (section~\ref{chap:hadronization}) and
hadronic expansion, we are concerned here with the
average proper time at which the partonic phase ends. After
consideration of the duration widths of these latter expansion
phases \cite{86,87} one arrives at an estimate for the average
time, spent before hadronization, of $\Delta t=3-4 \:fm/c$, again in
agreement with the parton cascade model mentioned above \cite{86}.
This model also leads to the conclusion that parton thermal
equilibrium is, at least, closely approached locally in these central Pb+Pb
collisions as far as mid-rapidity hadron production is concerned (at
forward-backward rapidity the cascade re-scattering processes do not
suffice, however).

This finding agrees with earlier predictions of $\tau_{relax}=1-2 \:
fm/c$ at top SPS energy \cite{91}. However we note that all such
calculations employ perturbative QCD methods, implying the
paradoxical consequence that equilibrium is closely approached toward
{\it the end} of the partonic phase, at such low $\sqrt{s}$, i.e.
in a QGP state at about $T=200 \: MeV$ which is, by definition, of
non-perturbative nature. We shall return to the question of partonic
equilibrium attainment at SPS energy in the discussion of the
hadronization process in nuclear collisions (section~\ref{chap:hadronization}).

Equilibrium conditions should set in earlier at top RHIC energy. As
transverse partonic expansion should set in after the proper time
interval $0.3 \:fm/c \le t_0 \le 1 \: fm/c$ (which is now resolved
by the early dynamics, unlike at top SPS energy), we take guidance
from the Bjorken estimate of primordial energy density which is
based on transverse energy production {\it data}. Conservatively
interpreting the result in equation~\ref{eq:equation19} we conclude that $\epsilon$ is
about four times higher than at $\sqrt{s}=17.3 \: GeV$ in the above proper
time interval. As the binary partonic
collision frequency scales with the square of with the square of the density $\rho$ (related to the 
energy density $\epsilon$ via the relation $\epsilon$ = $\left<E\right>\rho$ = $T\rho$), and is
inversely proportional to the relaxation time $\tau_{relax}$ we
expect
\begin{equation}
\tau_{relax}\propto (1/\rho)^2 \approx (T/\epsilon)^{2}
\label{eq:equation20}
\end{equation}
which implies that $\tau_{relax}(RHIC) \approx 0.25 \: \tau_{relax}(SPS)
\approx 0.5 \: fm/c$ if we employ the estimate $T(RHIC) \: = \: 2T(SPS)$. This crude estimate is, 
however, confirmed by the parton transport model of Molar and Gyulassy \cite{92}.

Partonic equilibration at $\sqrt{s}=200 \: GeV$ should thus set in
at a time scale commensurate to the (slightly smaller) formation
time scale, at which the to be participant partons are
resolved from the initial nucleon structure functions and enter shower multiplication. 
Extrapolating to the conditions expected at LHC energy ($\sqrt{s}=5.5 \: TeV$ for
A+A collisions), where the initial parton density of the structure
functions in Fig.~\ref{fig:Figure14} is even higher ($x \approx 10^{-3}$ at
mid-rapidity), and so is the initial energy density, we may expect
conditions at which the resolved partons are almost ''born into
equilibrium''.

Early dynamical local equilibrium at RHIC is required to understand the observations 
concerning elliptic flow. 
This term refers to a collective anisotropic azimuthal emission pattern of bulk hadrons in
semi-peripheral collisions, a hydrodynamical phenomenon that
originates from the initial geometrical non-isotropy of the
primordial interaction zone \cite{93,94}. 
A detailed hydrodynamic model analysis of the
corresponding elliptic flow signal at RHIC \cite{95} 
leads to the conclusion that local equilibrium (a prerequisite to 
the hydrodynamic description) sets in at $t_0 \approx 0.5 \: fm/c$. This conclusion agrees with the
estimate via equation~\ref{eq:equation20} above, based on Bjorken energy density and
corresponding parton collisions frequency.

We note that the concept of a hydrodynamic
evolution appears to be, almost necessarily ingrained in the physics
of a system born into (Hubble-type) expansion, with a primordial
correlation between coordinate and momentum space, and at extreme
initial parton density at which the partonic mean free path length
$\lambda$ is close to the overall spatial resolution resulting from
the saturation scale, i.e. $\lambda \approx 1/Q_s$.

The above considerations suggest that a quark-gluon plasma state should be created early
in the expansion dynamics at $\sqrt{s}=200 \: GeV$, at about $T= 300
\: MeV$, that expands hydrodynamically until hadronization is
reached, at $T \approx 165-170 \: MeV$. Its manifestations will be
considered in section~\ref{chap:hadronization}. 
At the lower SPS energy, up to $17.3 \: GeV$, we can conclude, with some caution, that a deconfined hadronic
matter system should exist at $T \approx 200 \: MeV$, in the closer
vicinity of the hadronization transition. It may closely resemble
the QGP state of lattice QCD, near $T_c$.

\subsection {Bulk hadron transverse spectra and radial expansion flow} 
\label{subsec:Bulk_hadron_transverse_spectra}
In this section we analyze bulk hadron transverse momentum
spectra obtained at SPS and RHIC energy, confronting the data with
predictions of the hydrodynamical model of collective expansion
matter flow that we have suggested in the previous section, to arise, almost
necessarily, from the primordial Hubble-type coupling between
coordinate and momentum space that prevails at the onset of the
dynamical evolution in A+A collisions at high $\sqrt{s}$. As all
hadronic transverse momentum spectra initially follow an
approximately exponential fall-off (see below) the bulk hadronic
output is represented by thermal transverse spectra at $p_T \le 2 \: GeV/c$.

Furthermore
we shall focus here on mid-rapidity production in near central A+A
collisions, because hydrodynamic models refer to an
initialization period characterized by Bjorken-type longitudinal
boost invariance, which we have seen in Figs.~\ref{fig:Figure7} and~\ref{fig:Figure9} to be
restricted to a relatively narrow interval centered at mid-rapidity.
Central collisions are selected to exploit the azimuthal symmetry of
emission, in an ideal impact parameter $b \rightarrow 0$ geometry.
We thus select the predominant, relevant hydrodynamic ''radial
flow'' expansion mode, from among other, azimuthaly oriented
(directed) flow patterns that arise once this cylindrical symmetry
(with respect to the beam direction) is broken in finite impact
parameter geometries.

In order to define, quantitatively, the flow phenomena mentioned
above, we rewrite the invariant cross section for production of
hadron species $i$ in terms of transverse momentum, rapidity, impact
parameter $b$ and azimuthal emission angle $\varphi_p$ (relative to
the reaction plane),
\begin{equation}
\frac{dN_i(b)}{p_Tdp_T dy d\varphi_p} = \frac{1}{2 \: \pi}\:
\frac{dN_i(b)}{p_Tdp_Tdy} \: {\left[1 + 2v_1^i \: (p_T,b) cos \varphi_p +
2v_2^i \: (p_T,b) cos (2 \varphi_p)+ . . .\right]}
\label{eq:equation21}
\end{equation}
where we have expanded the dependence on $\varphi_p$ into a Fourier
series. Due to reflection symmetry with respect to the reaction
plane in collisions of identical nuclei, only cosine terms appear.
Restricting to mid-rapidity production all odd harmonics vanish, in
particular the ''directed flow'' coefficient $v^i_1$, and we have
dropped the  y-dependence in the flow coefficients $v^i_1$ and
$v^i_2$. The latter quantifies the amount of ''elliptic flow'' as discussed above. In the following, we will restrict to
central collisions which we shall idealize as near-zero impact
parameter processes governed by cylinder symmetry, whence all
azimuthal dependence (expressed by the $v^i_1, \: v^i_2, . . .$
terms) vanishes, and the invariant cross section reduces to the
first term in equation~\ref{eq:equation21}, which by definition also corresponds to all
measurements in which the orientation of the reaction plane is not
observed.

\begin{figure}
\begin{center}
\vspace{0.2cm}
\includegraphics[scale=0.4]{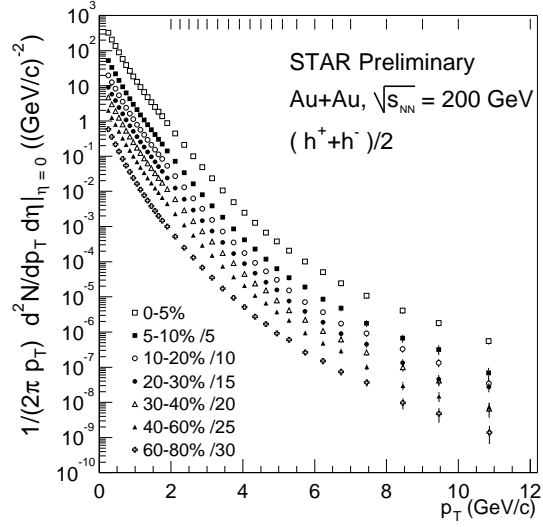}
\caption{Transverse momentum spectra of charged hadrons in Au+Au collisions
at $\sqrt{s}=200 \: GeV$, in dependence of collision centrality
\cite{97} (offset as indicated), featuring transition from
exponential to power law shape.}
\label{fig:Figure20}
\end{center}
\end{figure}

Typical transverse momentum spectra of the latter type are shown in
Fig.~\ref{fig:Figure20}, for charged hadron production in Au+Au collisions at
$\sqrt{s}=200 \: GeV$, exhibiting mid-rapidity data at various
collision centralities \cite{97}. We observe a clear-cut transition,
from bulk hadron emission at $p_T \le 2 \: GeV/c$ featuring a
near-exponential cross section (i.e. a thermal spectrum), to a high
$p_T$ power-law spectral pattern. Within the context of our previous
discussion (section~\ref{subsec:Gluon_Satu_in_AA_Coll}) we tentatively identify the low $p_T$ region
with the QCD physics near saturation scale. Hadron production at
$p_T \rightarrow 10 \: GeV/c$ should, on the other hand, be the
consequence of primordial leading parton fragmentation originating
from ''hard'', high $Q^2$ perturbative QCD processes.

We thus identify bulk hadron production at low $p_T$ as the
emergence of the initial parton saturation conditions that give rise
to high energy density and small equilibration time scale, leading
to a hydrodynamical bulk matter expansion evolution. Conversely, the
initially produced hard partons, from high $Q^2$ processes, are not
thermalized into the bulk but traverse it, as tracers, while being
attenuated by medium-induced rescattering and gluon radiation, the
combined effects being reflected in the high $p_T$ inclusive hadron
yield, and in jet correlations of hadron emission. We can not treat 
the latter physics observables in detail here, but will remain in the field of
low $p_T$ physics, related to hydrodynamical expansion modes,
focusing on radially symmetric expansion.\\
\\
In order to infer from the spectral shapes of the hadronic species
about the expansion mechanism, we first transform to the transverse
mass variable, $m_T=(p^2_T + m^2)^{1/2}$, via
\begin{equation}
\frac{1}{2\pi} \: \frac{dN_i}{p_T dp_T dy} = \frac{1}{2 \pi} \:
\frac{dN_i}{m_T dm_T dy}
\label{eq:equation22}
\end{equation}
because it has been shown in p+p collisions \cite{98} near RHIC
energy that the $m_T$ distributions of various hadronic species
exhibit a universal pattern (''$m_T$ scaling'') at low $m_T$:
\begin{equation}
\frac{1}{2\pi} \: \frac{dN_i}{m_T dm_T dy} = A_i \: exp (-m^i_T / T)
\label{eq:equation23}
\end{equation}
with a universal inverse slope parameter $T$ and a species dependent
normalization factor $A$. Hagedorn showed \cite{99} that this
scaling is characteristic of an adiabatic expansion of a fireball at
temperature $T$. We recall that, on the other hand, an ideal
hydrodynamical expansion is isentropic.\\
\begin{figure}
\begin{center}
\includegraphics[scale=0.5]{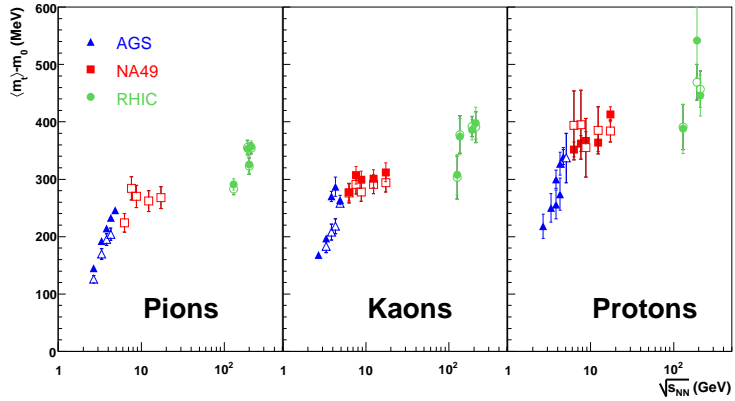}
\caption{The average transverse kinetic energy $\left<m_T\right> - m_0$ for pions, kaons
and protons vs. $\sqrt{s}$ in central Au+Au/Pb+Pb collisions
\cite{54}. Open symbols represent negative hadrons.}
\label{fig:Figure21}
\end{center}
\end{figure}\\
Fig.~\ref{fig:Figure21} shows the $\sqrt{s}$ dependence of the average transverse
kinetic energy $\left<m^i_T\right> -m^i$ for pions, kaons and protons observed
at mid-rapidity in central Au+Au/Pb+Pb collisions \cite{54}.
Similarly, the inverse slope parameter $T$ resulting from a fit of
equation~\ref{eq:equation23} to $K^+$ and $K^-$ transverse mass spectra  (at $p_T \le 2
\: GeV/c$) is shown in Fig.~\ref{fig:Figure22}, both for nuclear and p+p collisions
\cite {100}. We see, first of all, that $m_T$ scaling does not apply
in A+A collisions, and that the kaon inverse slope parameter, $T \approx
230 \: MeV$ over the SPS energy regime, can not be identified with
the fireball temperature at hadron formation which is $T_h \approx
165 \: MeV$ from Fig.~\ref{fig:Figure1}. The latter is seen, however, to be well represented
by the p+p spectral data exhibited in the left panel of Fig.~\ref{fig:Figure22}. There is, thus, not 
only thermal energy present in A+A transverse expansion, but also hydrodynamical radial
flow.

We note that the indications in Figs.~\ref{fig:Figure21} and~\ref{fig:Figure22}, of a plateau in both
$\left<m_T\right>$ and $T$, extending over the domain of SPS energies, $6 \le
\sqrt{s} \le 17 \: GeV$, have not yet been explained by any
fundamental expansive evolution model, including hydrodynamics.
Within the framework of the latter model, this is a consequence of
the {\it initialization problem} \cite{96} which requires a detailed
modeling, both of primordial energy density vs. equilibration time
scale, and of the appropriate partonic matter equation of state
(EOS) which relates expansion pressure to energy density. At top
RHIC energy, this initialization of hydro-flow occurs, both, at a
time scale $t_0 \approx 0.5 \: fm/c$ which is far smaller than the
time scale of eventual bulk hadronization ($t \approx 3 \: fm/c$),
and at a primordial energy density far in excess of the critical QCD
confinement density. After initialization, the partonic plasma phase
thus dominates the overall expansive evolution,over a time interval
far exceeding the formation and relaxation time scale. \\
\begin{figure}
\begin{center}
\includegraphics[scale=0.25]{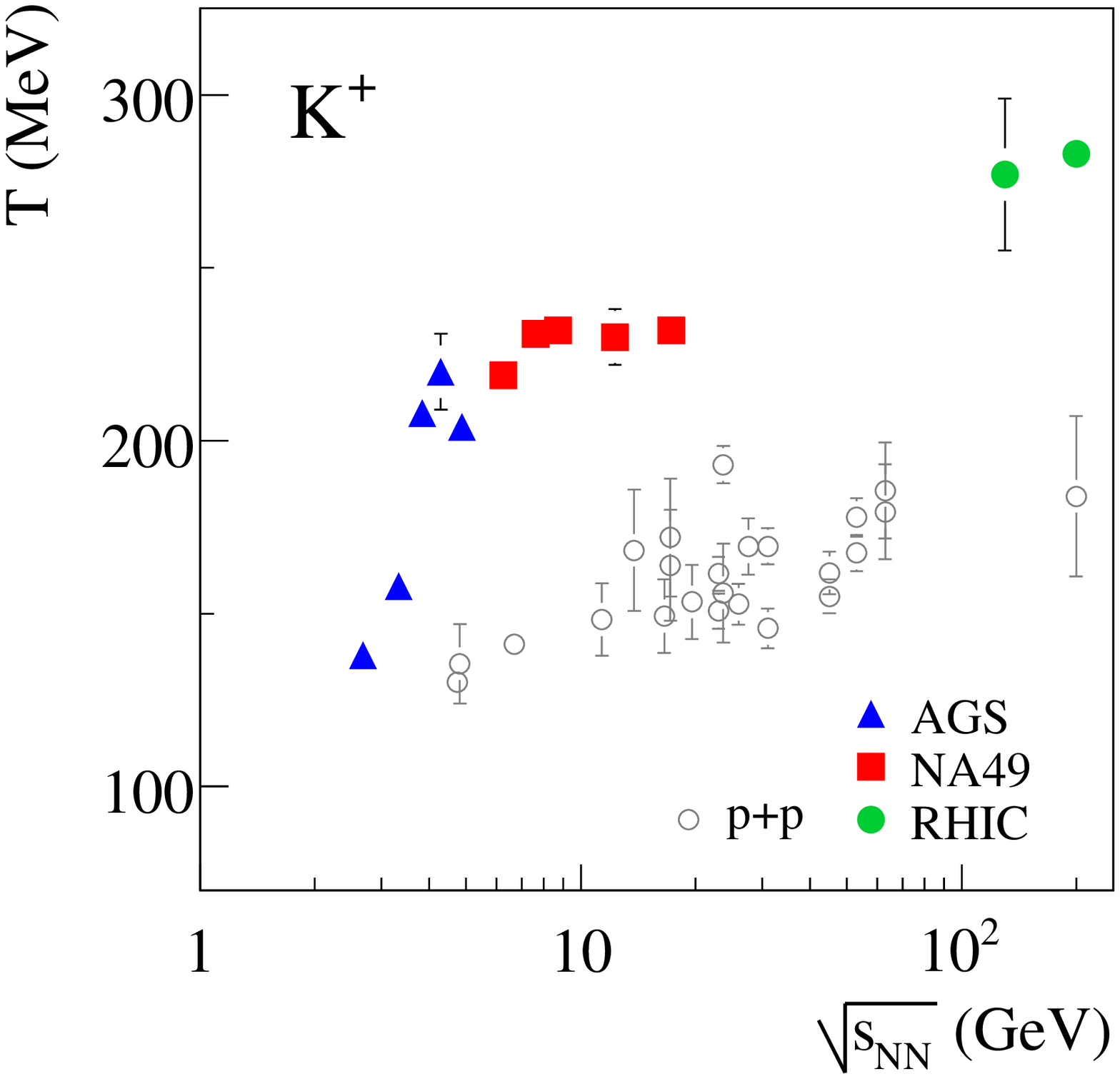}
\includegraphics[scale=0.25]{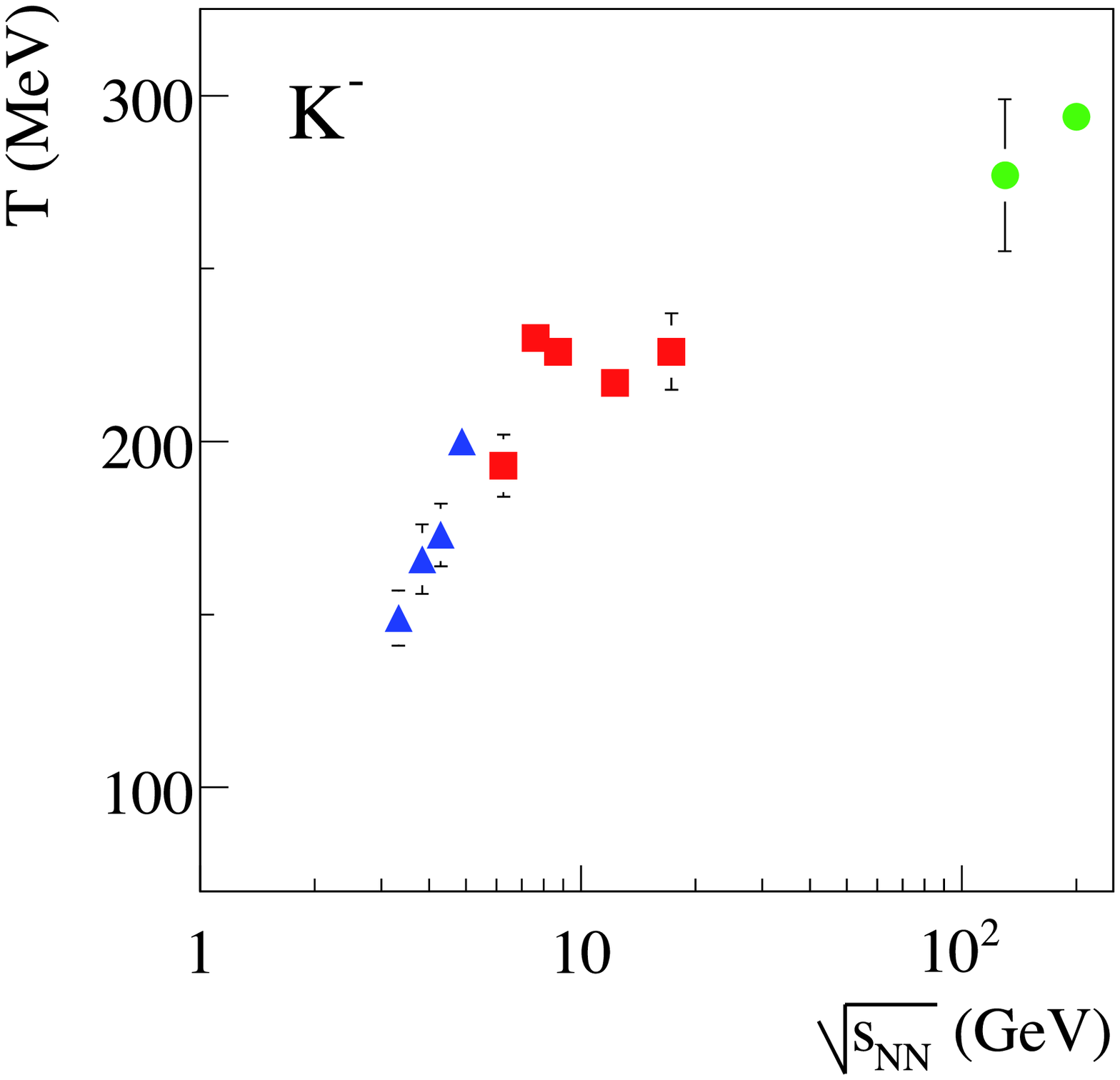}
\caption{The inverse slope parameter $T$ of equation~\ref{eq:equation23} for $K^+ \mbox{and} \:
K^-$ transverse mass spectra at $p_T < 2 \: GeV/c$ and mid-rapidity
in central A+A, and in minimum bias p+p collisions \cite{100}.}
\label{fig:Figure22}
\end{center}
\end{figure}
Thus, at RHIC energy, parton transport \cite{92} and relativistic hydrodynamic \cite{95, 96} 
models establish a well developed expansion mode that survives the
subsequent stages of hadronization and hadronic expansion. This is
reflected in their success in describing elliptic flow. On the other
hand, the hydrodynamical model far overestimates elliptic flow at
SPS energy \cite{96} at which, as we have shown in section\ref{subsec:Transvers_phase_space}, the
initialization period may be not well separated from the confinement
(hadronization) stage. Thus, whereas the expansion evolution at
$\sqrt{s}=200 \: GeV$ (occuring at near-zero baryo-chemical
potential in Fig.~\ref{fig:Figure1}) ''races'' across the parton-hadron phase
boundary with fully established flow patterns, near $\mu_B=0$ where
lattice QCD predicts the phase transformation to be merely a soft
cross-over \cite{16}, the dynamics at $\sqrt{s}=10-20 \: GeV$ may
originate from only slightly above, or even at the phase boundary,
thus sampling the domain $200 \le \mu_B \le 500 \: MeV$ where the
equation of state might exhibit a ''softest point'' \cite{96}. The hydrodynamic model thus faces 
formidable uncertainties regarding initialization at SPS energy. 

The plateau in Figs.~\ref{fig:Figure21},~\ref{fig:Figure22} may be the consequence of the fact
that not much flow is generated in, or transmitted from the partonic
phase, at SPS energies, because it is initialized close to the phase
boundary \cite{100} where the expected critical point \cite{9,10}
(Fig.~\ref{fig:Figure1}), and the corresponding adjacent first order phase transition
might focus \cite{101} or stall \cite{96} the expansion trajectory,
such that the observed radial flow stems almost exclusively from the
hadronic expansion phase. The SPS plateau, which we shall
subsequently encounter in other bulk hadron variables (elliptic
flow, HBT radii) might thus emerge as a consequence of the critical
point or, in general, of the flatness of the parton-hadron
coexistence line. RHIC dynamics, on the other hand, originates from
far above this line.

Hadronic expansion is known to proceed isentropically \cite{102}:
commensurate to expansive volume increase the momentum space volume
must decrease, from a random isotropic thermal distribution to a
restricted momentum orientation preferentially perpendicular to the fireball
surface, i.e. radial. The initial thermal energy, implied by the
hadron formation temperature $T_H=165 \: MeV$, will thus fall down
to a residual $T_F$ at hadronic decoupling from the flow field
(''thermal freeze-out'') plus a radial transverse kinetic energy
term $m_i \left< \beta _T\right>^2$ where $m_i$ is the mass of the considered
hadron species and $\left<\beta_T\right>$ the average radial velocity. We thus
expect \cite{103} for the slope of equation~\ref{eq:equation23}:
\begin{equation}
T=T_F+m_i\left<\beta _T\right>^2, \: p_T \le 2 \: GeV/c
\label{eq:equation24}
\end{equation}
and
\begin{equation}
T=T_F \left( \frac{1+\left<v_T\right>}{1-\left<v_T\right>} \right)^{1/2}, \: p_T \gg m_i
\label{eq:equation25}
\end{equation}
the latter expression valid at $p_T$ larger than hadron mass
scale ($T$ then is the ''blue-shifted temperature'' at decoupling
\cite{104} and $\left<v_T\right>$ the average transverse velocity). The
assumption that radial flow mostly originates from the hadronic
expansion phase is underlined by the proportionality of flow energy
to hadron mass (equation~\ref{eq:equation24}). \\
\begin{figure}
\begin{center}
\includegraphics[scale=0.8]{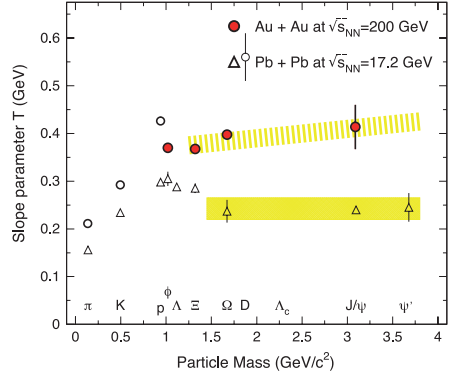}
\caption{Hadron slope parameters $T$ at mid-rapidity as a function of mass.
For Pb+Pb at $\sqrt{s}=17.3 \: GeV$ (triangles) and Au+Au at
$\sqrt{s}=200 \: GeV$ (circles); from \cite{103}.}
\label{fig:Figure23}
\end{center}\vspace{-0.6cm}
\end{figure}
Fig.~\ref{fig:Figure23} illustrates this proportionality, by a recent compilation \cite{103} of 
RHIC results for central Au+Au collisions at $\sqrt{s}= 200 \: GeV$, and SPS results for 
central Pb+Pb collisions at top SPS energy, $\sqrt{s}=17.3 \: GeV$. At the
latter energy the slope parameter of the $\Phi$ meson is seen to be
close to that of the similar mass baryons $p$ and $\Lambda$,
emphasizing the occurence of $m_i$  scaling as opposed to valence
quark number scaling that we will encounter in RHIC elliptic flow
data \cite{94}. As is obvious from Fig.~\ref{fig:Figure23} the multi-strange hyperons
and charmonia exhibit a slope saturation which is usually explained
\cite{103} as a consequence of their small total cross sections of
rescattering from other hadrons, leading to an early decoupling from
the bulk hadron radial flow field, such that $\left<\beta_T\right> _{\Omega} \:
< \: \left<\beta_T\right>_p$. \\
\begin{figure}
\begin{center}
\includegraphics[scale=0.8]{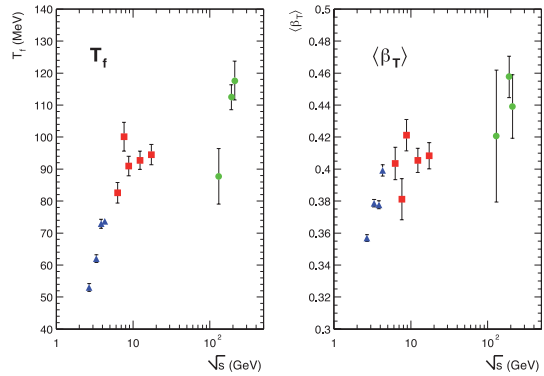}
\caption{Hadron decoupling temperature $T_f$, and average radial flow
velocity $\left<\beta_T\right>$ extracted from blast wave model (see equation~\ref{eq:equation26}) fits of
$m_T$ spectra vs. $\sqrt{s}$ \cite{54}.}
\label{fig:Figure24}
\end{center}
\end{figure}
According to our observations with equation~\ref{eq:equation24} a hydrodynamical ansatz
for the transverse mass spectrum of hadrons should thus contain the
variables ''true temperature'' $T_F$ at decoupling from the flow
field, and its average velocity $\left<\beta_T\right>$, common to all hadrons.
This is the case for the blast wave model \cite{104} developed as an
approximation to the full hydrodynamic formalism
\cite{96}, assuming a common decoupling or ''freeze-out'' from flow,
for all hadronic species, and a boost-invariant longitudinal
expansion:
\begin{equation}
\frac{dN_i}{m_T dm_T dy} = A_i \: m_T \: K_1 \: \left(\frac{m_T cosh
\rho}{T_F} \right) \: I_0 \left( \frac{p_T sinh \rho}{T_F} \right)
\label{eq:equation26}
\end{equation}
where $\rho=tanh ^{-1}\beta_T$. In an extended version of this model
a function is included that describes the radial profile of the
transverse velocity field, $\beta_T(r)=\beta_T^{max} \: r/R$,
instead of employing a fixed $\beta_T$ at decoupling \cite{106}.
Fig.~\ref{fig:Figure24} shows \cite{54} the resulting energy dependence of $T_F$ and
$\left<\beta_T\right>$, for the same set of data as implied already in Figs.~\ref{fig:Figure21}
and~\ref{fig:Figure22}. The ''true'' decoupling temperature rises steeply at the AGS and less so at
SPS energy (as does $\left<\beta_T\right>$), to a value of about $95 \:
MeV$ at top SPS energy, which is considerably lower than the
chemical freeze-out temperature, $T_H = 165 \: MeV$, at which the
hadronic species relative yield composition of the hadronic phase
becomes stationary (see section~\ref{chap:hadronization}, and Fig.~\ref{fig:Figure1}). Chemical decoupling
thus occurs early, near the parton-hadron phase boundary, whereas
hadronic radial flow ceases after significant further expansion and
cooling, whence the surface radial velocity (its average value given
by $\left<\beta_T\right>$ in Fig.~\ref{fig:Figure24}) approaches $\beta_T \approx 0.65$. Both
data sets again exhibit an indication of saturation, over the interval toward top SPS energy: the SPS plateau.
This supports our above conjecture that radial flow is,
predominantly, a consequence of isentropic bulk hadronic expansion
in this energy domain, which sets in at $T_H$. At RHIC energy, both
parameters exhibit a further rise, suggesting that primordial
partonic flow begins to contribute significantly to radial flow.

In summary we conclude that hadronic hadro-chemical freeze-out occurs near the parton-hadron boundary line of Fig.~\ref{fig:Figure1}, at high $\sqrt{s}$. This fixes the hadronic species equilibrium distribution at $T = T_H$. Onward from hadronic freeze-out, the expanding system is out of equilibrium as far as its hadronic species composition is concerned. However the momentum space distributions are subject to further change during the hadronic expansion phase, predominantly due to elastic scattering and development of radial flow. This process ends in final decoupling from strong interaction, at $T_f < T_H$. Analysis of $p_T$ spectra exhibits a collective radial flow pattern at $T_f$, as characterized by a radial flow velocity $\beta_T$.

\section{Hadronization and hadronic freeze-out in A+A collisions}
\label{chap:hadronization}

Within the course of the global expansion of the primordial reaction volume 
the local flow ''cells'' will hit the parton-hadron phase boundary
as their energy density approaches $\epsilon_{crit} \approx 1 \:
GeV/fm^3$.  Hadronization will thus occur, not at an instant over
the entire interaction volume, but within a finite overall time
interval \cite{86} that results from the spread of proper time at
which individual cells, or coherent clusters of such cells (as
developed during expansion) arrive at the phase boundary. However,
irrespective of such a local-temporal occurence, the hadronization
process (which is governed by non perturbative QCD at
the low $Q^2$ corresponding to bulk hadronization) universally
results in a novel, {\it global} equilibrium property that concerns
the relative abundance of produced hadrons and resonances. This
so-called ''hadrochemical equilibrium state'' is directly
observable, in contrast to the stages of primordial parton
equilibration that are only indirectly assessed, via dynamical model
studies.

This equilibrium population of species occurs both in elementary and
nuclear collisions \cite{107}. We have seen in Fig.~\ref{fig:Figure17} a first
illustration, by $e^+e^-$ annihilation data at $\sqrt{s}=91.2 \:
GeV$ LEP energy, that are well reproduced by the partition functions
of the statistical hadronization model (SHM) in its canonical form
\cite{84}. The derived hadronization temperature, $T_H = 165 \:
MeV$, turns out to be universal to all elementary and nuclear
collision processes at $\sqrt{s} \ge 20 \: GeV$, and it agrees with
the limiting temperature predicted by Hagedorn \cite{38} to occur in
any multi-hadronic equilibrium system once the energy density
approaches about $0.6 \: GeV/fm^3$. Thus, the upper limit of
hadronic equilibrium density corresponds, closely, to the lower
limit, $\epsilon_{crit} = 0.6 - 1.0 \: GeV/fm^3$ of partonic
equilibrium matter, according to lattice QCD \cite{48}. In
elementary collisions only about 20 partons or hadrons participate:
there should be no chance to approach thermodynamic equilibrium of
species by rescattering cascades, neither in the partonic nor in the
hadronic phase. The fact that, nevertheless, the hadron formation
temperature $T_H$ coincides with the Hagedorn limiting temperature
and with the QCD confinement temperature, is a consequence of the
non-perturbative QCD hadronization process itself \cite{85}, which
''gives birth'' to hadrons/resonances in canonical equilibrium, at
high $\sqrt{s}$, as we shall see below. This process also governs
A+A collisions but, as it occurs here under conditions of high
energy density extended over considerable volume, the SHM
description now requires a {\it grand} canonical ensemble, with
important consequences for production of strange hadrons
(strangeness enhancement).

The grand canonical order of hadron/resonance production in central
A+A collisions, and its characteristic strangeness enhancement
shows that a state of extended matter that is quantum mechanically
coherent must exist at hadronization \cite{87, 88, 107}. Whether or not it
also reflects partonic equilibrium properties (including flavor
equilibrium), that would allow us to claim the direct observation of
a quark gluon plasma state near $T_c$, can not be decided on the
basis of this observation alone, as the hadronization
process somehow generates, by itself, the observed hadronic
equilibrium. This conclusion, however, is still the subject of
controversy \cite{107}.\\
\begin{figure}
\begin{center}\vspace{-0.8cm}
\includegraphics[scale=0.4]{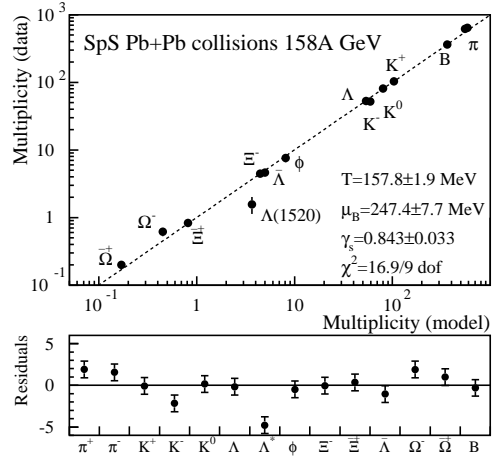}
\caption{Total hadron multiplicities in central Pb+Pb collisions at
$\sqrt{s}=17.3 \: GeV$ \cite{100} versus prediction of the grand
canonical statistical hadronization model \cite{19}.}
\label{fig:Figure25}
\end{center}\vspace{-0.6cm}
\end{figure} \\
Two typical examples of grand canonical SHM application are
illustrated in Figs.~\ref{fig:Figure25} and~\ref{fig:Figure26}, the first showing total hadron
multiplicities in central Pb+Pb collisions at $\sqrt{s}=17.3 \: GeV$
by NA49 \cite{100} confronted with SHM predictions by Becattini et
al.\ \cite{19}.
This plot is similar to Fig.~\ref{fig:Figure17} in which $e^+ e^-$ annihilation to hadrons is confronted with a SHM prediction derived from the \emph{canonical} ensemble \cite{84}.
Central Au+Au collision data at $\sqrt{s}=200 \: GeV$
from several RHIC experiments are compared to grand canonical model
predictions by Braun-Munzinger et al. \cite{108} in Fig.~\ref{fig:Figure26}. The key
model parameters, $T_H$ and the baryo-chemical potential $\mu_B$
result as $159 \: MeV$ ($160 \: MeV)$, and $247 \: MeV$ ($20 \:
MeV)$ at $\sqrt{s}=17.3 \: (200) \: GeV$, respectively. The
universality of the hadronization temperature is obvious from
comparison of the present values with the results of the canonical
procedure employed in $e^+e^-$ annihilation to hadrons at
$\sqrt{s}=91.2 \: GeV$ (Fig.~\ref{fig:Figure17}), and in canonical SHM fits
\cite{109} to p+p collision data at $\sqrt{s}=27.4 \: GeV$ where
$T_H=159$ and $169 \: MeV$, respectively.\\
\begin{figure}
\begin{center}
\includegraphics[scale=0.5]{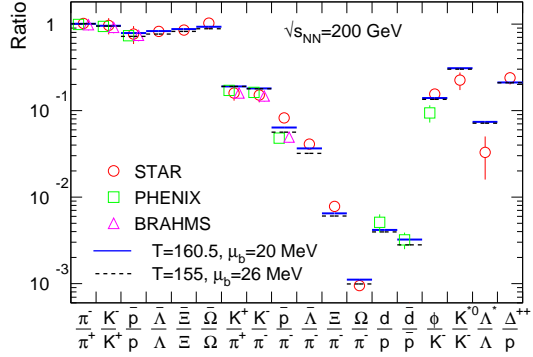}
\caption{Hadron multiplicity ratios at mid-rapidity in central Au+Au
collisions at $\sqrt{s}=200 \: GeV$ from RHIC experiments STAR,
PHENIX and BRAHMS, compared to predictions of the grand canonical
statistical model \cite{108}.}
\label{fig:Figure26}
\end{center}
\end{figure} \\
Figs.~\ref{fig:Figure25} and~\ref{fig:Figure26} illustrate two different approaches 
employed in grand canonical SHM application, the former addressing the values of the hadronic multiplicities as obtained in approximate
full $4 \pi$ acceptance (within limitations implied by detector
performance), the latter employing a set of multiplicity {\it
ratios} obtained in the vicinity of mid-rapidity as invited, at RHIC
energy, by the limited acceptance of the STAR and PHENIX
experiments. The latter approach is appropriate, clearly, in the
limit of boost-invariant rapidity distributions where hadron
production ratios would not depend on the choice of the observational
rapidity interval. We have shown in section~\ref{subsec:Rap_Distribution} that such conditions
do, in fact, set in at top RHIC energy, as referred to in Fig.~\ref{fig:Figure26}.
However, at low $\sqrt{s}$ the y-distributions are far from
boost-invariant, and the total rapidity gap $\Delta y$ may become
comparable, in the extreme case, to the natural rapidity widths of
hadrons emitted in the idealized situation of a single,
isotropically decaying fireball positioned at mid-rapidity. Its
rapidity spectra, equation~\ref{eq:equation5}, resemble Gaussians with widths $\Gamma_i
\approx 2.35 \: (T/m_i)^{1/2}$ for hadron masses $m_i$. Clearly, the
particle ratios $(dN_i/dy)/(dN_j/dy)$ then depend strongly on the
position of the rapidity interval $dy$: away from $y=0$ heavy
hadrons will be strongly suppressed, and particle yields in narrow
rapidity intervals are useless for a statistical model analysis
unless it is known a priori that the radiator is a single stationary
spherical fireball \cite{110}. This is not the case toward top SPS
energy (see Fig.~\ref{fig:Figure10}), due to significant primordial longitudinal
expansion of the hadron emitting source. Given such conditions, the
total multiplicity per collision event (the invariant yield divided
by the total overall inelastic cross section) should be employed in
the SHM analysis, as is exemplified in Fig.~\ref{fig:Figure25}.

\subsection{Hadronic freeze-out from expansion flow}
\label{sec:Hadronic_freeze_out}
The hadronic multiplicities result from integration of the
invariant triple differential cross section over $p_T$ and $y$. Instrumental, 
experiment-specific conditions tend to result in incomplete $p_{T}$ and/or $y$
acceptances. It is important to ascertain that the effects of hydrodynamic transverse
and longitudinal flow do not blast a significant part of the total hadron yield to outside the acceptance,
and that they, more generally, do not change the relative hadron yield composition, thus basically affecting 
the SHM analysis. To see that hadronization incorporates only the internal energy in the co-moving
frame \cite{110}, we first assume that hadrochemical freeze-out
occurs on a sharp hypersurface $\Sigma$, and write the total yield
of particle species $i$ as
\begin{equation}
N_i=\int \frac{d^3p}{E} \int_\Sigma \: p^{\mu}d^3 \sigma_{\mu} (x)\:
f_i (x,p)= \int_{\Sigma}d^3 \sigma_{\mu} (x) j_i^{\mu} (x)
\label{eq:equation27}
\end{equation}
where $d^3 \sigma$ is the outward normal vector on the surface, and
\begin{equation}
j_i^{\mu}(x) = g_i \int d^{4}p 2 \Theta (p^0) \delta(p^2-m^2_i) \:
p^{\mu}(exp \: [p \cdot u(x)-\mu_i]/T \pm 1)^{-1}
\label{eq:equation28}
\end{equation}
is the grand canonical number current density of species $i$, $\mu_i$
the chemical potential, $u(x)$ the local flow velocity, and $g_i$
the degeneracy factor. In thermal equilibrium it is given by
\begin{eqnarray}
j_i^{\mu}(x)& = & \rho_i(x) u^{\mu}(x) \: \: \mbox{with} \nonumber\\
 \rho_i(x) & = & u_{\mu}(x) j_i^{\mu}(x) = \int d^4 p 2 \Theta
(p^0)\delta (p^2-m_i^2) \: p \cdot u(x) \: f_i (p \cdot u (x); T;
\mu_i) \nonumber \\
 & = & \int d^3 p' \: f_i(E_{p'}; T, \mu_i)=\rho_i(T, \mu_i).
\label{eq:equation29}
\end{eqnarray}
Here $E_{p'}$ is the energy in the local rest frame at point $x$.
The total particle yield of species $i$ is therefore
\begin{equation}
N_i=\rho_i (T, \mu_i) \int_\Sigma d^3 \sigma_{\mu}(x) u^{\mu}(x) =
\rho_i (T,\mu_i)\: V_{\Sigma}(u^{\mu})
\label{eq:equation30}
\end{equation}
where only the total comoving volume $V_{\Sigma}$ of the freeze-out
hypersurface $\Sigma$ depends on the flow profile $u^{\mu}$. $V$ is
thus a common total volume factor at hadronization (to be determined separately), and the flow pattern
drops out from the yield distribution over species in $4 \pi$ acceptance \cite{110}.
For nuclear collisions at SPS energies and below one thus should
perform a SHM analysis of the total, $4 \pi$-integrated hadronic
multiplicities, as was done in Fig.~\ref{fig:Figure25}.

We note that the derivation above illustrates the termination
problem of the hydrodynamic description of A+A collisions, the
validity of which depends on conditions of a short mean free path,
$\lambda < 1 \: fm$. A precise argumentation suggests
that two different free paths are relevant here, concerning hadron
occupation number and hadron spectral freeze-out, respectively. As hadrochemical
freeze-out occurs in the immediate vicinity of $T_c$ (and $T_H
\approx 160-165\: MeV$ from Figs.~\ref{fig:Figure25},~\ref{fig:Figure26}), the hadron species
distribution stays constant throughout the ensuing hadronic phase,
i.e the ''chemical'' mean free path abruptly becomes infinite at
$T_H$, whereas elastic and resonant rescattering may well extend far
into the hadronic phase, and so does collective pressure and flow.
In fact we have seen in section~\ref{subsec:Bulk_hadron_transverse_spectra} that the decoupling from flow
occurs at $T_{F}$ as low as $90-100 \: MeV$ (Fig.~\ref{fig:Figure24}). Thus the hydrodynamic
evolution of high $\sqrt{s}$ collisions has to be, somehow
artificially, stopped at the parton-hadron boundary in order to get
the correct hadron multiplicities $N_i$, of equations~\ref{eq:equation27} to~\ref{eq:equation30}, which then stay frozen-out during the subsequent hadronic expansion.

The equations (\ref{eq:equation27}-\ref{eq:equation30}) demonstrate the application of the
Cooper-Frye prescription \cite{111} for termination of the
hydrodynamic evolution. The hyper-surface $\Sigma$ describes the
space-time location at which individual flow cells arrive at the
freeze-out conditions, $\epsilon = \epsilon_c$ and $T=T_c$, of hadronization. At this point, the resulting hadron/resonance spectra (for
species $i$) are then given by the Cooper-Frye formula
\begin{equation}
E \frac{dN_i}{d^3p} = \frac{dN_i}{dy p_Td p_T}  = \frac{g_i}{(2
\pi)^3} \: \int_{\Sigma} f_i (p \cdot u(x),x) p \cdot d^3 \sigma
(x),
\label{eq:equation31}
\end{equation}
where $p^{\mu} f_i d^3 \sigma_{\mu}$ is the local flux of particle
$i$ with momentum $p$ through the surface $\Sigma$. For the phase
space distribution $f$ in this formula one takes the local
equilibrium distribution at hadronic species freeze-out from the grand canonical SHM
\begin{equation}
f_i (E,x) = [exp \{(E_i - \mu_i(x))/T \} \pm 1]^{-1}
\label{eq:equation32}
\end{equation}
boosted with the local flow velocity $u^{\mu}(x)$ to the global
reference frame by the substitution $E \rightarrow p \cdot u(x)$.
Fixing $T=T_c$ (taken e.g. from lattice QCD) the hadron
multiplicities $N_i$ then follow from equation~\ref{eq:equation30}, and one compares to
experiment, as in Figs.~\ref{fig:Figure25},~\ref{fig:Figure26}. In order now to follow the further
evolution, throughout the hadronic rescattering phase, and to
finally compare predictions of equation~\ref{eq:equation31} to the observed flow data as represented by
the various Fourier-terms of equation~\ref{eq:equation21} one has to re-initialize (with
hadronic EOS) the expansion from $\Sigma$($T_{c}$ = 165 MeV) until final decoupling
\cite{96}, at $T \approx 100 \: MeV$, thus describing e.g. radial and elliptic flow.

\begin{figure}
\begin{center}
\includegraphics[scale=0.45]{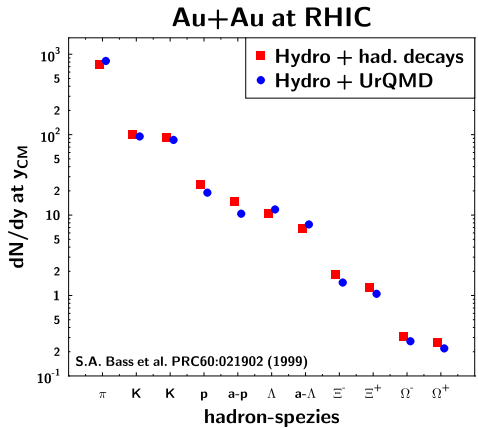}
\caption{Modification of mid-rapidity hadron multiplicities in central Au+Au
collisions at $\sqrt{s}=200 \: GeV$ after chemical freeze-out at
$T=T_c$. Squares show a hydrodynamic model prediction at $T=T_c$
(without further interaction); circles show the result of an
attached UrQMD hadronic cascade expansion calculation \cite{114}.}
\label{fig:Figure27}
\end{center}
\end{figure}

Alternatively, one might end the hydrodynamic description at
$T=T_c$ and match the thus obtained phase space distribution of
equation~\ref{eq:equation31} to a microscopic hadron transport model of the hadronic
expansion phase \cite{95,112}. This procedure is illustrated in
Fig.~\ref{fig:Figure27} by an UrQMD \cite{113} calculation of Bass and Dumitru
\cite{114} for central Au+Au collisions at top RHIC energy. We
select here the results concerning the survival of the hadronic
multiplicities $N_i$ throughout the dynamics of the hadronic
expansion phase, which we have postulated above, based on the
equality of the hadronization temperatures, $T_H \approx 160 \:
MeV$, observed in $e^+e^-$ annihilation (Fig.~\ref{fig:Figure17}), where no hadronic
expansion phase exists, and in central collisions of $A \approx 200$
nuclei (Figs.~\ref{fig:Figure25},~\ref{fig:Figure26}). In fact, Fig.~\ref{fig:Figure27} shows that the $\{N_i\}$
observed at the end of the hadronic cascade evolution agree,
closely, with the initial $\{N_i\}$ as derived from a Cooper-Frye
procedure (equation~\ref{eq:equation30}) directly at hadronization. On the other hand, $p_{T}$ spectra and 
radial flow observables change, drastically, during the hadronic cascade expansion phase. \\
 \\
The hadronic multiplicity distribution  $\{N_i\}$,
arising from the hadronization process at high $\sqrt{s}$,
freezes-out instantaneously also in A+A collisions, and is thus
preserved throughout the (isentropic) hadronic expansion phase. {\it
It is thus directly measurable} and, moreover, its hadrochemical
equilibrium features lend themselves to an analysis within the
framework of Hagedorn-type statistical, grand canonical models. As
we shall show below, the outcome of this analysis is contained in a
$[T_H, \mu_B]$ parameter pair that reflects the conditions of QCD
matter prevailing at hadronization, at each considered $\sqrt{s}$.
In fact, the $[T, \mu]$ points resulting from the SHM analysis
exhibited in Figs.~\ref{fig:Figure25},~\ref{fig:Figure26} (at $\sqrt{s} = 17.3$ and $200 \: GeV$,
respectively) have been shown in the QCD matter phase diagram of Fig.~\ref{fig:Figure1} to
approach, closely, the parton-hadron phase coexistence line
predicted by lattice QCD. Thus, $T_H \approx T_c$ at high
$\sqrt{s}$: hadrochemical freeze-out occurs in the immediate
vicinity of QCD hadronization, thus providing for a location of the QCD phase boundary.

\subsection{Grand canonical strangeness enhancement}
\label{sec:Grand_canonical_strange}
The statistical model analysis \cite{19,107,108} of the
hadronization species distribution ${N_i}$ in A+A collisions is
based on the grand canonical partition function for species $i$,
\begin{equation}
ln Z_i= \frac{g_i V}{6 \pi^2 T} \: \int^{\infty}_0 \:
\frac{k^4dk}{E_i (k) exp \left\{(E_i (k) - \mu_i)/T \right\} \pm 1}
\label{eq:equation33}
\end{equation}
where $E_i^2=k^2+m^2_i$, and $\mu_i \equiv \mu_B B_i + \mu_s S_i +
\mu_I I_3^i$ is the total chemical potential for baryon number $B$,
strangeness $S$ and isospin 3-component $I_3$. Its role in equation~\ref{eq:equation33} is
to enforce, {\it on average} over the entire hadron source volume,
the conservation of these quantum numbers. In fact, making use of
overall strangeness neutrality $(\sum_i N_i S_i = 0)$ as well as of
conserved baryon number (participant Z+N) and isospin (participant
(N-Z)/Z) one can reduce $\mu_i$ to a single effective potential
$\mu_b$. Hadronic freeze-out is thus captured in three parameters,
$T,V$ and $\mu_b$. The density of hadron/resonance species $i$ then
results as
\begin{equation}
n_i=\frac{T}{V} \frac {\delta}{\delta_{\mu}} \: ln Z_i
\label{eq:equation34}
\end{equation}
which gives
\begin{equation}
N_i=Vn_i=\frac{g_i V}{(2 \pi)^2} \: \int^{\infty}_0 \frac{k^2dk}{exp
\left\{(E_i(k)- \mu_i)/T \right\} \pm 1}.
\label{eq:equation35}
\end{equation}

We see that the common freeze-out volume parameter is canceled if one considers hadron
multiplicity ratios, $N_i/N_j$, as was done in Fig.~\ref{fig:Figure26}. Integration
over momentum yields the one-particle function
\begin{equation}
N_i= \frac{VTg_i}{2 \pi^2} \: m_i^2 \: \sum_{n=1}^{\infty} \:
\frac{(\pm 1)^{n+1}}{n} \: K_2 \left(\frac{nm_i}{T}\right) exp
\left( \frac {n \mu_i}{T} \right)
\label{eq:equation36}
\end{equation}
where $K_2$ is the modified Bessel function. At high $T$ the effects
of Bose or Fermi statistics (represented by the $\pm 1$ term in the
denominators of equation~\ref{eq:equation33} and equation~\ref{eq:equation35}) may be ignored
, finally leading to the Boltzmann approximation
\begin{equation}
N_i= \frac{VTgi}{2 \pi^2} \: m_i^2 \: K_2 \left(\frac{m_i}{T}\right)
exp \left( \frac {\mu_i}{T} \right)
\label{eq:equation37}
\end{equation}
which is the first term of equation~\ref{eq:equation36}. This approximation is employed
throughout the SHM analysis. It describes the {\it primary} yield of
hadron species $i$, directly at hadronization. The abundance of
hadronic resonance states is obtained convoluting equation~\ref{eq:equation35} with a
relativistic Breit-Wigner distribution \cite{19}. Finally, the
overall multiplicity, to be compared to the data, is determined as
the sum of the primary multiplicity (cf.\ equation~\ref{eq:equation37}) and the contributions
arising from the unresolved decay of heavier hadrons  and resonances:
\begin{equation}
N_i^{observed}= N_i^{primary} + \sum_j Br (j \rightarrow i) \:N_j.
\label{eq:equation38}
\end{equation}

After having exposed the formal gear of grand canonical ensemble
analysis we note that equation~\ref{eq:equation37} permits a simple, first orientation
concerning the relation of $T$ to $\mu_B$ in A+A collisions by
considering, e.g., the antiproton to proton production ratio. From 
equation~\ref{eq:equation37} we infer the simple expression
\begin{equation}
N (\overline{p})/N(p) = exp (-2 \mu_B/T).
\label{eq:equation39}
\end{equation}
Taking the mid-rapidity value 0.8 for $\overline{p}/p$ (from Fig.~\ref{fig:Figure26})
at top RHIC energy, and assuming that hadronization occurs directly
at the QCD phase boundary, and hence $T \approx T_c \approx 165 \:
MeV$, we get $\mu_B \simeq 18 \: MeV$ from equation~\ref{eq:equation39}, in close agreement
with the result, $\mu_B = 20 \: MeV$, obtained \cite{108} from the
full SHM analysis.
Equation \ref{eq:equation39} illustrates the role played by $\mu_B$ in the grand canonical ensemble. It logarithmically depends on the ratio of newly created quark-antiquark pairs (the latter represented by the $\bar{p}$ yield), to the total number of quarks including the net baryon number-carrying valence quarks (represented by the $p$ yield).
\\
\begin{figure}
\begin{center}
\includegraphics[scale=0.4]{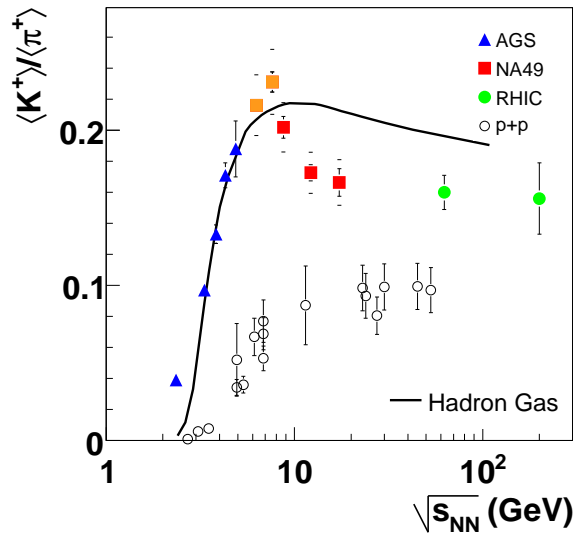}
\caption{The ratio of total $K^+$ to  total $\pi^+$ multiplicity as a
function of $\sqrt{s}$, in central Au+Au and Pb+Pb collisions and in
p+p minimum bias collisions \cite{100}.}
\label{fig:Figure28}
\end{center}
\end{figure} \\
The most outstanding property of the hadronic multiplicities
observed in central A+A collisions is the enhancement of all strange
hadron species, by factors ranging from about 2 to 20, as compared
to the corresponding production rates in elementary hadron-hadron
(and $e^+e^-$ annihilation) reactions at the same $\sqrt{s}$. I.e.
the nuclear collision modifies the relative strangeness output by a
''nuclear modification factor'', $R_s^{AA}=N_s^{AA}/0.5 \:
N_{part}\:  \cdot \: N_s^{pp}$, which depends on $\sqrt{s}$ and
$N_{part}$ and features a hierarchy with regard to the strangeness
number $s=1,2,3$ of the considered species, $R^{AA}_{s=1} <
R^{AA}_{s=2} < R^{AA}_{s=3}$. These properties are illustrated in
Figs.~\ref{fig:Figure28} and~\ref{fig:Figure29}. The former shows the ratio of total $K^+$ to
positive pion multiplicities in central Au+Au/Pb+Pb
collisions, from lower AGS to top RHIC energies, in comparison to
corresponding ratios from minimum bias p+p collisions \cite{100}. We
have chosen this ratio, instead of $\left<K^+\right>/N_{part}$, because it
reflects, rather directly, the ''Wroblewski ratio'' of produced
strange to non-strange quarks \cite{107}, contained in the produced hadrons, 
\begin{equation}
\lambda_s \equiv \frac
{2(\left<s\right>+\left<\overline{s}\right>)}{\left<u\right>+\left<d\right>+\left<\overline{u}\right>+\left<\overline{d}\right>}
\approx \begin{cases} 0.2\ \text{in } \text{pp}  \\ 0.45\ \text{in } \text{AA}. \end{cases}
\label{eq:equation40}
\end{equation}
The low value of $\lambda_s$ in $pp$ (and all other elementary)
collisions reflects a quark population far away from $u,d,s$ flavor
equilibrium, indicating {\it strangeness suppression} \cite{109}.\\
\begin{figure}
\begin{center}
\includegraphics[scale=0.53]{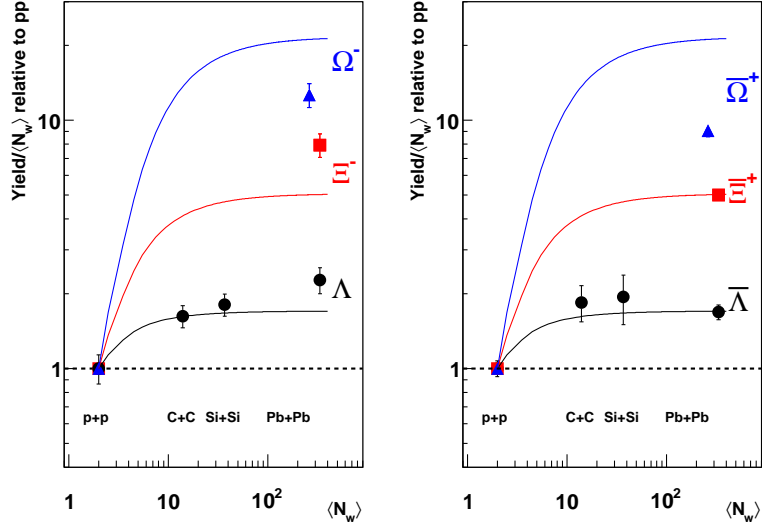}\vspace{-0.3cm}
\caption{The nuclear modification factors $R^{AA}_{s=1,2,3}$ for hyperon and
anti-hyperon production in nucleus-nucleus collisions at
$\sqrt{s}=17.3 \: GeV$, relative to the p+p reference at the same
energy scaled by $N_W(=N_{part})$. The NA49 data refer to total $4
\pi$ yields \cite{116}. Closed lines represent the inverse
strangeness suppression factors from ref. \cite{119}, at
this energy.}
\label{fig:Figure29}
\end{center}
\end{figure} \\
The so-called {\it strangeness enhancement} property of A+A
collisions (obvious from Figs.~\ref{fig:Figure28} and \ref{fig:Figure29}) is, thus, seen as the removal of
strangeness suppression; it is also referred to as a {\it
strangeness saturation}, in SHM analysis \cite{107,108}, for the
reason that $\lambda_s \approx 0.45$ corresponds to the grand
canonical limit of strangeness production, implicit in the analysis
illustrated in Figs.~\ref{fig:Figure25} and~\ref{fig:Figure26}. The average $R^{AA}_{s=1}$ at
$\sqrt{s} \ge 10 \: GeV$ thus is about 2.2, both in the data of
Fig.~\ref{fig:Figure28} and in the statistical model. It increases (Fig.~\ref{fig:Figure29}) toward about 10 in
s = 3 production of $\Omega$ hyperons. \\ \\
In order to provide for a first guidance concerning the above facts
and terminology regarding strangeness production we propose an
extremely naive argument, based on the empirical fact of a universal
hadronization temperature (Figs.~\ref{fig:Figure17},~\ref{fig:Figure25},~\ref{fig:Figure26}) at high $\sqrt{s}$. Noting that $\left<s\right>=\left<\overline{s}\right>$ and $\left<u\right> \approx \left<\overline{u}\right> \approx\left<d\right>
\approx \left<\overline{d}\right>$ in a QGP system at $\mu_b$ near zero, and
$T=165 \: MeV$, just prior to hadronization, $\lambda_s \approx
\left<s\right>/\left<u\right> \approx exp \{(m_u - m_s)/T\} = 0.45$  at $p_T \rightarrow
0$ if we take current quark masses, $m_s - m_u \approx 130 \: MeV$.
I.e. the value of $\lambda_s$ in A+A collisions at high $\sqrt{s}$
resembles that of a grand canonical QGP at $\mu_b \rightarrow 0$, as
was indeed shown in a 3 flavor lattice QCD calculation \cite{115} at
$T \approx T_c$. On the contrary, a p+p collision features no QGP
but a small fireball volume, at $T \approx T_c$, within which local
strangeness neutrality, $\left<s\right>=\left<\overline{s}\right>$ has to be strictly
enforced, implying a canonical treatment \cite{109}. In our naive
model the exponential penalty factor thus contains twice the strangeness quark
mass in the exponent, $\lambda_s \:  \mbox {in pp collisions}
\approx exp \{2(m_u-m_s)/T \} \approx 0.2$, in agreement with the
observations concerning strangeness suppression, which are thus
referred to as canonical suppression. In a further extension of our
toy model, now ignoring the $u,d$ masses in comparison to $m_s
\approx 135 \: MeV$, we can estimate the hierarchy of hyperon
enhancement in A+A collisions,
\begin{equation}
R^{AA}_s \propto N_s^{AA}/N_s^{pp} \cdot 0.5 \ N_{part} \approx exp \{(-sm_s + 2sm_s)/T
\} = 2.2, \: 5.1, \:  11.6
\label{eq:equation41}
\end{equation}
for $s=1,2,3$, respectively. Fig.~\ref{fig:Figure29} shows that these estimates
correspond well with the data \cite{116} for $R^{AA}$ derived in $4
\pi$ acceptance for $\Lambda,\: \Xi \: \:  \mbox{and} \:  \: \Omega$
as well as for their antiparticles, from central Pb+Pb collisions at
$\sqrt{s}=17.3 \: GeV$. The p+p reference data, and C+C, Si+Si
central collisions (obtained by fragmentation of the SPS Pb beam)
refer to separate NA49 measurements at the same energy.

The above, qualitative considerations suggest that the relative
strangeness yields reflect a transition concerning the fireball
volume (that is formed in the course of a preceding dynamical
evolution) once it enters hadronization. Within the small volumes,
featured by elementary collisions (see section~\ref{sec:Origin_of_hadro}), phase space is
severely reduced by the requirement of {\it local} quantum number
conservation \cite{109,117} including, in particular, local
strangeness neutrality. These constraints are seen to be removed
in A+A collisions, in which extended volumes of high primordial
energy density are formed. Entering the hadronization stage, after
an evolution of expansive cooling, these extended volumes will decay
to hadrons under conditions of global quantum mechanical coherence,
resulting in quantum number conservation occuring, non-locally, and
{\it on average} over the entire decaying volume. This large
coherent volume decay mode removes the restrictions, implied by
local quantum number balancing. In the above, naive model we have
thus assumed that the hadronization of an Omega hyperon in A+A
collisions faces the phase space penalty factor of only three $s$
quarks to be gathered, the corresponding three $\overline{s}$ quarks
being taken care of elsewhere in the extended volume by global
strangeness conservation. In the framework of the SHM this situation
is represented by the grand canonical ensemble (equations~\ref{eq:equation35},~\ref{eq:equation37}); the
global chemical potential $\mu_b$ expresses quantum number
conservation {\it on average}. Strict, local conservation is
represented by the canonical ensemble.

The grand canonical (GC) situation can be shown to be the large
collision volume limit (with high multiplicities $\{N_i\}$) of the
canonical (C) formulation \cite{118,119}, with a continuous
transition concerning the degree of canonical strangeness
suppression \cite{119}. To see this one starts from a system that is
already in the GC limit with respect to baryon number and charge
conservation whereas strangeness is treated canonically. Restricting
to $s=1 \: \: \mbox {and} \: \: -1$ the GC strange particle
densities can be written (from equation~\ref{eq:equation37}) as
\begin{equation}
n_{s=\pm 1}^{GC} = \frac {Z_{s=\pm 1}}{V} \: \lambda ^{\pm 1}_s
\label{eq:equation42}
\end{equation}
with
\begin{equation}
Z_{s=\pm 1}= \frac {Vg_s}{2 \pi^2} \: m_s^2 \: K_2 \:
(\frac{m_s}{T}) \: exp \Big\{(B_s \mu_B + Q_s \mu_Q)/T \Big\}
\label{eq:equation43}
\end{equation}
and a ''fugacity factor'' $\lambda_s^{\pm 1}= exp \: (\mu_s/T)$. The
canonical strange particle density can be written as \cite{119}
\begin{equation}
n_s^C = n_s^{GC} \cdot (\tilde{\lambda_s})
\label{eq:equation44}
\end{equation}
with an effective fugacity factor
\begin{equation}
\tilde{\lambda_s} \: = \: \frac{S_{\pm 1}}{\sqrt{S_1 S_{-1}}} \:
\frac{I_1(x)}{I_0(x)}
\label{eq:equation45}
\end{equation}
where $S_{\pm 1}=\sum_{s=\pm 1} \: Z_{s=\pm1}$ is the sum over all
created hadrons and resonances with $s=\pm 1$, the $I_n(x)$ are
modified Bessel functions, and $x=2 \sqrt{S_1 S_{-1}}$ is
proportional to the total fireball volume $V$. In the limit $x
\approx V \rightarrow \infty$ the suppression factor $I_1(x)/I_0(x)
\rightarrow 1 $, and the ratio $S_{\pm1}/\sqrt{S_1 S_{-1}}$
corresponds exactly to the fugacity $\lambda_s$ in the GC
formulation (see equation~\ref{eq:equation42}). Thus the C and GC formulations are equivalent in
this limit, and the canonical strangeness suppression effect
disappears. Upon generalization to the complete strange hadron
spectrum, with $s=\pm 1, \pm 2, \pm 3,$ the strangeness suppression
factor results \cite{119} as
\begin{equation}
\eta(s) = I_s(x)/I_0(x).
\label{eq:equation46}
\end{equation}

In particular for small $x$ (volume), $\eta(s) \rightarrow (x/2)^s$,
and one expects that the larger the strangeness content of the
particle the smaller the suppression factor, and hence the larger
the enhancement in going from elementary to central A+A collisions.
This explains the hierarchy addressed in equation~\ref{eq:equation41}, and apparent from
the data shown in Fig.~\ref{fig:Figure29}. In fact, the curves shown in this figure
represent the results obtained from equation~\ref{eq:equation46}, for $s=1,2,3$ hyperon
production at $\sqrt{s}=17.3 \: GeV$ \cite{119}. They are seen to be
in qualitative agreement with the data. However the scarcity
of data, existing at top SPS energy for total hyperon yields,
obtained in $4 \pi$ acceptance (recall the arguments in section~\ref{sec:Hadronic_freeze_out})
both for A+A and p+p collisions does not yet permit to cover
the SHM strangeness saturation curves in detail, for $s>1$.

This saturation is seen in Fig.~\ref{fig:Figure29}, to set in already at modest
system sizes, but sequentially so, for ascending hyperon
strangeness. Note that SHM saturation is sequentially approached, from equation~\ref{eq:equation46},
with increasing fireball {\it volume} $V$. In order to make contact
to the experimental size scaling with centrality, e.g. $N_{part}$,
the model  of ref. \cite{119}, which is illustrated in Fig.~\ref{fig:Figure29}, has converted the genuine volume
scale to the $N_{part}$ scale by assuming a universal eigenvolume of
$7 \: fm^3$ per participant nucleon. I.e. $N_{part}=10$ really means
a coherent fireball volume of $70 \: fm^3$, in Fig.~\ref{fig:Figure29}. Within this
definition, saturation of $s=1,2,3$ sets in at fireball volumes at
hadronization of about 60, 240 and 600 $fm^3$, respectively: this is
the {\it real} message of the SHM curves in Fig.~\ref{fig:Figure29}.\\
\begin{figure}
\begin{center}
\includegraphics[scale=1.05]{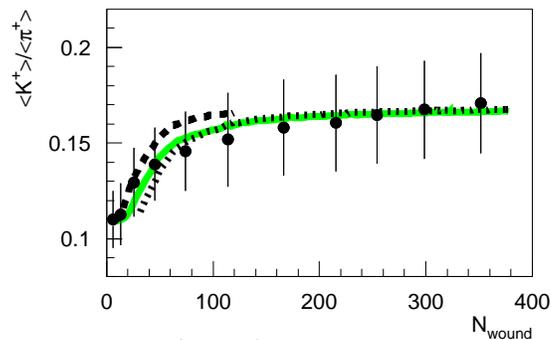}\vspace{-0.3cm}
\caption{The mid-rapidity $K^+$ to $\pi^+$ ratio vs. $N_{part}$ in minimum
bias Au+Au collisions at $\sqrt{s}=200 \: GeV$, compared to the
percolation model \cite{120} (solid line); a prediction of which for
Cu+Cu at similar energy is given by the long dashed line (see text
for detail).}
\label{fig:Figure30}
\end{center}
\end{figure} \\
The above direct translation of coherent fireball volume to
participant number is problematic \cite{120} as it assumes that all
participating nucleons enter into a single primordially coherent
fireball. This is, however, not the case \cite{120} particularly in
the relative small scattering systems that cover the initial, steep
increase of $\eta(s)$, where several local high density clusters are
formed, each containing a fraction of $N_{part}$. This is revealed
by a percolation model \cite{120} of cluster overlapp attached to a Glauber calculation
of the collision/energy density. At each $N_{part}$ an average
cluster volume distribution results which can be transformed by equation~\ref{eq:equation46}
to an average $\{\eta(s,V)\}$ distribution whose weighted mean is
the appropriate effective canonical suppression factor corresponding
to $N_{part}$. On the latter scale, the SHM suppression curve thus
shifts to higher $N_{part}$, as is shown in Fig.~\ref{fig:Figure30} for the $K^+/\pi^+$ ratio
vs. $N_{part}$, measured at mid-rapidity by  PHENIX in Au+Au
collisions at $\sqrt{s}=200 \: GeV$, which is reproduced by the
percolation model \cite{120}. Also included is a prediction for
Cu+Cu at this energy which rises more steeply on the common
$N_{part}$ scale because the collision and energy density reached in
central Cu+Cu collisions, at $N_{part} \approx 100$, exceeds that in
peripheral Au+Au collisions (at the same $N_{part}$) which share a
more prominent contribution from the dilute surface regions of the
nuclear density profile. We note, finally, that this deviation from
 universal $N_{part}$ scaling does not contradict the observations
of a perfect such scaling as far as overall charged particle
multiplicity densities are concerned (recall Fig.~\ref{fig:Figure12}) which are
dominated by pions, not subject to size dependent canonical
suppression.

\subsection{Origin of hadro-chemical equilibrium}
\label{sec:Origin_of_hadro}
The statistical hadronization model (SHM) is
{\it not} a model of the QCD confinement process leading to hadrons,
which occurs once the dynamical cooling evolution of the system
arrives at $T_c$. At this stage the partonic reaction volume, small
in elementary collisions but extended in A+A collisions, will decay
(by whatever elementary QCD process) to on-shell hadrons and
resonances. This coherent quantum mechanical decay results in a
de-coherent quasi-classical, primordial on-shell hadron-resonance
population which, at the instant of its formation, lends itself to a
quasi-classical Gibbs ensemble description. Its detailed modalities
(canonical for small decaying systems, grand canonical for extended
fireballs in A+A collisions), and its derived parameters [$T,
\mu_B$] merely recast the conditions, prevailing at hadronization.
The success of SHM analysis thus implies that the QCD hadronization
process ends in statistical equilibrium concerning the
hadron-resonance species population.

\begin{figure}
\begin{center}
\includegraphics[scale=0.3]{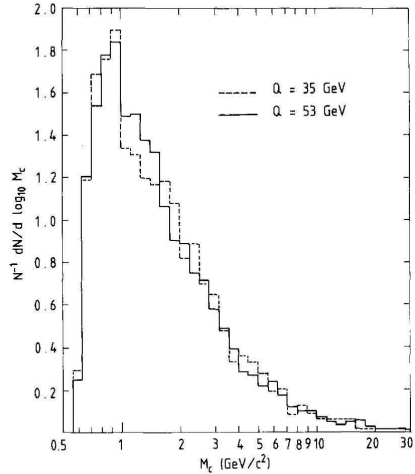}\vspace{-0.3cm}
\caption{Invariant mass spectrum of color neutralization clusters in the
Veneziano-Webber hadronization model \cite{83,121}.}
\label{fig:Figure31}
\end{center}
\end{figure}

In order to identify mechanisms in QCD hadronization that
introduce the hadro-chemical equilibrium we refer to jet
hadronization in $e^+e^-$ annihilation reactions, which we showed in
Fig.~\ref{fig:Figure17} to be well described by the canonical SHM. In di-jet
formation at LEP energy, $\sqrt{s}=92 \: GeV$, we find a charged
particle multiplicity of about 10 per jet, and we estimate that,
likewise, about 10 primordial partons participate on either side of
the back-to-back di-jet \cite{85}. There is thus no chance for
either a partonic or hadronic, extensive rescattering toward
chemical equilibrium. However, in the jet hadronization models
developed by Amati and Veneziano \cite{83}, Webber \cite{121} and
Ellis and Geiger \cite{85} the period of QCD DGLAP parton shower
evolution (and of perturbative QCD, in general) ends with local
color neutralization, by formation of spatial partonic singlet clusters.
This QCD ''color pre-confinement'' \cite{83} process reminds of a
coalescence mechanism, in which the momenta and the initial virtual
masses of the individual clustering partons get converted to
internal, invariant virtual mass of color neutral, {\it spatially
extended} objects. Their mass spectrum \cite{121} extends from about
0.5 to 10 $GeV$. This cluster mass distribution, shown in Fig.~\ref{fig:Figure31}, represents the first
stochastic element in this hadronization model.\\
 \\
The clusters are then re-interpreted within non-perturbative QCD:
their internal, initially perturbative QCD vacuum energy gets
replaced by non-perturbative quark and gluon condensates, making the
clusters appear like hadronic resonances. Their subsequent quantum
mechanical decay to on-shell hadrons is governed by the phase space
weights given by the hadron and resonance spectrum \cite{85,121}.
I.e. the clusters decay under ''phase space dominance'' \cite{85},
the outcome being a micro-canonical or a canonical hadron and
resonance ensemble \cite{84,107}. The apparent hadro-chemical
equilibrium thus is the consequence of QCD color neutralization to
clusters, and their quantum mechanical decay under local quantum
number conservation and phase space weights. We note that the
alternative description of hadronization, by string decay
\cite{122}, contains a quantum mechanical tunneling
mechanism, leading to a similar phase space dominance \cite{123}.

Hadronization in $e^+e^-$ annihilation thus occurs from local
clusters (or strings), isolated in vacuum, of different mass but
similar energy density corresponding to QCD confinement. These
clusters are boosted with respect to each other but it was shown
\cite{124} that for a Lorentz invariant scalar, such as
multiplicity, the contributions of each cluster (at similar $T$) can
be represented by a single canonical system with volume equal to the
sum of clusters. In the fit of Fig.~\ref{fig:Figure17} this volume sum amounts to
about 25 $fm^3$ \cite{84}; the individual cluster volumes are thus
quite small, of magnitude a few $fm^3$ \cite{85}. This implies
maximum canonical strangeness suppression but may, in fact, require
a micro-canonical treatment of strangeness \cite{109}, implying a
further suppression. These MC effects are oftentimes included
\cite{125} in the canonical partition functions by an extra
strangeness fugacity parameter $\gamma_s < 1$ which suppresses
$s=1,2,3$ in a hierarchical manner, $\left<N_i(s)\right> \approx
(\gamma_s)^{s_i}$. The fit of Fig.~\ref{fig:Figure17} requires $\gamma_s=0.66$, a
value typical of canonical multiplicity analysis in $p+p, \:
p+\overline{p}$ and $e^+e^-$ annihilation collisions \cite{109} at
$\sqrt{s} \ge 30 \: GeV$.

The above picture, of hadrochemical equilibrium resulting from the
combined stochastic features of QCD color neutralization by cluster
formation, and subsequent quantum mechanical decay to the on-shell
hadron and resonance spectrum (under phase space governance) lends
itself to a straight forward extension to A+A collisions. The
essential new features, of grand canonical hadronization including
strangeness enhancement, should result from the fact that extended
space-time volumes of $\epsilon > \epsilon_{crit}$ are formed in the
course of primordial partonic shower evolution, an overlap effect increasing
both with $\sqrt{s}$ and with the size of the primordial interaction
volume. As the volume of the elementary hadronization clusters
amounts to several $fm^3$ it is inevitable that the clusters
coalesce, to form extended ''super-cluster'' volumes prior to
hadronization \cite{120}. As these super-clusters develop toward
hadronization via non perturbative QCD dynamics, it is {\it
plausible} to assume an overall quantum mechanical coherence to
arise over the entire extended volume, which will thus decay to
hadrons under global quantum number conservation, the decay products
thus modeled by the GC ensemble.

Our expectation that space-time coalescence of individual hadronization clusters
will lead to a global, quantum mechanically coherent extended
super-cluster volume, that decays under phase space dominance, appears as an analogy to the dynamics and
quantum mechanics governing low energy nuclear fission from a preceding ``compound nucleus"~\cite{126}.
Note that the observation of a smooth transition from
canonical strangeness suppression to grand canonical saturation
(Figs.~\ref{fig:Figure29},~\ref{fig:Figure30}) lends further support to the above picture of a
percolative growth \cite{120} of the volume that is about to undergo
hadronization. 

An extended, coherent quark gluon plasma state would,
of course, represent an ideal example of such a volume \cite{127}
and, in fact, we could imagine that the spatial extension of
the plasma state results from a percolative overlap of primordial
zones of high energy density, which becomes more prominent with
increasing $\sqrt{s}$ and $N_{part}$. A QGP state preceding
hadronization will thus lead to all the observed features. However,
to be precise: the hadronizing QCD system of extended matter
decaying quantum coherently, could still be a non-equilibrium
precursor of the ideal equilibrium QGP, because we have seen above
that hadrochemical equilibrium also occurs in $e^+e^-$ annihilation,
where no partonic equilibrium exists. It gets established in the
course of hadronization, irrespective of the degree of equilibrium
prevailing in the preceding partonic phase.

\subsection{Hadronization vs. rapidity and $\sqrt{s}$}
\label{sec:Hadronization_vs_rap}
We have argued in section~\ref{sec:Hadronic_freeze_out} that, at relatively low $\sqrt{s}$, the
total rapidity gap $\Delta y$ does not significantly exceed the
natural thermal rapidity spreading width $\Gamma_i \approx 2.35 \:
(\Gamma/m_i)^{1/2}$ of a single, isotropically decaying fireball,
centered at mid-rapidity and emitting hadrons of mass $m_i$
\cite{110}.
However, this procedure involves an idealization
because in the real Pb+Pb collision the  intersecting dilute surface
sections of the nuclear density profiles will lead to a significant
contribution of single-scattering NN collisions, outside the central
high density fireball. The leading hadron properties of such
''corona collisions'' result in wider proper rapidity distributions,
quite different from those of the central fireball decay hadrons.
Their contribution will thus be prominent near target/projectile
rapidity, and will feature a canonically suppressed strangeness. The
one-fireball assumption, although inevitable at small $\Delta y$,
does not quite resemble the physical reality. This may explain the
need for an additional strangeness suppression factor in the GC
one-particle partition function (equation~\ref{eq:equation33}) that has, unfortunately,
also been labeled $\gamma_s$ but expresses physics reasons quite
different from the extra suppression factor that reflects
micro-canonical phase space constraints in elementary collisions. It turns out that all GC
analysis of central A+A collisions at low $\sqrt{s}$, and addressed
to total $4 \pi$ multiplicities, requires a $\gamma_s$ of 0.7 - 0.85
\cite{19}; in the fit of Fig.~\ref{fig:Figure25} $\gamma_s=0.84$.

\begin{figure}
\begin{center}
\includegraphics[scale=0.55]{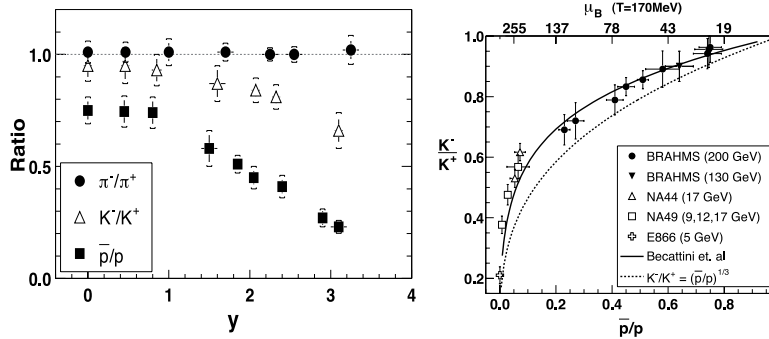}\vspace{-0.3cm}
\caption{(left) Anti-hadron to hadron ratios as a function of rapidity in
central Au+Au collisions at $\sqrt{s}=200 \: GeV$. The right panel
shows an interpretation of the correlation between $\overline{p}/p$
and $K^-/K^+$ in terms of baryo-chemical potential $\mu_B$ variation
in the grand canonical statistical model. From \cite{128}.}
\label{fig:Figure32}
\end{center}
\end{figure}

At RHIC, $\Delta y \approx 11 \gg \Gamma_i$, and such difficulties
disappear: $\gamma_s \approx 1$ at midrapidity and, moreover, the wide gap permits
a SHM analysis which is differential in $y$. Fig.~\ref{fig:Figure32} shows the
$y$-dependence of the ratios $\pi^-/\pi^+, \: K^-/K^+$ and
$\overline{p}/p$ as obtained by BRAHMS \cite{128} in central Au+Au
collisions at $\sqrt{s}=200 \: GeV$. The figure shows a dramatic
dependence of the $\overline{p}/p$ ratio, which reflects the local
baryochemical potential according to equation~\ref{eq:equation39}. At $y_{CM} > 1$ the
$\overline{p}/p$ ratio drops down steeply, to about 0.2 at $y
\approx 3.5$, thus making close contact to the top SPS energy value
obtained by NA49 \cite{129}. The $K^-/K^+$ ratio follows a similar
but weaker drop-off pattern, to about 0.65 again matching with the
top SPS energy value of about 0.6 \cite{130}. The deviation from
unity of these ratios reflects the rapidity densities of initial
valence $u,d$ quarks, relative to the densities of newly created
light and strange quark-antiquark pairs, i.e. the $y$ distribution
of the net baryon number density, and of the related baryo-chemical
potential of the GC ensemble. Thus, in analyzing successive bins of
the rapidity distributions in Fig.~\ref{fig:Figure32}, the major variation in the GC
fit concerns the baryo-chemical potential $\mu_B$(y) which increases
from about 20 $MeV$ (Fig.~\ref{fig:Figure26}) at mid-rapidity, to about 150 $MeV$ at
$y \ge 3$ while the hadronization temperature stays constant, at
$T=160 \: MeV$. This interplay between $K^-/K^+, \: \overline{p}/p$
and $\mu_B$ is illustrated \cite{128} in the right hand panel of
Fig.~\ref{fig:Figure32}, and shown to be well accounted for by the GC statistical
model \cite{131}.\\
 \\
These considerations imply that hadronization at RHIC (and LHC)
energy occurs {\it local} in $y$-space and {\it late} in time. The density distribution of net baryon
number results from the primordial pQCD shower evolution (c.f.
section~\ref{subsec:Gluon_Satu_in_AA_Coll}), and is thus fixed at formation time, $t_0 \le 0.5 \:
fm/c$ at RHIC. Hadronization of the bulk partonic matter occurs
later, at $t \ge 3 \: fm/c$ \cite{86,95}, and transmits the local
conditions in rapidity space by preserving the local net baryon quantum number
density. Most importantly we conclude that hadronization occurs, not
from a single longitudinally boosted fireball but from a succession
of ''super-clusters'', of different partonic composition depending
on $y$, and decaying at different time due to the Lorentz-boost that
increases with $y$, in an ''inside-outside'' pattern (c.f. Fig.~\ref{lightcone}).
We are thus whitnessing at hadronization a Hubble expanding system
of local fireballs. The detailed implications of this picture have
not been analyzed yet. Note that a central RHIC collision
thus does not correspond to a single hadronization ''point'' in the
[$T, \: \mu$] plane of Fig.~\ref{fig:Figure1} but samples $\{T,\: \mu\}$ along the
QCD parton-hadron coexistence line \cite{132}.

\begin{figure}
\begin{center}
\includegraphics[scale=0.35]{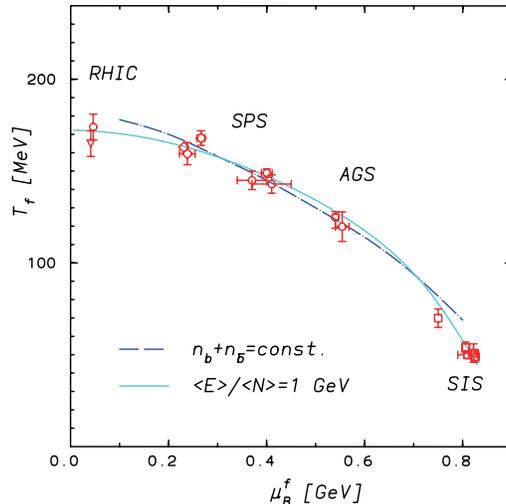}\vspace{-0.3cm}
\caption{Energy dependence of the hadro-chemical freeze-out points obtained
by grand canonical statistical model analysis in the plane [$T, \:
\mu_B$], with interpolating curve at fixed energy per particle of
about 1 $GeV$ \cite{107,139}.}
\label{fig:Figure33}
\end{center}
\end{figure}

Throughout this section we have discussed hadronic freeze-out at
high $\sqrt{s}$ only (top SPS to RHIC energy), because of the
proximity of the chemical freeze-out parameters [$T, \: \mu_b$] to
the QCD phase boundary from lattice QCD, which suggests an overall
picture of hadronization, to occur directly from a partonic cluster
or super-cluster. Our discussion of the GC statistical hadronization
model has been explicitly or implicitly based on the assumption that hadronic 
freeze-out coincides with hadronization. However, the GC model has also been applied successfully to
hadro-chemical freeze-out at $\sqrt{s}$ down to a few $GeV$
\cite{19,107,108} where it is not expected that the dynamical
evolution traverses the phase boundary at all, but grand canonical
multiplicity distributions, and their characteristic strangeness
enhancement pattern, are observed throughout. Toward lower
$\sqrt{s}, \: T$ decreases while $\mu_b$ increases, as is shown in
Fig.~\ref{fig:Figure33} which presents a compilation of all reported freeze-out
parameters \cite{108}.\\
 \\
These points have also been included in the phase diagram of Fig.~\ref{fig:Figure1}
which shows that they are gradually branching away from the phase
separation boundary line that could recently be predicted by lattice
QCD once new methods had been developed to extend the theory to
finite $\mu_B$ \cite{9,10}. At $\sqrt{s} \ge 20 \: GeV$ we see that
\begin{equation}
\epsilon_c (QCD) \approx \epsilon_H \approx \epsilon_{GC}
\label{eq:equation47}
\end{equation}
where $\epsilon_{GC}$ is the freeze-out density inferred from GC
analysis \cite{19,107,108}.

In turn, the GC hadronic freeze-out points drop below the lattice
QCD coexistence line at lower $\sqrt{s}$, implying that chemical
freeze-out now occurs within the hadronic expansion phase. This
requires a model of freeze-out, now governed by the properties of a
high density hadronic medium, upon expansive cooling and dilution.
Holding on to the model of a quantum mechanical de-coherence decay
to on-shell hadrons that we discussed in section~\ref{sec:Origin_of_hadro}, we argue that an
initial, extended high density hadronic fireball, given sufficient
life-time at $T$ smaller, but not far below $T_c$, could also be
seen as a quantum mechanically coherent super-cluster, as governed
by effective mean fields \cite{133}. In such a medium hadrons, at
$T$ near $T_c$, acquire effective masses and/or decay widths far off
their  corresponding properties in vacuum: they are off-shell,
approaching conditions of QCD chiral symmetry restoration as $T
\rightarrow T_c$ \cite{134}. This symmetry is inherent in the
elementary QCD Langrangian, and ''softly'' broken within the light
quark sector by the small non-zero current quark masses, but
severely broken at $T \rightarrow 0$ by the high effective
constituent quark masses that get dressed by non perturbative QCD
vacuum condensates. Pictorially speaking, hadrons gradually loose
this dressing as $T \rightarrow T_c$ \cite{135}, introducing a
change, away from in vacuum properties, in the hadronic mass and
width spectrum. Such in-medium chiral restoration effects have, in
fact, been observed in relativistic A+A collisions, by means of
reconstructing the in-medium decay of the $\rho$ vector meson to an
observed $e^+e^-$ pair \cite{136}.

A dense, high $T$ hadronic system, with mean-field
induced off-shell constituents is also, clearly, quantum
mechanically coherent. At a certain characteristic density,
$\epsilon < \epsilon_c$, and temperature $T<T_c$, as reached in the
course of overall hadronic expansion, this extended medium will
undergo a decoherence transition to classical on-shell hadrons. Its
frozen-out hadronic multiplicity distribution should be, again,
characterized by the phase space weights of a grand canonical
ensemble at $T<T_c$. Theoretical studies of such a mean field
hadronic expansion mode \cite{137} have also shown that such
mechanisms play essentially no role at $\sqrt{s} \ge 20 \: GeV$
because the expanding system is already in rapid flow once it
traverses the phase boundary, with an expansion time scale shorter
than the formation time scale of mean field phenomena. At lower
energies, on the other hand, the system might not even dive into the
deconfined phase but spend a comparatively long time in its direct
vicinity, at the turning point between compression and re-expansion
where all dynamical time constants are large, and the hadron density
is high, such that the inelastic hadronic transmutation rate becomes
high (particularly in collisions of more than two hadronic
reactants, with reaction rates \cite{138} proportional to
$\epsilon^n$), and sufficiently so for maintaining hadronic chemical
equilibrium after it is first established at maximum hadron density, in low $\sqrt{s}$
systems that do not cross the phase boundary at all.

The GC freeze-out parameters [$T, \mu$] at various $\sqrt{s}$ in
Fig.~\ref{fig:Figure33} permit a smooth interpolation in the $T, \mu$ plane
\cite{139}, which, in turn, allows for GC model predictions which
are continuous in $\sqrt{s}$. Such a curve is shown in Fig.~\ref{fig:Figure28}
compared to the $4 \pi$ data points for the $K^+/\pi^+$ multiplicity
ratio in central collisions Au+Au/Pb+Pb, at all $\sqrt{s}$
investigated thus far. It exhibits a smooth maximum, due to the
interplay of $T$ saturation and $\mu_B$ fall-off to zero, but does
not account for the sharp peak structure seen in the data at
$\sqrt{s} \approx 7 \: GeV$ and $\mu_B \approx 480 \: MeV$. This
behavior is not a pecularity of the $K^+$ channel only; it also is
reflected in an unusually high Wroblewski ratio (see equation~\ref{eq:equation40}) obtained
at $\sqrt{s} = 7.6 \: GeV$, of  $\lambda_s = 0.60$ \cite{19}. This
sharp strangeness maximum is unexplained as of yet. It implies
that hadron formation at this $\sqrt{s}$ reflects influences that
are less prominent above and below, and most attempts to understand
the effect \cite{141,142,143} are centered at the assumption that at
this particular $\sqrt{s}$ the overall bulk dynamics will settle
directly at the phase boundary where, moreover, finite $\mu_B$
lattice theory also expects a QCD critical point \cite{9,10,11}.
This would cause a softest point to occur in the equation of state,
i.e. a minimum in the relation of expansion pressure vs. energy
density, slowing down the dynamical evolution \cite{144,145}, and
thus increasing the sensitivity to expansion modes characteristic of
a first order phase transition \cite{143}, which occurs at $\mu_b \ge
\mu_b^{crit}$. Such conditions may modify the K/$\pi$ ratio (Fig.~\ref{fig:Figure28}) \cite{143}.

It thus appears that the interval from top AGS to lower SPS energy,
$5 \le \sqrt{s} \le 10 \: GeV$, promises highly interesting
information regarding the QCD phase diagram (Fig.~\ref{fig:Figure1}) in the direct
vicinity of the parton-hadron coexistence line. In particular, the
physics of a critical point of QCD matter deserves further study.
Observable consequences also comprise so-called ''critical
fluctuations'' \cite{146,147} of multiplicity density, mean
transverse momentum and hadron-chemical composition \cite{148}, the
latter in fact being observed near $\sqrt{s}=7 \: GeV$ in an event
by event study of the $K/\pi$ ratio in central Pb+Pb collisions
\cite{149}.

%
%

%
%



\printindex
\end{document}